\documentclass{aa} 
\usepackage{graphicx}
\usepackage{txfonts}
\usepackage{longtable,lscape}
 \usepackage{natbib}
\bibpunct{(}{)}{;}{a}{}{,} 
\begin{document}

\title{Evolution of Cluster Red-Sequence Galaxies
 from   redshift  0.8   to   0.4:   ages, metallicities       and
 morphologies\thanks{Based  on observations obtained   at the ESO Very
 Large Telescope (VLT) as part of the  Large Programme 166.A-0162 (the
 ESO Distant Cluster Survey).}}          

\author{P. S\'anchez-Bl\'azquez\inst{1,2} 
\and Pascale Jablonka\inst{2,3,4} 
\and Stefan Noll\inst{5,6} 
\and Bianca M. Poggianti\inst{7}
\and John Moustakas \inst{8} 
\and Bo Milvang-Jensen\inst{9,10}
\and Claire  Halliday\inst{11}  
\and Alfonso  Arag\'on-Salamanca\inst{12} 
\and Roberto P.    Saglia\inst{5} 
\and Vandana Desai\inst{13}
\and Gabriella De Lucia\inst{14}
\and Douglas  I.    Clowe\inst{15}
\and Roser Pell\'o\inst{16}       
\and Gregory Rudnick\inst{17}   
\and Luc Simard\inst{18}      
\and Simon   D.M.  White\inst{14}  
\and  Dennis Zaritsky\inst{19}}

\institute{Center for Astrophysics, University of Central Lancashire, 
   Preston, PR1 2HE, UK
\and
Laboratoire d'Astrophysique,  Ecole   Polytechnique  F\'ed\'erale de
Lausanne (EPFL), Observatoire de Sauverny CH-1290 Versoix, Switzerland
\and
Universit\'e de  Gen\`eve, Observatoire  de Sauverny CH-1290  Versoix,
Switzerland
\and
GEPI, CNRS-UMR8111, Observatoire de  Paris, section de Meudon, 5 Place
Jules Janssen, F-92195 Meudon Cedex, France
\and
Max-Planck-Institut f\"ur extraterrestrische Physik,
Giessenbachstrasse, D-85748 Garching bei M\"unchen, Germany
\and
Observatoire Astronomique de Marseille-Provence, 38 rue Fr\'ed\'eric
Joliot-Curie,
13388 Marseille cedex 13, France
\and
Osservatorio Astronomico di Padova, Vicolo dell'Osservatorio 5, 35122
Padova, Italy
\and
Center for Cosmology and Particle Physics, New York University, 4 Washington Place, New York, NY 10003
\and
Dark Cosmology Centre, Niels Bohr Institute, University of Copenhagen,
Juliane Maries Vej 30, DK-2100 Copenhagen, Denmark
\and
The Royal Library / Copenhagen University Library, Research Dept., Box 2149, DK--1016 Copenhagen K, Denmark
\and
INAF, Osservatorio Astrofisico di Arcetri, Largo E. Fermi 5, I-50125
Firenze, Italy
\and
School of Physics and Astronomy, University of Nottingham, University
Park, Nottingham NG7 2RD, UK
\and
{\it Spitzer} Science Center, Caltech, Pasadena CA 91125, USA 
\and
Max-Planck-Institut f\"ur Astrophysik, Karl-Schwarzschild-Strasse 1,
D-85748 Garching bei M\"unchen, Germany 
\and
Ohio University, Department of Physics and Astronomy, Clippinger Labs
251B, Athens, OH 45701, USA
\and
Laboratoire d'Astrophysique de Toulouse-Tarbes, CNRS, Universit\'e de Toulouse, 14
Avenue Edouard Belin, F--31400 Toulouse, France
\and
NOAO, 950 N. Cherry Avenue, Tucson, AZ 85719, USA
\and
Herzberg Institute of Astrophysics, National Research Council of
Canada, 5071 West Saanich Road, Victoria, Canada BC V9E 2E7
\and
Steward Observatory, University of Arizona, 933 North Cherry Avenue,
Tucson, AZ 85721, USA
} 
            
\date{ }
\abstract{

We present a comprehensive analysis of the stellar population
properties (age, metallicity and the alpha-element enhancement [E/Fe])
and morphologies of red-sequence galaxies in 24 clusters and groups
from $z\sim0.75$ to $z\sim0.45$.  The dataset, consisting of 215
spectra drawn from the ESO Distant Cluster Survey, constitutes the
largest spectroscopic sample at these redshifts for which such an
analysis has been conducted.  Analysis reveals that the evolution of the stellar
population properties of red-sequence galaxies depend on their mass: 
while the properties of most massive are well described by
passive evolution and high-redshift formation, the less massive
galaxies require a more extended star formation history.  We show that
these scenarios reproduce the index-$\sigma$ relations as well as the
galaxy colours.   The two main results of this work are (1)
the evolution of the line-strength
indices for the red-sequence galaxies can be reproduced if 
40\% of the galaxies with $\sigma<$ 175
kms$^{-1}$ entered the red-sequence between $z=0.75$ to $z=0.45$, in
agreement with the fraction derived in studies of the luminosity
functions, and   
 (2) the percentage the red-sequence galaxies exhibiting early-type
morphologies (E and S0) {\it decreases}
by 20\% from $z=0.75$ to $z=0.45$. This can be understood
if the red-sequence gets populated at later times with disc galaxies
whose star formation has been  quenched.
We conclude that the processes quenching star
formation do not necessarily produce a simultaneous morphological
transformation of the galaxies entering the red-sequence.
 }

\titlerunning{Evolution of Cluster Red-Sequence Galaxies}

\maketitle

\section{Introduction}
Early-type galaxies are strongly clustered \citep[e.g.,][]{Lov95,
  Her96, Will98, Shep01}, making galaxy clusters an ideal place for
their study.  
Stellar population studies have shown that both, local and high-redshift 
early-type galaxies follow  tight correlations between the colors and line-strength
features and other properties of the galaxies  (mostly related with their masses)
\citep[e.g.][]{Bow92, K00, T00b, Ber05, SB06a}
The mere existency of these correlations, as well as their evolution with redshift,
seem to be  consistent with a very early (z$\ge$2) and coordinated formation  
of their stars \citep{VS77, Bow92,
  Ell97,KA98, SED98, Kel01, Ben98, Zie01}. 
 However, the evolution of the
cluster red-sequence luminosity function with redshift challenges this
view (De Lucia et al.\ 2004; De Lucia et al.\ 2007; Kodama et
al.\ 2004, Rudnick et al. 2008, in preparation; although see
\citet{And08} for contradictory results), and so does the
morphological evolution since $z=0.4$ \citep[e.g.,][]{Dres97}, and the
evolution of the blue/star-forming fraction (Butcher \& Oemler 1984;
Poggianti et al.\ 2006) in clusters.

A deeper understanding of the nature of the evolution of the cluster
red-sequence requires to go beyond colors (affected by the age-metallicity degeneracy) 
and derive the stellar
populations parameters (age and chemical abundances) with time, using stellar population models.  
However, breaking the age-metallicity
degeneracy requires a combination of indices with different
sensitivities to both parameters (see, e.g., Rabin 1982)\nocite{Rab82}.  
\footnote{Combination of optical and near-IR colours have also been probed useful
to break the age-metallicity degeneracy (e.g., Peletier, Valentijn \& Jameson 1990;
MacArthur et al. 2004; James  et al. 2005, among others), although dust reddening is
still a problem.}
Unfortunately, instrumental limitations have
long prevented accurate measurements of absorption line indices at
high redshift.  Recent observational advances, on the ground and
in space, now allow such measurements in intermediate and high
redshift galaxies, both in clusters and in the field \citep[][]{Zie01,
  Barr05, Sch06, J05}.  The analysis of these data have revealed that, indeed, the
age-metallicity degeneracy is confusing the interpretation of the
scaling relations: large spreads in galaxy luminosity weighted ages
and metallicities at high redshift have been found in datasets showing
very tight Faber-Jackson, Mgb-$\sigma$ and Fundamental plane (FP)
relations, challenging the classical interpretation of the tightness
of the scaling relations.

High redshift spectroscopic samples  are, however, still restricted to a
maximum of  $\sim$30 galaxies.  Moreover,  they often target  a single
cluster \citep[][]{Kel01,J05, Tran07, Kel06}.  This could be biasing the
results, as studies at low redshift have shown that the star formation
histories of early-type galaxies might depend on cluster properties
\citep[e.g., compare][]{KD98, CRC03, SB03, J99, Nel05, T08}.  
Therefore,  a study based on a large sample of galaxies in 
clusters covering a large range in   cluster masses is  necessary to obtain
a complete picture of galaxy evolution.

The present work is based on 24 clusters with  
redshifts between 0.39 and 0.8 from the ESO Distant Cluster Survey
(hereafter, EDisCS). The major novelty of the present work is that we
do not only consider very massive structures; our sample spans a large
range in cluster velocity dispersions.  Therefore, we minimize
possible biases due to the relationship between galaxy properties and
environment.  

It is now common to
study red-sequence galaxies rather than morphologically classified
early-type galaxies, as 
colours are easier to measure in large datasets than
morphology, and they appear to be  more correlated with environment
\citep[][]{Kauff04, Blan05, MM06}.  Nonetheless, observations have also revealed the
existence of an intrinsic spread in morphology at given colour
\citep[e.g.,][]{Con06, Bal04, Cross04}. This stress the need for
investigating the relation between morphology and colour, and their
evolution with redshift and, therefore, this is the strategy we adopt.

 Our sample of
red-sequence galaxies is much larger than those used in previous
efforts.  It encompasses 337 galaxies, distributed in 24 clusters and
groups, for which we study stellar populations and morphologies.
Our sample of red-sequence galaxies is not restricted to the most massive, 
but span a  wide range of internal velocity dipsersion (100-350 km/s),
comparable to the samples analysed  at low-redshift.

Throughout the paper, we adopt a concordance cosmology with $\Omega_{\rm
M}=0.3$, $\Omega_{\Lambda}=0.7$, H$_0$=70~km~s$^{-1}$ Mpc$^{-1}$.  All
magnitudes are quoted in the Vega system.

\section{The sample}
\label{sec.sample}
The ESO Distant Cluster Survey
(EDisCS) is a photometric and spectroscopic survey of galaxies
in 20 fields containing clusters with redshifts between 0.39 and 0.96.
These fields were selected from the Las Campanas Distant Cluster Survey
\citep{Gon01}, specifically from  the 30 highest surface
brightness candidates. A full description  of the sample selection can
be found in  \citet{White05}.  EDisCS  includes structures
with velocity dispersions from $\sim$150 to $\sim$1100~km~s$^{-1}$,
i.e. from small groups to clusters.

Deep   optical   photometry with VLT/FORS2   \citep{White05},  near-IR
photometry  with  SOFI on the ESO/NTT   (Arag\'on-Salamanca et al., in
preparation), and   deep-multi-slit  spectroscopy with  VLT/FORS2 were
acquired for each field.  The same high-efficiency grism was used
in all observing runs  (grism  600RI+19, $\lambda_0$ = 6780\AA~).  
The wavelength range varies with the field and 
the x-location of the slit on the mask \citep[see][for details]{MJ08}, but 
it was chosen to cover, at least, a rest-frame wavelength range 
from 3670 to 4150~\AA~(in order to include [OII] and H$\delta$ lines)
for the assumed cluster redshift.

The  spectral resolution is $\sim$6 \AA~(FWHM), corresponding to
rest-frame 3.3  \AA~ at  $z=0.8$  and 4.3 \AA~at  $z=0.4$.  Typically,
four-  and  two-hour  exposures were  obtained  for the  high$-z$  and
mid$-z$ samples, respectively. The
spectroscopic selection,  observations, data reduction and  catalogs are presented
in detail in \citet{Hall04} and \citet{MJ08}. 
In brief, standard reduction procedures (bias subtraction, flat fielding, cosmic-ray removal,
geometrical distortion corrections, wavelength calibration, sky subtraction and flux calibration)
were performed with IRAF. Particular attention was paid to the sky-subtraction that was performed
before any interpolation or rebinning of the data \citep[see][for details]{MJ08}.
This dataset has also been complemented with 80 orbits of HST/ACS imaging in F814W of the highest
redshift clusters \citep{Des07}, H$\alpha$ narrow-band imaging
\citep{Finn05}, and XMM-Newton/EPIC X-ray observations \citep{Jon06}.

 26 structures (groups or clusters) were identified in the 20 EDisCS fields 
\citep{Hall04, MJ08}.  Two of these structures 
were not considered in this paper. The first 
one, cl1103.7-1245,
because its spectroscopic redshift (0.96) is too far away from the
redshift targeted by the photo-z-based selection (0.70). This can 
introduce observational biases.
The second one, cl1238.5-1144, because it could only be observed
for 20 minutes and the S/N of the spectra is not sufficient for our analysis.

To define the red-sequence, we use total-I magnitudes, estimated
by adding, to the Kron magnitudes,  a correction appropriate
for a point source measured within an aperture equal to the galaxy's Kron 
aperture. These corrections are obtained empirically from 
unsaturated and isolated stars on each convolved I-band image
(see White et al.\ 2005 for details).

\begin{figure*}
\resizebox{\textwidth}{!}{\includegraphics[angle=0.]{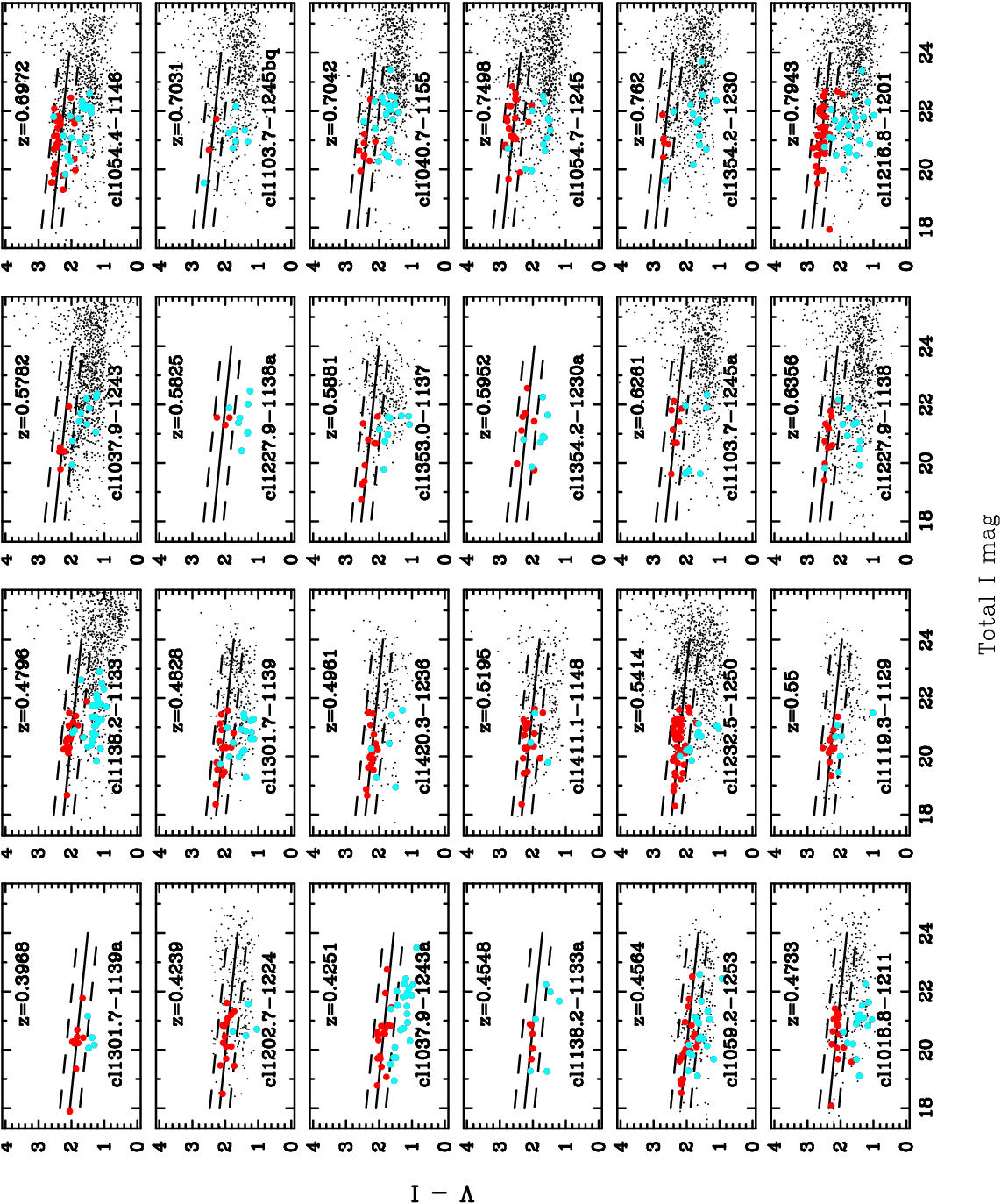}}
\caption{$V$, $I$ colour-magnitude diagrams for the 24 EDisCS
structures considered in this study.  Galaxies with no- or weak-
emission are represented by red circles, while galaxies with strong
emission are marked as blue circles.  The fit, with a fixed slope of
$-0.09$, of the secure cluster members with no or only weak emission
is shown as a solid line.  The dashed lines trace the $\pm$ 0.3 mag
boundary taken into account to select the red-sequence galaxies.  This
corresponds to the measured dispersions of cl1232.5-1250 and
cl1216.8-1201, our two most populous galaxy clusters.  When photo-z
membership is available, we show the photo-z selected cluster galaxies
as small black dots.}
\label{colo-mag}
\end{figure*}

A galaxy is considered a member of a given cluster (or group) if its
redshift falls within 3$\times \sigma_{\rm cluster}$ of the cluster
redshift, where $\sigma_{\rm cluster}$ is the cluster velocity
dispersion presented in \citet{Hall04} and \citet{MJ08}.  Similarly to
all other EDisCS analyses, galaxy groups are identified by
$\sigma_{\rm cluster} < 400 $ km/s.  We define red-sequence galaxies
as those with {\it secure} redshifts, and colours between $\pm$0.3~mag
from the best linear fit (with slope fixed to $-0.09$) to the colour-magnitude
relation (V-I vs. I) of the  objects without emission lines. This 
definition coincides with that of   \citet{White05} and De Lucia et al.\ (2004,
2007).  The width of the region corresponds to $\sim$3 times the
RMS-dispersion of the red-sequence colours of our two most populated
clusters, cl1216.8-1201 and cl1232.5-1250. 
Table \ref{tb:clusterlist} lists the structures selected for the present work
and the total number of galaxies on the red-sequence.
 Figure~\ref{colo-mag}
displays the $V-I$ vs $I$ colour-magnitude diagrams of the 24 EDisCS
structures studied in the present paper.

\begin{figure}
\resizebox{0.5\textwidth}{!}{\includegraphics[angle=-90]{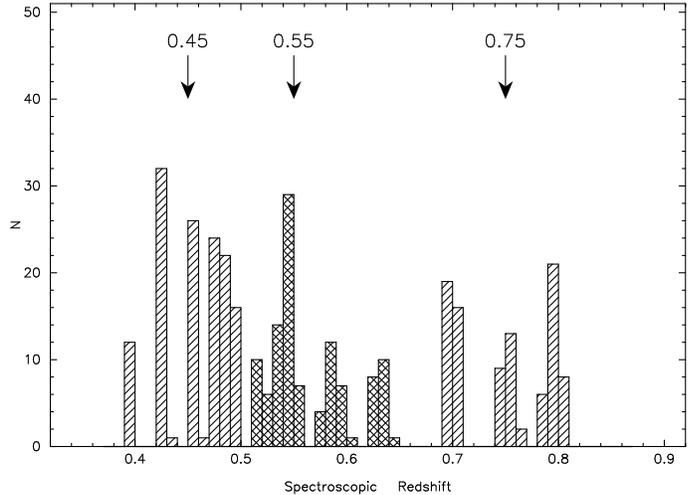}}
\caption{Redshift distribution of the red-sequence galaxies in 
the 3 redshifts-bins  with no or negligible emission lines, before
any signal-to-noise selection.}
\label{Groups}
\end{figure}

  To study the evolution of the galaxy population properties with redshift,
 we divide our galaxies in 3 different redshift intervals : (1) $0.39 < z < 0.5$; (2)
$ 0.5 \leq z < 0.65$; and (3) $0.65 \leq z < 0.81$.  For convenience, we
refer to these intervals by their median redshifts, $z=0.45$,
$z=0.55$, and $z=0.75$.  Figure~\ref{Groups} shows the redshift
distribution of our sample of galaxies.  

\begin{table*}[h]
\begin{tabular}{clcrrrcr}
Group & Cluster Name       &    z       & ACS & N$_{reds}$ & N$_{N+W}$ & $\sigma\pm err$ & R$_{200}$\\
      &                    &            &     &           &           &  (kms$^{-1}$)   & Mpc     \\
\hline
\hline
0.45 &cl1018.2-1211  &   0.4734  &   & 17  &  9 & 486$^{+59}_{-63}$  & 0.93\\
     &cl1037.9-1243a &   0.4252  & x & 18  &  9 & 537$^{+46}_{-48}$  & 1.06\\
     &cl1059.2-1253  &   0.4564  &   & 26  & 18 & 510$^{+52}_{-56}$  & 0.99 \\
     &cl1138.2-1133  &   0.4796  & x & 17  & 8  & 732$^{+72}_{-76}$  & 1.40 \\
     &cl1138.2-1133a &   0.4548  & x &  8  & 3  & 542$^{+63}_{-71}$  & 1.05 \\
     &cl1202.7-1224  &   0.4240  &   & 18  & 7  & 518$^{+92}_{-104}$ & 1.02 \\
     &cl1301.7-1139  &   0.4828  &   & 19  & 10 & 687$^{+81}_{-86}$  & 1.31 \\
     &cl1301.7-1139a &   0.3969  &   & 13  & 6  & 391$^{+63}_{-69}$  & 0.78 \\
     &cl1420.3-1236  &   0.4962  &   & 18  & 13 & 218$^{+43}_{-50}$  & 0.41 \\
\hline
0.55 &cl1037.9-1243  &   0.5783  &  x &  7  & 5  & 319$^{+53}_{-52}$  & 0.57\\
     &cl1103.7-1245a &   0.6261  &  x &  9  & 6  & 336$^{+36}_{-40}$  & 0.59 \\
     &cl1119.3-1129  &   0.5500  &    & 17  &  9 & 166$^{+27}_{-29}$  & 0.30 \\
     &cl1227.9-1138  &   0.6357  &  x & 14  &  8 & 574$^{+72}_{-75}$  & 1.00 \\
     &cl1227.9-1138a &   0.5826  &  x &  4  &  1 & 341$^{+42}_{-46}$  & 0.61 \\
     &cl1232.5-1250  &   0.5414  &  x & 41  & 20 &1080$^{+119}_{-89}$ & 1.99 \\
     &cl1353.0-1137  &   0.5882  &    & 10  & 7  & 666$^{+136}_{-139}$& 1.19 \\
     &cl1354.2-1230a &   0.5952  &  x &  8  & 5  & 433$^{+95}_{-104}$ & 0.77 \\
     &cl1411.1-1148  &   0.5195  &   & 18  & 12 & 710$^{+125}_{-133}$& 1.32 \\
\hline
 0.75&cl1040.7-1155  &   0.7043  & x & 14  & 8  & 418$^{+55}_{-46}$  & 0.70 \\
     &cl1054.4-1146  &   0.6972  & x & 36  & 16 & 589$^{+78}_{-70}$  & 1.06 \\
     &cl1054.7-1245  &   0.7498  & x & 20  & 6  & 504$^{+113}_{-65}$ & 0.70\\
     &cl1103.7-1245b &   0.7031  & x &  3  & 2  & 252$^{+65}_{-85}$  & 0.42 \\
     &cl1216.8-1201  &   0.7943  & x & 38  & 23 &1018$^{+73}_{-77}$  & 1.61 \\
     &cl1354.2-1230  &   0.7620  & x &  7  & 5  & 648$^{+105}_{-110}$& 1.05 \\

\hline
\end{tabular}
\caption{Col. (1) Redshift interval in which the galaxy structure falls
(see text for details). 
Col. (2): Cluster name.  Following \citet{MJ08}, we  label the secondary
structures  with letters "a"  and "b".   Col.   (3): Cluster  redshift.
Col. (4): marked with x if ACS images are
available.  Col.(5):  Number of secure spectroscopically confirmed red-sequence 
galaxies.  Col.  (6) Number of red-sequence galaxies with no or
negligible  emission, full-filling our signal-to-noise ratio criterion.
Col.  (7) Velocity  dispersion of the  cluster.   Col.  (8) R$_{200}$,
calculated as in \citet{Finn05}.}
\label{tb:clusterlist}
\end{table*}

\begin{figure*}
\resizebox{0.8\textwidth}{!}{\includegraphics[angle=-90]{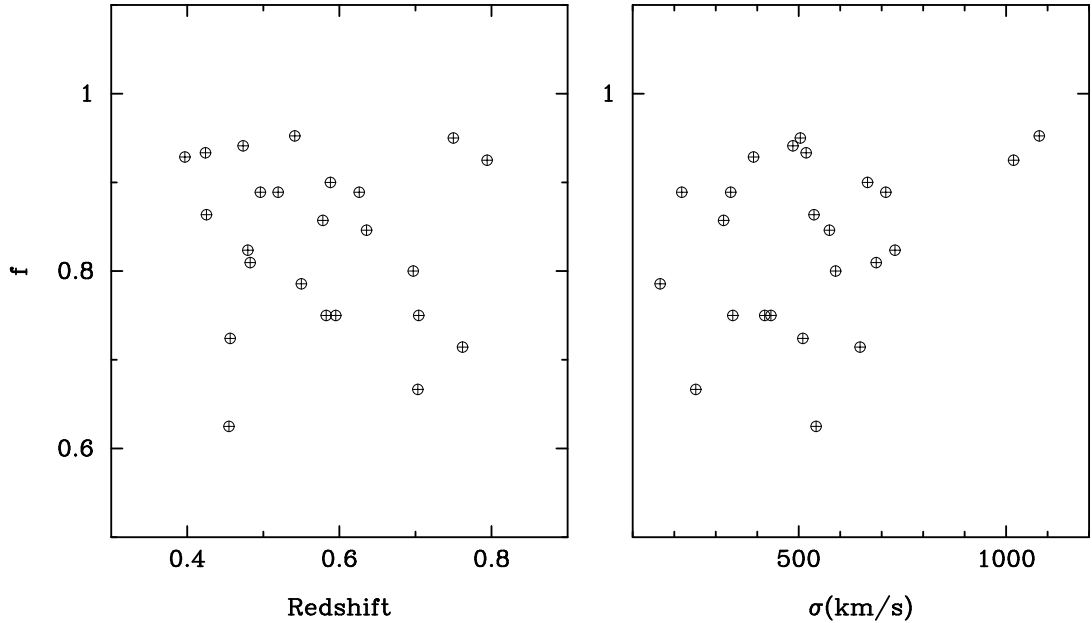}}

\caption{Fraction of red-sequence galaxies that have no or only weak
emission lines (EW[OII] $<$7$\AA$) as a function of redshift and
group/cluster velocity dispersion. This fraction is calculated as the
ratio between the N+W red-sequence galaxies and the total number of
red-sequence galaxies in the spectroscopic sample.}
\label{STypes}
\end{figure*}

 Dust reddened star-forming galaxies 
are estimated to make up $\sim$20\% of galaxies in low-redshift
clusters \citep{Strat01, Gav02, Bell04, Fran07}.  This contamination
is likely to be more important in the strong-emission line regime 
To avoid contamination from these galaxies that do not have red stellar populations,
and due to the difficulty of measuring reliable absorption line indices
in galaxies with emission lines, we discard, for our absorption line
analysis,  galaxies showing
evidence of emission lines. 
 Our selection is based both on visual
inspection of the 2D- spectra and on measurements of the line
equivalent widths (EWs).  These measurements were taken from
\citet{Pog06}. We also measured the [OII] and H$\beta$ emission-line
EWs after carefully subtracting the underlying stellar continuum (see
Moustakas \& Kennicutt \citeyear{2006ApJS..164...81M} for details).
After comparing the visual inspection with the quantitative measurements, 
we keep galaxies with equivalent widths of
[OII]$\lambda$3727 $  < 7$\AA.  
 In Fig.~\ref{colo-mag} we distinguish between galaxies with strong
emission lines and those with no- or weak- emission lines (N+W,
hereafter) in the spectroscopic sample of red-sequence galaxies.
Figure~\ref{STypes} shows the fraction of N+W red-sequence galaxies as
a function of redshift and structure velocity dispersion.  The scatter
is large and there is no trend with either $\sigma_{\rm cluster}$
or redshift.  On average, 24\% of the red-sequence galaxies show
EW[OII]3727$>$7\AA.  Section~\ref{sec.emiss} discusses the nature of
the emission in these galaxies. 

 We also remove the brightest cluster
galaxies (BCGs) from the sample, as their position in the cluster may
lead to a different evolution from the rest of the red population (De
Lucia \& Blaizot 2007)\nocite{dLB07}.  A photometric study of the BCGs
in the EDisCS clusters is presented in Whiley et
al.\ (2008)\nocite{Whiley08}.

Our  initial  sample  includes a  total of   337  N+W red-sequence
galaxies.  Their mean spectral  signal-to-noise ratio per \AA~ (S/N), measured
between 4000 and 4500  \AA~ (rest-frame) is $\sim$ 17,  19 and  12 per
\AA~ at redshift  0.45,   0.55 and 0.75, respectively.   We  
consider all  galaxies independently of their  distance to the cluster
centers except for the morphological analysis. The radial cuts, when 
applied, will be stated
explicitly. However,  to measure line-strength indices,  we require a  minimum  signal-to-noise ratio  of  10 per
\AA~, measured between 4000  and  4500\AA~ (rest-frame). Thie leave us, for our spectroscopic analysis, with a 
total of 215  galaxies  satisfying our  selection  criteria, i.e, 70\% of the initial sample (see Table
\ref{tb:clusterlist} for their distribution). 


\section{Source of ionization for the emission line red-sequence galaxies}
\label{sec.emiss}

As inferred from Fig.~\ref{STypes},  a non negligible fraction  of
red-sequence galaxies exhibit  emission lines, with equivalent widths
reaching   $\sim$30\AA~  for   [OII] and  $\sim$12\AA~   for H$\beta$.
Although these galaxies are discarded  from further investigation,  we
are  interested in trying to unveil their  nature.   We investigate below the
various possible ionization sources and  look for evidence of
evolution with redshift.

We perform a similar analysis than  \citet{Yan06}.
\citet{Yan06} studied 
a large sample of red galaxies from the 
Sloan Digital Sky Survey (SDSS) finding emission in 52.2\% 
of them. They  showed that they could be classified in two
main groups: galaxies with  high [OII]$\lambda 3727\AA$/H$\alpha$
(LINER-like   and  quiescent   galaxies, representing 20.6\% of the 
total red populution with emission) and low-[OII]/H$\alpha$ galaxies 
(mostly  dusty-starforming ones, with a small fraction of Seyferts, accounting 
for 9\% of the red-sequence galaxies with emission).  
  Counting galaxies
with  only  [OII]   seen  in  emission (instead    of both [OII]   and
H$\alpha$) increases the fraction of LINERS to 28.8\%.  Finally, 14\%
of the red galaxies have H$\alpha$ detected  but no [OII], making them
difficult to classify. Nevertheless, Yan et al.\  (2006) show that the
majority  of  these galaxies  have  [NII]/H$\alpha >  0.6$ suggesting a
non-starforming origin for their emission lines.

\begin{figure}
\resizebox{0.5\textwidth}{!}{\includegraphics[angle=-90]{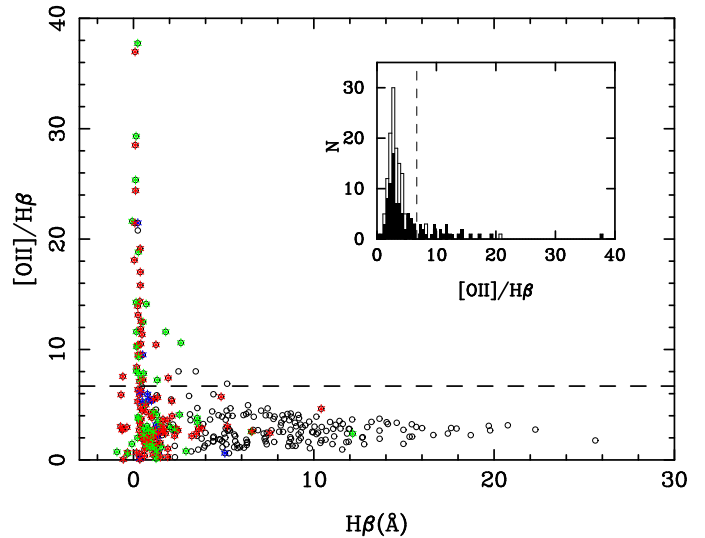}}
\caption{Ratio between [OII] and H$\beta$ for the full EDisCS sample
of cluster members (black dots).  Red, green, and blue stars indicate
the red-sequence galaxies in the 0.45, 0.55, and 0.75 redshift bins,
respectively. The[OII]/H$\beta$ distribution is also presented in the
form of a histogram: the black-filled zone represents the red-sequence
galaxies, to be compared to the full sample. Dashed lines indicate the
[OII]/H$\beta$ = 6.7 limit above which the galaxy emission is 
considered to be due to mechanisms other than star formation.}
\label{OIIoverHbeta}
\end{figure}

In the present case, the determination of the ionization sources is
hampered by the fact that the wavelength coverage of the spectra does
not include H$\alpha$.  We consider H$\beta$ instead.  For each
galaxy, the stellar continuum has been fitted and the [OII] and
H$\beta$ equivalent widths subsequently measured.
Figure~\ref{OIIoverHbeta} illustrates the distribution of [OII] and
H$\beta$ EWs for the red-sequence galaxies and for  the  full EDisCS spectroscopic sample.

Out of the EDisCS full sample of red-sequence galaxies (393), 243 have
spectra with a wavelength range appropriately centered and wide enough
to include both [OII]$\lambda\lambda 3727$\AA~ and H$\beta$. The
redshift distribution of these galaxies has a mean of 0.52 and a
standard deviation of 0.09.  We count 57 galaxies with a 2-$\sigma$
detection (threshold identical to \citet{Yan06}) of both [OII] and
H$\beta$, 68 with only one of the two lines detected, and 118 with no
detection at all (at the 2-$\sigma$ level).  Hence, very similarly to
\citet{Yan06}, we obtain 51.4\% of emission line red-sequence
galaxies.

 Following \citet{Kew04}, we consider that star formation cannot
 induce [OII]/H$\alpha$ larger than 1.5. This value is derived before
 any reddening correction and was independently confirmed by
 \citet{2006ApJS..164...81M}.  We assume a mean value of
 H$\alpha$/H$\beta$=4.46, which was derived by \citet{Yan06} before
 any reddening correction, and which corresponds to a median
 extinction of A$_{\rm V}$=1.40, assuming Type B recombination.  Under
 this condition, [OII]/H$\beta$= 6.7 draws a boundary between star
 forming galaxies and those with emission powered by an AGN.  As shown
 in figure~11 of \citet{Yan06}, this limit is very conservative.
 Indeed, one hardly finds any star forming galaxies with
 [OII]/H$\beta$ $>$ 5.  However, it does eliminate most Seyferts and
 selects mostly LINER-type galaxies.  Interestingly, the limit of 6.7
 seems to represent a 
 natural upper limit for the bulk of the whole cluster galaxy
 population (see the histogram part of Figure~\ref{OIIoverHbeta}).
 For our statistics below, we use the empirical bimodal demarcation
 advised by \citet{Yan06}, i.e., EW([OII])=18EW(H$\beta$)$-$6 rather
 than a constant [OII]/H$\beta$ ratio. We have checked, however, that
 both methods lead to identical results (within 1-2\%).

Table~\ref{emission.lines} shows the emission-line properties of the 
red-sequence galaxies with spectra wide enough to cover H$\beta$ and [OII].  To go beyond upper and lower limits we would
need, at least, another set of two lines to build more diagnostic
diagrams (e.g., Kewley et~al.\ 2006). However, we believe that our counts 
give reliable hints on the major trends for the ionization processes
at play in our sample.  The fraction of quiescent red galaxies is very
much the same as the percentage reported by \citet{Yan06} in the local
Universe.  The obvious difference between the two studies resides in
the nature of the galaxies which do have detections of both H$\beta$
and [OII].  While \citet{Yan06} find that only 9\% of those galaxies
have low-[OII]/H$\beta$ ratios, we obtain more than twice this percentage, 19\%.  
In addition, only 4\% of
our red-sequence galaxies have high-[OII]/H$\beta$ ratios
characteristic of LINERS.  This fraction might increases up to 8\% if we
include galaxies with [OII] detected and H$\beta$ undetected. However, this 
is still a factor of 3 smaller than 
the fraction of high-[OII]/H$\beta$ galaxies in the red-sequence reported
by \citet{Yan06}, 29\%.

The conclusion for this tentative identification of the ionizing
sources in the EDisCS emission line red-sequence galaxies is that most
of them are dusty star forming ones.  We can not disentangle, at this
stage, whether time or environment, or even more technically the use
of H$\beta$ instead of H$\alpha$, leads to the difference with the
results of \citet{Yan06}.  Further investigation, both in field
intermediate redshift galaxies and in low redshift cluster galaxies,
is required to shed light on this matter.

\begin{table}
\centering
\begin{tabular}{ll|lr}
[OII]  & H$\beta$ & [OII]/H$\beta$ & Fraction \\
\hline
N      & N        &                &  48.5 \% \\
N      & Y        &  uncertain     &  24 \% \\
Y      & N        &  high          &   4\% \\
Y      & Y        &  high          &   4\%    \\      
Y      & Y        &  low           &  19.5\%    \\       
\hline
\end{tabular}
\caption{Emission line properties of the red-sequence
galaxies for which the  spectra covers both  [OII] and H$\beta$.  The
first two columns indicate whether or not [OII] and H$\beta$ have been
detected (at 2-$\sigma$  level). High and low  [OII]/H$\beta$ stand for
EW([OII])$>$18EW(H$\beta$)$-$6 and EW([OII])$<=$18EW(H$\beta$)$-$6, respectively.
\label{emission.lines}}
\end{table}


\section{Stellar velocity dispersions}
\label{sec.sigma}

Velocity dispersions were measured in all our spectra using the IDL
routine PPXF developed by \citet{CE04}.  This routine, based on
maximum penalized likelihood  with optimal template, is specially suited to extract as much
information as possible from the spectra while suppressing the noise
in the solution and, therefore, is perfect to measure the kinematics
in low S/N data.   This algorithm estimates the best fit to a galaxy spectrum
 by combining stellar templates that are convolved with the appropiate
 mean galaxy velocity and velocity dispersion. 
 The final values of these parameters are sensitive to the template 
 missmatch and,  therefore, the use of this technique requires templates which 
 match closely the galaxy spectrum  under scrutiny.
 This is achieved with the use of extensive stellar library spanning a large range of 
metallicities and ages. We use 35 synthetic spectra from the library of single stellar population 
models from Vazdekis et al. (2009, in preparation), which makes use of the new stellar library MILES
\citep{SB06_miles}, degraded to the resolution of EDisCS spectra. The library contains spectra spanning 
an age range from 0.13 to 17 Gyr and metallicities from $[$Z/H$]=-0.68$ to $[$Z/H$]=+0.2$.
Errors were calculated by means of Monte
Carlo simulations, in which each point was perturbed with the typical
observed error, following a Gaussian distribution. Because the template mistmach 
affects the measure of the velocity and $\sigma$ determined with PPXF,   
 a new optimal template was derived in each simulation. The errors were obtained as the
standard deviation of a total of 50 simulations. The mean uncertainty
in the velocity dispersion calculated in our spectra with S/N $>$ 10
is $\sim$25\%.  Given the signal-to-noise ratio and the resolution of
our spectra, $\sigma$ values below 100 kms$^{-1}$ are considered
untrustworthy and, consequently, galaxies with $\sigma <100$
kms$^{-1}$ are eliminated in all analysis involving the comparison
with $\sigma$.

\section{Line-strength indices}

For all of the galaxies in the red sequence, we measure Lick/IDS
indices as defined by \citet{T98} and \citet{WO97}.  The strength of
the Lick/IDS as a function of age, metallicity, and individual
chemical abundances has been calibrated by many authors for a simple
stellar population, i.e. a population formed in a single,
instantaneous burst \citep[e.g.][TMB03 hereafter]{w94, V99, TMB03},
and for more complicated star formation histories
\citep[e.g.][]{BC03}.  These indices remain the most popular way to
extract information about the stellar ages and metallicities from the
integrated light of galaxies.  


The errors on the indices were estimated from the uncertainties caused
by photon noise using the formula by \citet{Car98} and from
uncertainties due to the wavelength calibration, derived using
Monte-Carlo simulations.  The variances were estimated from the
residuals between the observed spectrum and the $"$best template$"$,
which was obtained in the calculation of the velocity dispersion, and
which was previously broadened to match the line width of each galaxy.
For the clusters with $z> 0.6$, indices redward of 4500~\AA~ were
affected by telluric absorption in the atmosphere and by sky
subtraction residuals and were not measured.  For the clusters with $z
\sim 0.45$ all Lick/IDS indices from H$\delta_A$ to H$\beta$ were
measured, as well as D4000. Each individual spectrum was visually
examined to check for indices affected by sky residuals or telluric
absorption in the atmosphere.  All affected indices were discarded
from any subsequent analysis.

                                                                   
Line-strength indices are sensitive to the line broadening due to both
instrumental resolution and the stellar velocity dispersion.  In order
to compare the galaxy spectra with stellar population models and to
compare the line-strength indices of galaxies with different velocity
dispersion, the indices need to be corrected to identical levels of
intrinsic Doppler and instrumental broadening.

Because the models we are using (see Sec.~\ref{sec.indices}) predict,
not only line-strength indices, but the whole spectral distribution, 
we can degrade the synthetic
spectra to the resolution of the data.  We decided to broaden all of
the observed and synthetic spectra to a final resolution of
325~km~s$^{-1}$ (including the velocity dispersion of the galaxy and
the instrumental resolution) before measuring the indices. Galaxies
with a velocity dispersion higher than $\sim$ 315~km~s$^{-1}$ could
not be broadened, but the number of galaxies with velocity dispersions
above this limit is very small and, therefore, we decided to include
them on the plots, although they have not been included in the data
analysis.

  One of the problems of using Lick/IDS-based models  
(e.g., Worthey 1994; Thomas, Maraston \& Bender 2003)
is that the data need to be  transformed to the spectrophotometric system of 
the Lick/IDS stellar library.
To do this, it is common to observe stars from the Lick/IDS library using the 
same instrumental configuration than for the science objects and to derive
small offsets between the indices measured in those stars and the ones from Lick/IDS. When
analyzing data at high redshift, this is of course impossible.
In principle, no further corrections to the indices are required when
comparing data with stellar population models using flux-calibrated
libraries.
Although we do not have to apply the offsets to our indices, we have
computed them using the stars in common between MILES and the Lick/IDS
library because it may be useful for other studies.  The final offsets
and the comparison can be found in Appendix~\ref{lickoffsets}
\footnote{Note that these offsets will not correct for any systematic
effect due to a not-perfect calibration in the data. They are only
useful to correct the fitting-function-based models from the
non-perfect calibration of the Lick/IDS stars.}.

Some of the Lick/IDS indices are affected by emission lines. In
particular, emission, when present, fills the Balmer lines, lowering
the values of the indices and, hence, increasing the derived age.
High-order Balmer indices (H$\delta$ and H$\gamma$) are much less
affected by emission than the classic index H$\beta$ \citep{WO97}, but
nevertheless still affected. To correct for
emission, two different approaches are normally adopted: First, 
assume a correlation between the equivalent width of some other
emission line and the emission in the Balmer line
\citep[e.g.][]{T00b}.  Second,  fit an optimal template and
subtract the emission directly from the residuals.  \citet{Nel05} have
shown that the first approach suffers from significant uncertainties,
in part because there are almost certainly several competing sources
of ionization in early type galaxies. The second approach requires
that the fit of the optimal template to the individual spectrum is
very good. The S/N of our individual spectra simply do not allow us to
explore that option.  Therefore, instead of trying an emission
correction to the Balmer lines, we have analyzed the differences in
the results by first eliminating {\it all} the galaxies showing {\it
any} emission.  It is re-assuring to see that none of our conclusions
 change when the (weak)
emission line galaxies are excluded (we remind the reader than the
galaxies with strong emission lines have been excluded from the
analysis).

 Table \ref{line-strength}, available at the CDS,  lists the measured line-strength indices in our sample 
of galaxies with non-or weak emission lines. A portion of the table 
is shown here to show its content an structure.

\begin{table*}
\caption{Line-strength indices in our sample of red-galaxies with S/N(\AA)$>$10 and no- or weak-emission lines. The number between brackets after
the indices values indicates the presence (1) or not (0) of sky-subtraction residuals or telluric absorptions inside the definition passband of the
index. Only those indices with labels (0) are used in our analysis. Type: Indicate the presence (2) or not (1) of weak ([OII]$<$7\AA) emission 
lines. Last column lists the measured velocity dispersion and its error. 
This is only a portion of
the Table, shown for guidance regarding its format and content. The full table is electronically 
available at CDS and at http://www.ucm.es/info/Astrof/psb/ediscs.html.}\label{line-strength}
\begin{tabular}{lrrrrrrrrrrrrr}
\hline\hline
                     & \multicolumn{1}{c}{D4000}& \multicolumn{1}{c}{H$\delta_A$}&\multicolumn{1}{c}{H$\delta_F$}&
                       \multicolumn{1}{c}{CN$_2$}&\multicolumn{1}{c}{Ca4227}&\multicolumn{1}{c}{G4300}&\multicolumn{1}{c}{H$\gamma_A$}\\
\hline
1018471-1210513&1.83$\pm$0.01 (1)    &1.28$\pm$0.41 (0)       &2.11$\pm$0.27 (0)      &0.027$\pm$0.015 (0)&
                       1.27$\pm$0.23 (0)   &2.71$\pm$0.43 (0)       &$-1.53\pm$0.47 (0)\\
1018464-1211205&2.22$\pm$0.02 (1)    &$-0.18\pm$0.69 (0)      &0.84$\pm$0.47 (0)      &0.038$\pm$0.024 (0)&
                      0.87$\pm$0.38 (0)    &4.86$\pm$0.65 (0)       &$-4.75\pm$0.81 (0)\\
1018467-1211527&2.13$\pm$0.01 (0)    &$-0.97\pm$0.29 (0)      &0.63$\pm$0.19 (0)      &0.108$\pm$0.010 (0)&
                      0.48$\pm$0.16 (0)    &5.10$\pm$0.26 (0)       &$-5.55\pm$0.33 (0)\\
1018401-1214013&1.98$\pm$0.02 (0)    &$-0.57\pm$0.65 (0)      &0.40$\pm$0.45 (0)      &0.072$\pm$0.022 (0)&
                      0.53$\pm$0.37 (0)    &7.13$\pm$0.57 (0)       &$-4.71\pm$0.76 (0)\\
\hline\hline
                     &\multicolumn{1}{c}{H$\gamma_F$}&\multicolumn{1}{c}{Fe4383}&\multicolumn{1}{c}{Ca4455}&\multicolumn{1}{c}{Fe4531}&
                      type&\multicolumn{1}{c}{$\sigma\pm$err}&\\  
1018471-1210513      & 0.52$\pm$0.28 (0)    &3.71$\pm$0.68 (0)       &0.67$\pm$0.38 (0)      &3.28$\pm$0.58 (0)  &2    &245.9$\pm$15.8&\\
1018464-1211205      &$-2.01\pm$0.52 (0)    &4.10$\pm$1.09 (0)       &0.91$\pm$0.60 (0)      &3.47$\pm$0.93 (0)  &1    &279.9$\pm$27.6&\\
1018467-1211527      &$-1.73\pm$0.21 (0)    &3.27$\pm$0.44 (0)       &0.98$\pm$0.24 (0)      &3.03$\pm$0.38 (0)  &1    &211.1$\pm$14.8\\
1018401-1214013      &$-0.29\pm$0.45 (0)    &1.77$\pm$1.06 (0)       &0.71$\pm$0.57 (0)      &1.37$\pm$0.91 (0)  &1    &129.2$\pm$21.0\\
\hline
\end{tabular}
\end{table*}

\subsection{The local sample \& aperture correction}
\label{localsample}
To increase the time baseline of our analysis, we compare the EDisCS
spectra with  36 early-type galaxies (ellipticals and
lenticulars) in the Coma cluster at redshift $z=0.02$.  The
characteristics of this sample are described in \citet{SB06a}. Because 
the galaxies were selected morphologically we
checked first that all the objects belong to the red-sequence of this
cluster, using the colours and magnitudes from \citet{Mob01}.  We are
aware that the comparison with the Coma cluster is probably not the
most appropriate, as the the velocity dispersion of this cluster
exceeds that of all the clusters in our sample at intermediate
redshifts. For this reason, we will base all the conclusions of this paper in the
intercomparison of the EDisCS clusters.  The local sample is included here
to show that some of our results can be extrapolated to 
the local Universe, where higher S/N spectra can be obtained.  One of the
main problems when comparing observations at different redshifts is
that, for a fixed aperture, one is sampling different physical regions
within the galaxies.  Early-type galaxies show variations in their
main spectral characteristics with radius and, therefore, a correction
due to these aperture differences is necessary.  In order to compare
directly with the sample at medium- and high-redshift, we extracted
the 1D spectra in the same way as the EDisCS spectra, inside an
aperture equal to the FWHM of the spatial profile.  However, aperture
effects are not entirely mitigated via such an extraction, as the
EDisCS spectra were observed with a 1$^{\prime\prime}$-wide slit,
equivalent to a much larger physical aperture at the redshift of the
Coma cluster.  \citet{Jor95}, \citet{Jor97} and \citet{J05} derived
aperture corrections using mean gradients obtained in the
literature. 
The index corrected for aperture effects
can be obtained as  $\log (\rm index)_{\rm corr}=\log(\rm index)_{\rm ap}+
\alpha \log \frac{\rm r_{\rm ap}}{\rm r_{\rm nor}}$ for those indices
measured in \AA~ and ${\rm index}_{\rm corr}= {\rm index}_{\rm ap} + \alpha \log
\frac{\rm r_{\rm ap}}{\rm r_{\rm nor}}$ for molecular indices measured in
magnitudes, where ${\rm r_{\rm nor}}$ is the final normalised aperture,
and $r_{\rm ap}$ is the equivalent circular aperture, obtained 
as 2 ${\rm r}_{\rm ap} = 1.025\times 2(xy/\pi)^{1/2}$, being ${x}$ and ${y}$ 
the width and length of the rectangular aperture. 
 J{\o}rgensen (1995, 1997) calculated $\alpha$ parameters
for a large subset of Lick indices but, in some cases, using the gradients
measured in a very small sample of galaxies.
In this work, we have taken advantage of the large sample
of galaxies with measured line-strength gradients published in
\citet{SB06c}, \citet{SB07} and \citet{Jab07} to calculate new
aperture corrections. These also include corrections for the
higher-order Balmer lines that have not been published before, as far
as we are aware.  
Appendix~\ref{appendix.aperture}
lists the new $\alpha$ parameter calculated in this paper.
We correct all our indices to
mimic the physical aperture of the slit used to observe the galaxies
at $z=0.75$.  The aperture corrections ($\alpha \log \frac{\rm r_{\rm ap}}
{\rm r_{\rm nor}}$) for all the indices at the
different redshift bins are listed in Table
\ref{aperture.corrections}.

\begin{table*}
\centering
\begin{tabular}{rrrrrr}
Index       & Coma          & $z=0.45$         & $z=0.55$            & $z=0.75$            &correction type\\
\hline
\hline
H$\delta_A$  & 1.031     &  0.063  &  0.016 &  0.00  & additive\\
CN$_1$       & $-0.083$  &$-0.005$ &$-0.001$& $0.00$ & additive\\
Ca4227       &  0.856    &  0.990  &  0.997 &  1.0   & multiplicative\\
G4300        & 0.914     &  0.995  &  0.998 &  1.0   & multiplicative\\
H$\gamma_F$  & 0.623     &  0.038  &  0.009 &  0.0   & additiva\\
Fe4383       &  0.817    &  0.987  &  0.996 &  1.0   & multiplicative\\
Ca4455       & 0.726     &  0.980  &  0.995 &  1.0   & multiplicative\\
Fe4531       & 0.876     &  0.992  &  0.998 &  1.0   & multiplicative\\
C4668        & 0.676     &  0.976  &  0.994 &  1.0   & multiplicative\\
H$\beta$     & 1.179     &  1.010  &  1.000 &  1.0   & multiplicative\\
\hline
\end{tabular}
\caption{Aperture correction for Lick/IDS indices measured in our sample.
The correction for the molecular indices and for the higher-order Balmer lines
  is additive while the correction for 
the atomic indices is multiplicative.  The aperture correction was only applied 
to the Coma galaxies, as it is very small for the other redshifts. 
The type of correction is indicated
in the last column.\label{aperture.corrections}}
\end{table*}

Line-strength gradients show large variations between early-type
galaxies \citep[see, e.g.][among others]{Car93, Gor90, DSP93, Hall98, SB06c, Kun06, Jab07}, 
and the slope of the
gradients does not seem to correlate clearly with any of their other
properties.  For this reason, we use the mean gradient to compute the
aperture corrections for all the galaxies. However, in order to
calculate the error in the aperture correction due to the fact that
early-type galaxies show a large variation in the slope of their
gradients, we perform a series of Monte Carlo simulations where the
mean index-gradient was perturbed by a random amount, given by a
Gaussian distribution with $\sigma$ equal to the typical RMS-
dispersion in the gradients
The errors in the aperture correction calculated this way are also
indicated in Table~\ref{aperture.corrections}.  In some cases
(e.g., higher-order Balmer lines), these errors are very large. In fact, 
despite the large mean aperture correction for H$\delta$ and 
H$\gamma$ indices, the aperture correction is compatible, within 
those errors, with being null. These errors have been
added quadratically to the errors of the indices for the Coma
galaxies. 
 For the 0.45 and 0.55 bins, the error in the aperture
correction due to the scatter among the mean gradients is negligible
(less than 1\% of the total correction) and we do not list them.

Gradients might evolve with redshift. That is the reason why we
correct the indices in the Coma cluster to the same equivalent
aperture as the highest redshift bin instead of correcting the highest
redshift measurements as is commonly done. This way, the correction is
performed to the galaxies where the gradients used to derive the
aperture correction have been measured.
\subsection{Stacked spectra}
\label{sec.stacked}
 
The degeneracy between age and metallicity effects on galaxy
colours could well affect the interpretation of their colour-magnitude
diagrams. This degeneracy can  be partially  broken  by  using  a
combination of two or more  spectral indices  chosen for their
different sensitivities to age and metallicity. The S/N of our galaxy
spectra is  not high enough  to  do this  analysis  on an
individual basis. Therefore, in order to derive the  evolution with
time  of the  galaxy mean ages and  chemical abundances, we stacked the
galaxy spectra in each of the 0.45, 0.55 and 0.75 redshift bins.

\begin{figure}
\centering
\resizebox{0.4\textwidth}{!}{\includegraphics[angle=-90]{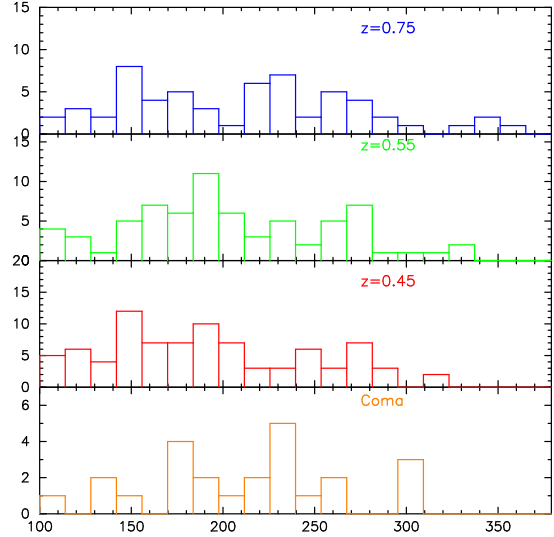}}
\caption{Velocity dispersion distribution of the individual spectra stacked
in the different redshift bins. From bottom to top:
$z$=0,0.45,0.55,0.75.}
\label{distribut.sigma.fig}
\end{figure}

We made sure that the distribution of galaxy velocity dispersions ($\sigma$) 
in all the different redshift intervals  is similar (Fig.~\ref{distribut.sigma.fig}). 
This a prerequisite to any further
comparison between the three redshift bins, as most line-strength
indices are strongly correlated with this parameter
\citep[e.g.][]{G93}.  To explore the possible dependence of star
formation history on galaxy mass, we distinguish between galaxies with
$\sigma$ higher and lower than 175km~s$^{-1}$ in all three redshift
groups.  We perform a Kolmorogov-Smirnov test comparing the
distribution of $\sigma_{\rm s}$ at different redshifts to look for
possible significant differences.  We performed the test separately
for galaxies with $\sigma $ larger and smaller than 175~km~s$^{-1}$ and found
them compatible with being drawn from the same distribution.

\begin{figure}
\resizebox{0.5\textwidth}{!}{\includegraphics[angle=-90]{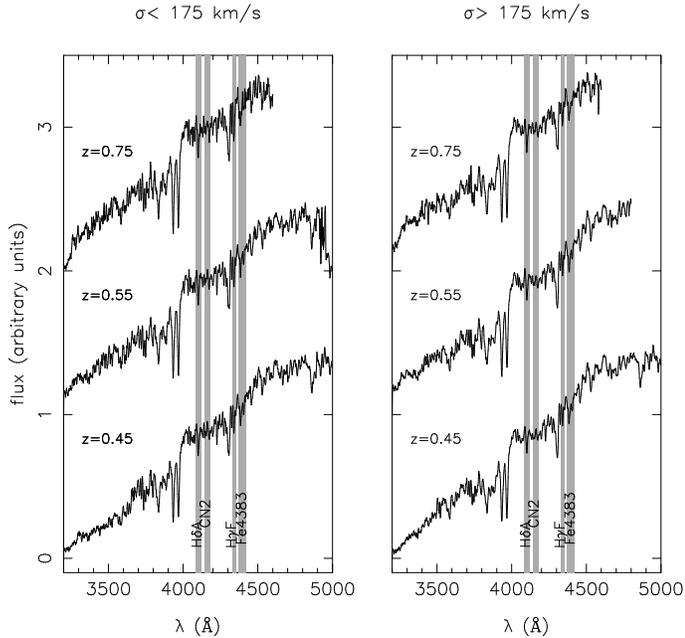}}
\caption{Final stacked spectra for our three redshift and velocity
dispersion bins. The shaded areas identify the central bands of the
four indices used in our spectral analysis.}
\label{StackedSpectra.fig}
\end{figure}

Before adding them, we normalize all spectra by their mean flux in the region between
3900\AA~ and 4400\AA, to avoid a bias towards the indices of the most
luminous galaxies.  Finally, we co-add them, clipping out all the
pixels deviating more than 2-$\sigma$ from the mean.
Figure~\ref{StackedSpectra.fig} displays the resulting 6 stacked
spectra for the three redshift bins and the two velocity dispersion
regimes.

\begin{figure}
\resizebox{0.4\textwidth}{!}{\includegraphics[angle=-90]{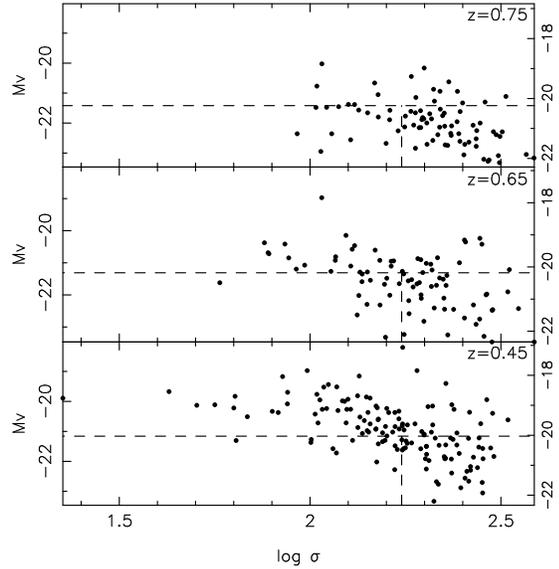}}
\caption{Relation between the absolute magnitude Mv and the velocity
dispersion for our sample of galaxies in the three considered redshift
bins.  The right label shows the absolute magnitude Mv evolved
passively to $z=0$ using \citet{BC03} models at solar
metallicity. The vertical dashed line indicates the position corresponding
to our $\sigma$-cut (175 km~s$^{-1}$) while the horizontal line marks
the mean magnitude for the galaxies with a sigma between 160 and 180
km~$^{-1}$.}
\label{faber.jackson}
\end{figure}

Earlier works have studied the evolution of red-sequence galaxies as a
function of magnitude (e.g.  De Lucia et~al.  2004; De Lucia et~al.\
2007; Rudnick et~al.\ 2008, in preparation). These studies have put
the boundary between faint and bright galaxies at Mv$=-20$.  Assuming
a formation redshift of 3 followed by passive evolution, our
$\sigma$=175~km~s$^{-1}$ cut corresponds to Mv$\sim-$20.0, $-20.2$ and
$-20.2$ mag at $z=0.45$, 0.55 and 0.75, respectively (see
Figure~\ref{faber.jackson}).  As can be seen, the magnitude cut is not
very different between our study and the above mentioned photometric
studies.  However, those studies reach much fainter magnitudes in
their $"$faint$"$ bin.

\begin{figure}
\centering
\resizebox{0.25\textwidth}{!}{\includegraphics[angle=-90]{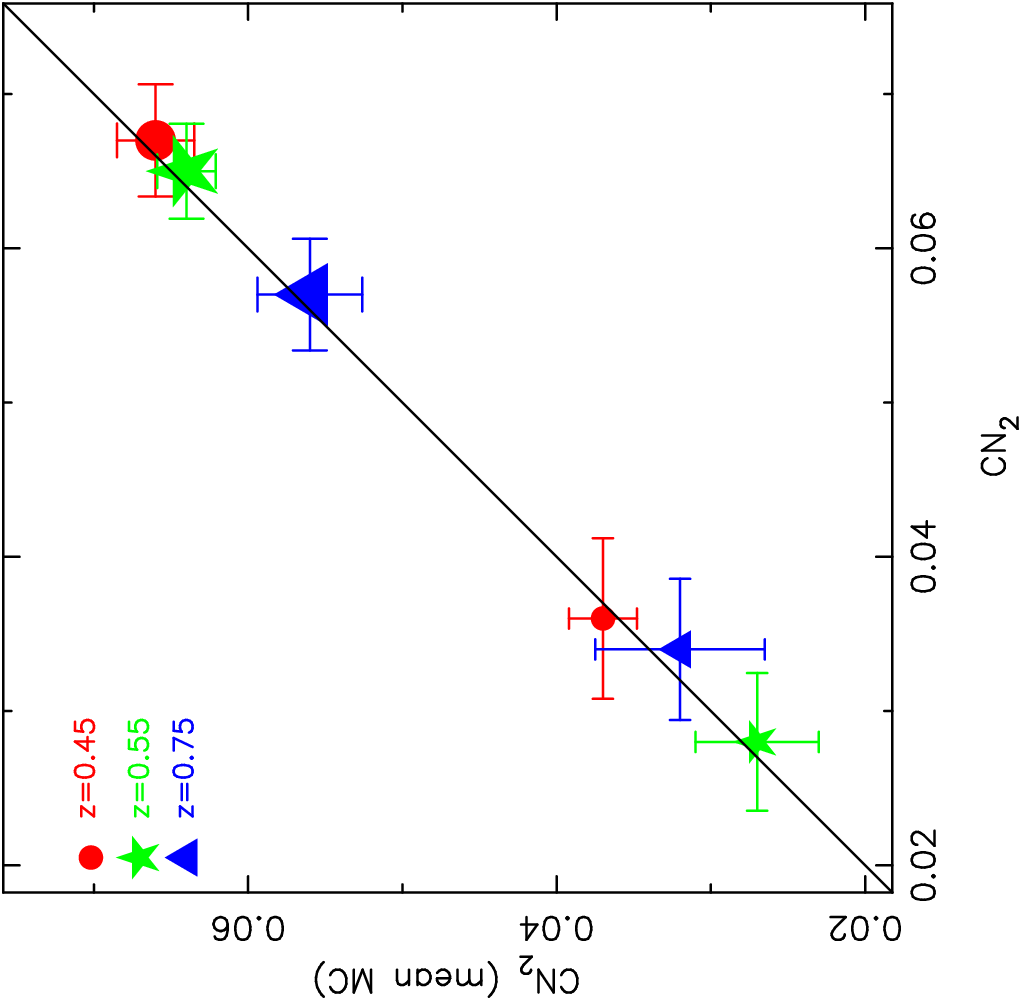}}
\resizebox{0.25\textwidth}{!}{\includegraphics[angle=-90]{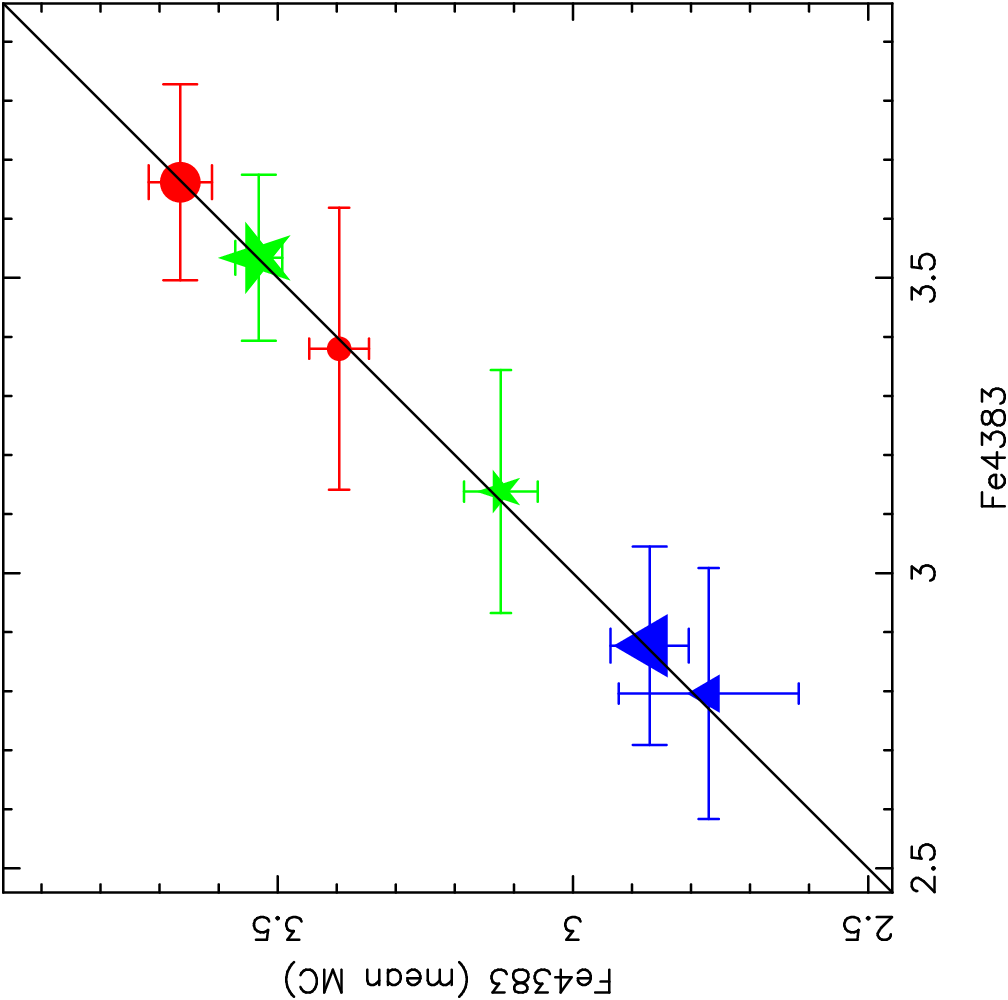}}
\resizebox{0.25\textwidth}{!}{\includegraphics[angle=-90]{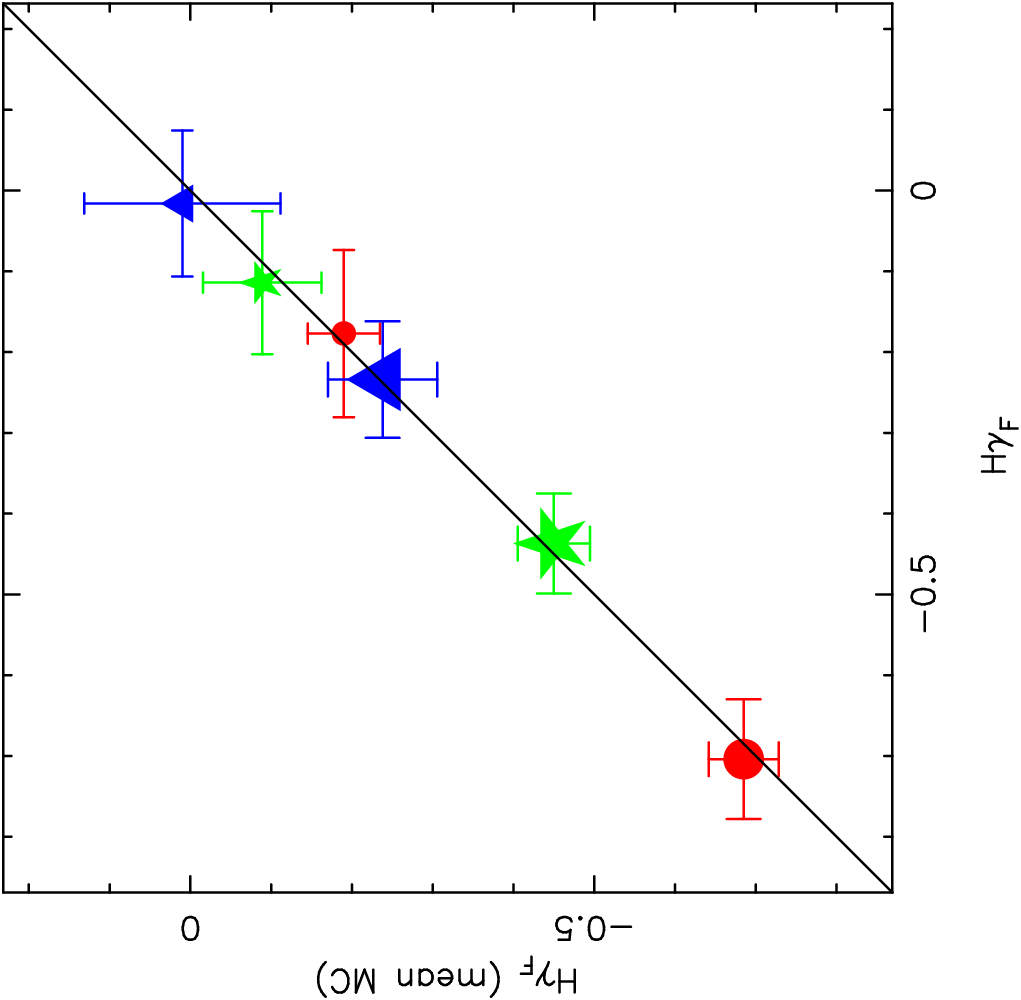}}
\caption{Comparison between the indices measured in the stacked
spectra and the mean index calculated in the 20 stacked spectra
obtained with Monte Carlo simulations as described in the text to
analyze the robustness of our separation of galaxies with $\sigma$ due
to errors in this parameter. The errors in the $y$-direction represent
the rms of all the 20 measured indices, while the errors in the
$x$-direction represent the errors in the index due to the S/N of the
stacked spectra. Big and small symbols show the indices measured in
the stack with $\sigma$ larger than 175 kms$^{-1}$ and smaller than $\sigma< 175$ kms$^{-1}$
respectively. The meaning of the different symbols is indicated in the
inset.}
\label{sim}
\end{figure}

Velocity dispersions and line-strength indices for the stacked spectra
were measured using the same techniques as for the individual spectra.
Errors on the indices were calculated with the formula of
\citet{Car98}, using the signal-to-noise of the stacked spectra.
These errors do not reflect the differences between the spectra added
in each bin, which are larger than the formal errors obtained using
the signal-to-noise ratio.  However we do not intend to study the
distribution of galaxy properties but rather their mean values.  To
explore the robustness of our results to errors in velocity
dispersion, we performed 20 Monte Carlo simulations in which the
individual galaxy velocities were randomly perturbed following a
Gaussian error distribution.  After each simulation, the spectra were
stacked again, indices were measured, and their mean and
RMS-dispersion were calculated. Fig.~\ref{sim} compares the indices
obtained in the original stacked spectra with the mean indices
obtained from the 20 simulations. The separation in low- and
high-$\sigma$ galaxies is robust to the errors in velocity dispersion.
When the RMS-dispersion for all the simulated galaxies is larger than
the error we have calculated for the individual indices, we add the
quadratic difference as a residual error to the original error in the
index.  We have also checked that the mean values of the individual
indices for all the stacked galaxies is compatible, within the errors,
with the index measured in the stacked spectra.

The Lick/IDS indices measured in the stacked spectra as well as their errors are listed in 
Table~\ref{Lick.stacked}.

\begin{table*}
\begin{tabular}{l|rrrrrrrrrrr}
\hline\hline
Redshift group &                    &\multicolumn{1}{c}{D4000}&\multicolumn{1}{c}{H$\delta_A$}&\multicolumn{1}{c}{H$\delta_F$}&
                                     \multicolumn{1}{c}{CN$_2$}&\multicolumn{1}{c}{Ca4227}&\multicolumn{1}{c}{G4300}&\multicolumn{1}{c}{H$\gamma_A$}&
                                     \multicolumn{1}{c}{H$\gamma_F$}&\multicolumn{1}{c}{Fe4383}&\multicolumn{1}{c}{$\sigma$} \\
\hline
               &                    &\multicolumn{1}{c}{\AA}&\multicolumn{1}{c}{\AA}&\multicolumn{1}{c}{\AA}&\multicolumn{1}{c}{mag}&
\multicolumn{1}{c}{\AA}&\multicolumn{1}{c}{\AA}&\multicolumn{1}{c}{\AA}&\multicolumn{1}{c}{\AA}&\multicolumn{1}{c}{\AA}&\multicolumn{1}{c}{kms$^{-1}$}\\
0.45           & $\sigma>175$ km/s  & 2.044 &$-0.433$ &0.894 &0.067&0.932 &4.589& $-4.114$  & $-0.704$ & 3.662   & 233.1\\
               &                    & 0.004 & 0.140   &0.094 &0.005&0.076 &0.133&   0.161   &  0.098   & 0.220   &   7.2\\
0.45           & $\sigma<175$ km/s  & 2.045 &0.103    &1.074 &0.036&0.841 &4.053& $-3.145$  & $-0.177$ & 3.380   & 122.3\\
               &                    & 0.004 &0.139    &0.094 &0.005&0.077 &0.136&   0.158   &  0.096   & 0.222   &  12.0\\
0.55           & $\sigma>175$ km/s  & 2.051 &$-0.075 $&1.143 &0.065&0.847 &4.600& $-3.779$  & $-0.437$ & 3.534   & 218.1 \\
               &                    & 0.002 & 0.087   &0.058 &0.003&0.048 &0.083&   0.100   &   0.061  & 0.137   &   9.0\\
0.55           & $\sigma<175$ km/s  & 1.995 & 0.045   &0.960 &0.028&0.757 &4.341& $-3.138$  & $-0.114$ & 3.138   & 134.9\\
               &                    & 0.004 & 0.140   &0.095 &0.005&0.078 &0.135&   0.159   &  0.096   & 0.224   &  10.1\\
0.75           & $\sigma> 175$ km/s & 2.008 & 0.699   &1.400 &0.057&0.824 &4.381& $-3.061$  & $-0.234$ & 2.877   & 233.2\\
               &                    & 0.003 & 0.104   &0.070 &0.004&0.059 &0.102&   0.120   &  0.073   & 0.171   &   6.2\\
0.75           & $\sigma< 175$ km/s & 1.987 & 0.453   &1.163 &0.034&0.826 &4.244& $-2.963$  & $-0.016$ & 2.796   & 154.2\\
               &                    & 0.003 & 0.130   &0.089 &0.004&0.073 &0.128&   0.149   & 0.091    & 0.213   &   8.1\\
\hline
\end{tabular}
\caption{Line-strength indices and velocity dispersion measured in the stacked spectra. The second row of
each line lists the errors. Last column show the velocity dispersion measured in the different stacked spectra.\label{Lick.stacked}}
\end{table*}

\section{Stellar population models}
\label{sec.indices}
In this study we compare our measured values with the models by 
Vazdekis et al.\  (2008, in preparation,   V08 hereafter\footnote{The  models  are
publicly                       available                             at
http://www.ucm.es/info/Astrof/miles/models/models.html}).        These
models are based on the MILES stellar library \citep{SB06_miles}.  The
spectra from this library have a resolution of 2.3\AA~, constant along
the whole wavelength range, and  are flux-calibrated.  Thanks to this,
the models predict not only individual indices, but the whole spectral
energy distribution  from 3500 to 7500\AA~  as  a function of  age and
metallicity, for ages between 0.1 and 17.78 Gyr and metallicities from
[Z/H]=$-1.68$ to +0.2 dex\footnote{In the normalization to the solar values
a solar metallicity Z$_{\odot}$=0.02 has been used}.

However, the  chemical composition of stars  in the  MILES library (as
in the rest of empirical libraries included in stellar population models)
matches  that of the  solar-neighborhood.\footnote{Although  the
detailed   chemical composition  of  all the   stars  in the MILES  is
unknown,  it is reasonable  to  assume that this  is  the case.}  This
means  that the models, in principle,  can only  predict accurate ages
and metallicities  for stellar   populations  with the  same  chemical
patterns  as  the   solar vicinity.   We  know that this
condition  does not  apply to  bright nearby  early-type galaxies
\citep[see review by][]{W98}),  where  Mg/Fe and maybe Ti/Fe,  Na/Fe, N/Fe
and C/Fe  are  enhanced with respect  to  the solar values.  This  may
result in systematics  errors  in the measured ages  and metallicities
\citep[see][] {W98, TMB03}.  In  order to overcome this problem,  we
calibrate  the models for   different chemical compositions using  the
method first introduced  by \citet{Tan98} and \citet{T00b}.   This method  uses  stellar
atmospheres to characterize the variation  of the different indices to
relative changes of different chemical species.  We first interpolate
the existing  models to make  a grid with 168 ages  from 1 Gyr to 17.8
Gyr  in steps of 0.1  Gyr and 209 metallicities from  $-1.68$ to +0.4 in
steps of  0.1.    Then, we  modify each  index   according to  their
fractional response  to variations of  different  elements using (from
Trager et al.):
\begin{equation}
\frac{\Delta I}{I}=\left (\prod_i (1+R_{0.3}(X_i))^{[X_i/H]/0.3}  \right) -1,
\end{equation}

where $R_{0.3}(X_i)$ is the response function for element $i$ at
$[X_i/H]=+0.3$ dex.  As in Trager et al.\ , we also consider that the
fractional light contribution from each stellar component is 53\% from
cool giant stars, 44\% from turnoff stars and 4\% from cool dwarfs
stars.  We use the calibrations by \citet{Korn05} instead of the ones
by \citet{TB95} (used by Trager et al.)  because they include the
higher order Balmer lines H$\delta$ and H$\gamma$, and they are
computed at other metallicities different from solar.  However, the
differences between the two studies are very small \citep{Korn05}.  Following the approach of several
previous authors, we do not change all the elements individually but
we assume than some of them are linked by nucleosynthesis and,
therefore, change them in lock-step. We built models where N, Ne, Na,
Mg, Si and S are enhanced by 0.3 dex with respect to Fe while C is
enhanced by +0.15 dex. The reason that C is only enhanced by +0.15
is that enhancing C by +0.3 dex brings the C/O ratio very close to
the values at which a carbon star is formed \citep[see][for a
  discussion]{Hou02, Korn05}.  Ca and the Fe peak elements are
depressed to keep the overall metallicity constant.
\footnote{Despite the fact that Ca is an $\alpha$-element, theoretically linked 
with the overabundant elements as Mg, measurements show that it may be
depressed  in  elliptical  galaxies (although see   \citet{Pro05} for a
different point  of view). We follow Trager  et al.\  2000 and include
this element  in the depressed  group.}   To obtain response functions
for enhancements different from +0.3 dex  we, again, follow  \citet{T00a}:
We   first    calculate $\Delta$[Fe/H]   and    $\Delta$[E/Fe] using
$\Delta$[Fe/H]=$-$A$\Delta$[E/Fe] = $-\frac{A}{1-A}\Delta$[E/H],  where  E
represents  the {\it  enhanced elements}.   $A$  is the   response of the
enhanced elements to changes in [Fe/H] at fixed  [Z/H]. For the chosen
enhanced model,  $A =0.929$ (Trager   et  al. 2000).  Then,  we  scale
exponentially the   response functions   by  the appropriate   element
abundance.
We built models with enhancements from [E/Fe]=$-0.1$ to $0.5$ in steps
of 0.05.

Because we are not working at the Lick/IDS resolution, in principle,
we should not be using \citet{Korn05} response functions to compute
the sensitivity of different indices to variations of chemical
abundance ratios, as they were computed at this resolution.  However,
\citet{Korn05} showed that the influence of the resolution on the
response functions is very small.

Note that the models used here (as well as most existing models in the
literature) {\it do not} include isochrones with chemical abundance
ratios different from solar. The effect of the different chemical
abundance ratios is only included in the atmospheres.  The influence
that the inclusion of $\alpha$-enhanced isochrones may have in the
final predictions is still unclear, especially at super-solar
metallicities, but some studies have shown 
that it is
much smaller than the one in the model atmospheres \citep{Coel07}, 
at least for line-strength indices at wavelengths shorter than 
$\sim$5750. 

We want to caution the reader that the absolute values of ages
obtained directly with stellar population models are subject, not only
to errors in the data (which are taken into account) but also to
errors in the stellar population models.  Stellar evolutionary models
are still affected by uncertainties that leave room for improvement
\citep[see, e.g.][]{Cas05}, as illustrated by the {\it non-negligible}
differences still existing among the results provided by different
theoretical groups.  Furthermore, it must be noted that the oldest
ages of the models we are using (17.78 Gyr) are older than the current
age of the Universe. The derivation of very old spectroscopic ages,
inconsistent with the ages derived from colour-magnitude diagrams for
globular clusters and with the age of the Universe, was first pointed
out by \citet{Gib99} and is a well-known problem in the community.
Some solutions have been proposed, such as the inclusion of atomic
diffusion in the stellar evolutionary models or the use of
$\alpha$-enhanced isochrones \cite[see][] {V01}, although none of
these have been implemented in publicly available stellar population
models. \citet{Sch02} showed, for the particular case of 47Tuc, that
the luminosity function of the red giant branch is underestimated in
the stellar evolutionary models and that the use of the observed
luminosity functions instead of theoretical ones results in derived
ages for this cluster consistent with the age of the Universe and with
those derived directly from the colour-magnitude diagram. A similar
effect in super-solar metallicity models would cause spectroscopic
ages to be 30\% too high (see Schiavon et~al.\ 2002 for details).  It
is beyond the scope of this paper to discuss these uncertainties in
stellar population models, but it is clear that conclusions based on
the absolute values of age are simply too preliminary.  Relative
differences in age and metallicity are more reliable than absolute
values, and we will therefore base our interpretations on relative
values.

Lick/IDS indices, as well as colours,
respond to variations in both age and chemical abundances, but the
relative sensitivity of the different indices to these two parameters
is not the same. In order to break the existing degeneracy between age
and metallicity, it is necessary to combine two or more indices.  In
what follows, for the main analysis, we use 4 indices: Fe4383, CN$_2$,
H$\delta_A$ and H$\gamma_F$.  We choose these because, of all the
Lick/IDS indices that could be measured in {\it all} galaxies, they
are the most sensitive to variations of age (H$\delta$ and H$\gamma$)
and total metallicity (CN$_2$, Fe4383).  TMB03 recommend CN and Fe4383
as the best blue indices to calculate $\alpha$/Fe abundance ratios.
Furthermore, we choose H$\gamma_F$ instead of H$\gamma_A$ because the
ages measured with H$\gamma_A$ are systematically younger than those
measured with H$\gamma_F$ or H$\beta$ when $\alpha$-enhanced models
are used \citep[][]{TMK04}.  As the origin of these differences is
unclear, we follow the advice by \citet{TMK04} and use H$\gamma_F$,
despite the higher photon noise in this index compared to its wider
version (H$\gamma_A$).

\section{Age \& Metallicity}
\begin{figure*}
\resizebox{0.4\textwidth}{!}{\includegraphics[angle=-90]{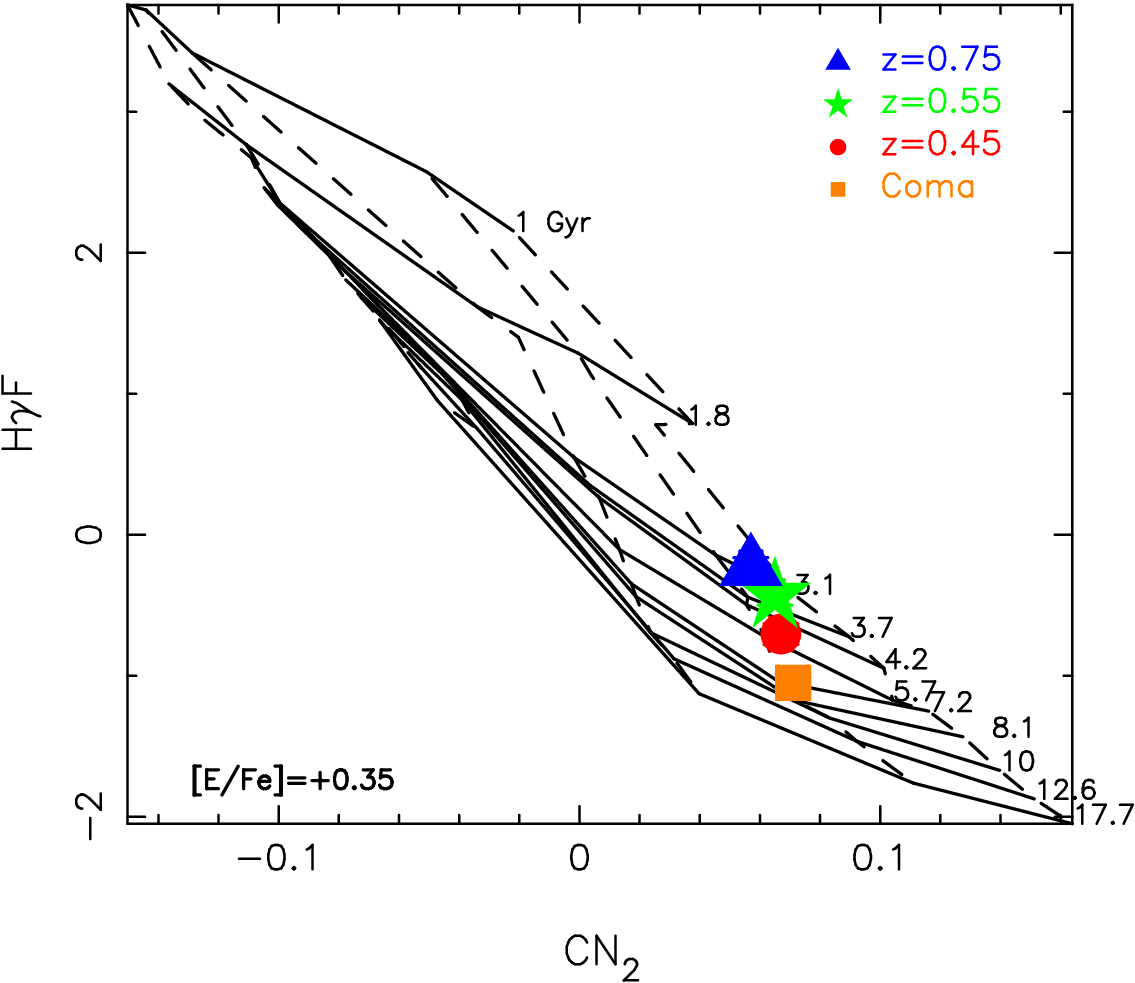}}
\resizebox{0.4\textwidth}{!}{\includegraphics[angle=-90]{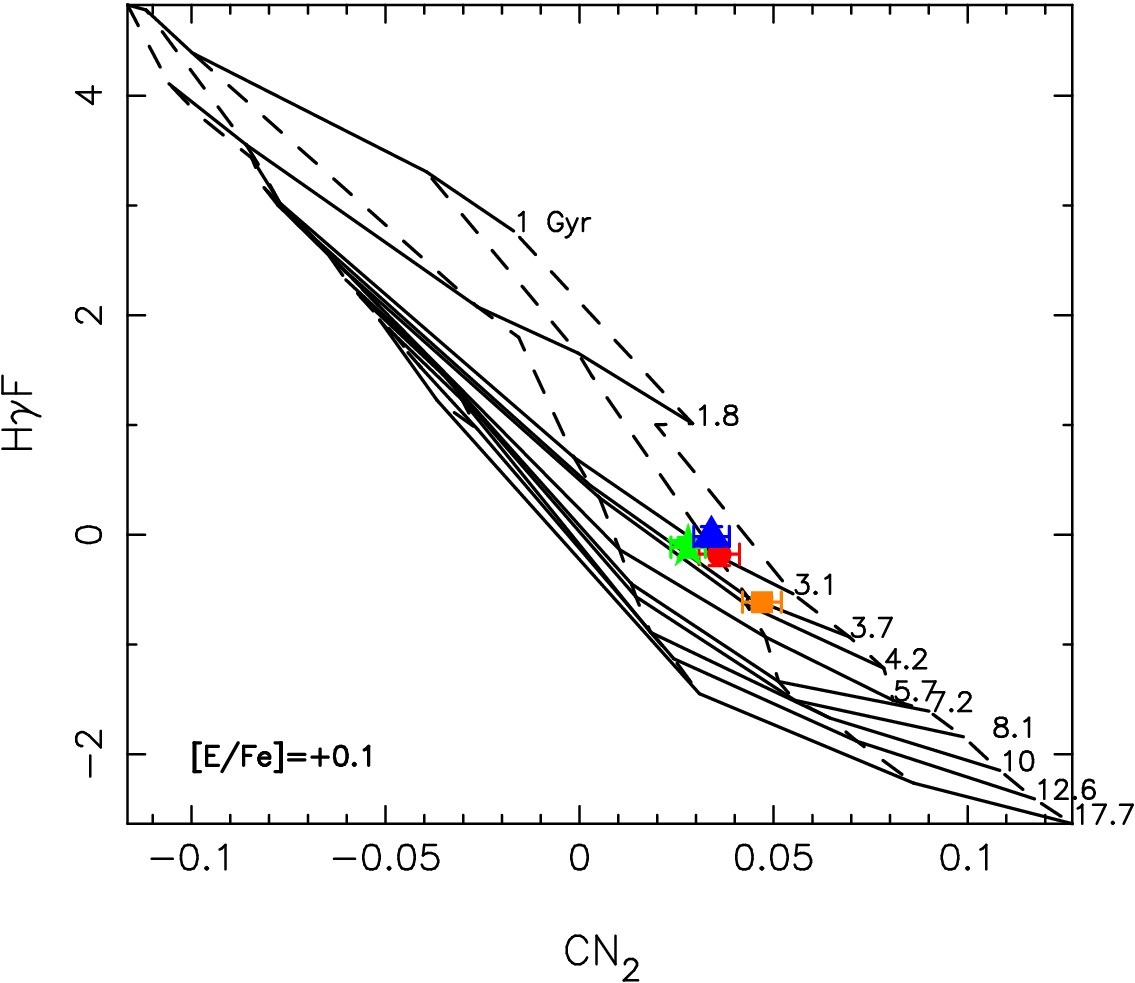}}
\resizebox{0.4\textwidth}{!}{\includegraphics[angle=-90]{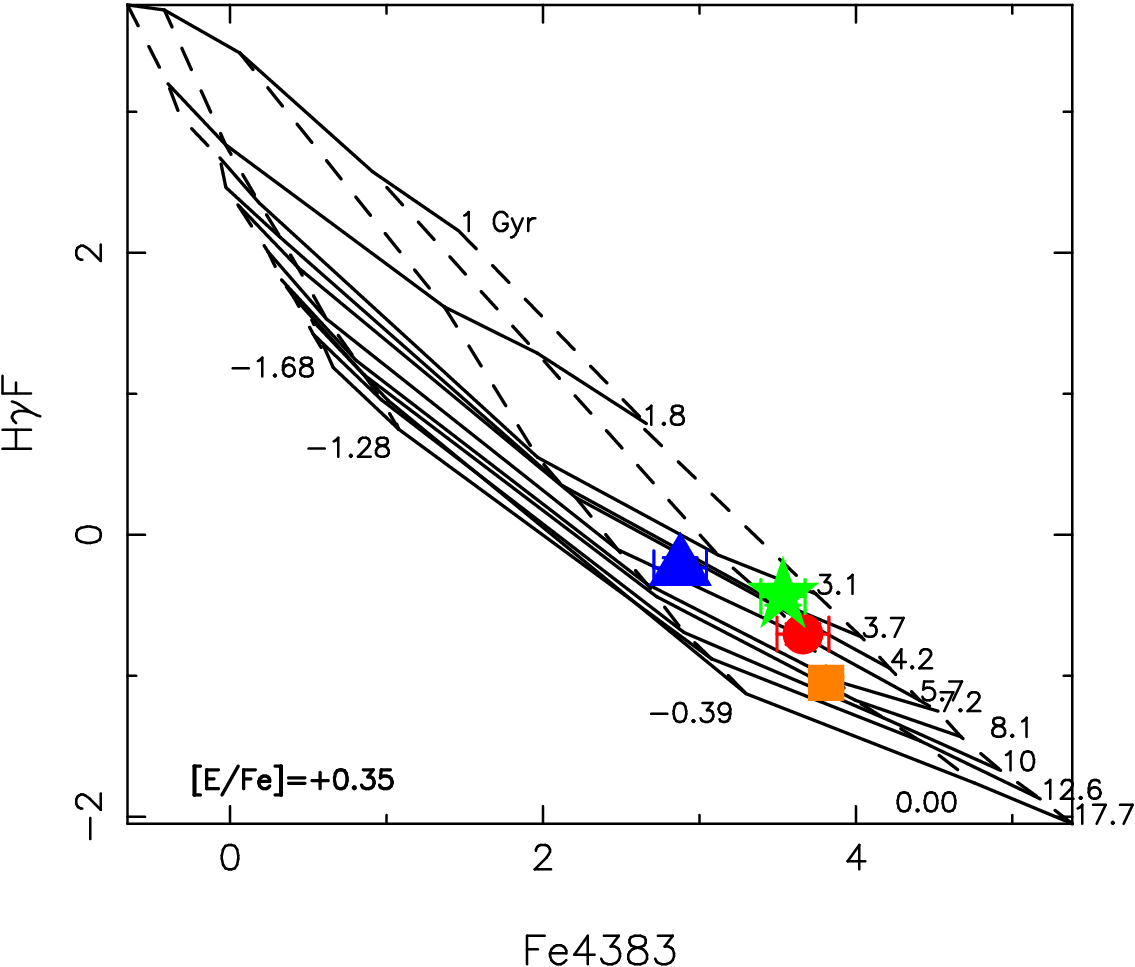}}
\hspace{3.3cm}
\resizebox{0.4\textwidth}{!}{\includegraphics[angle=-90]{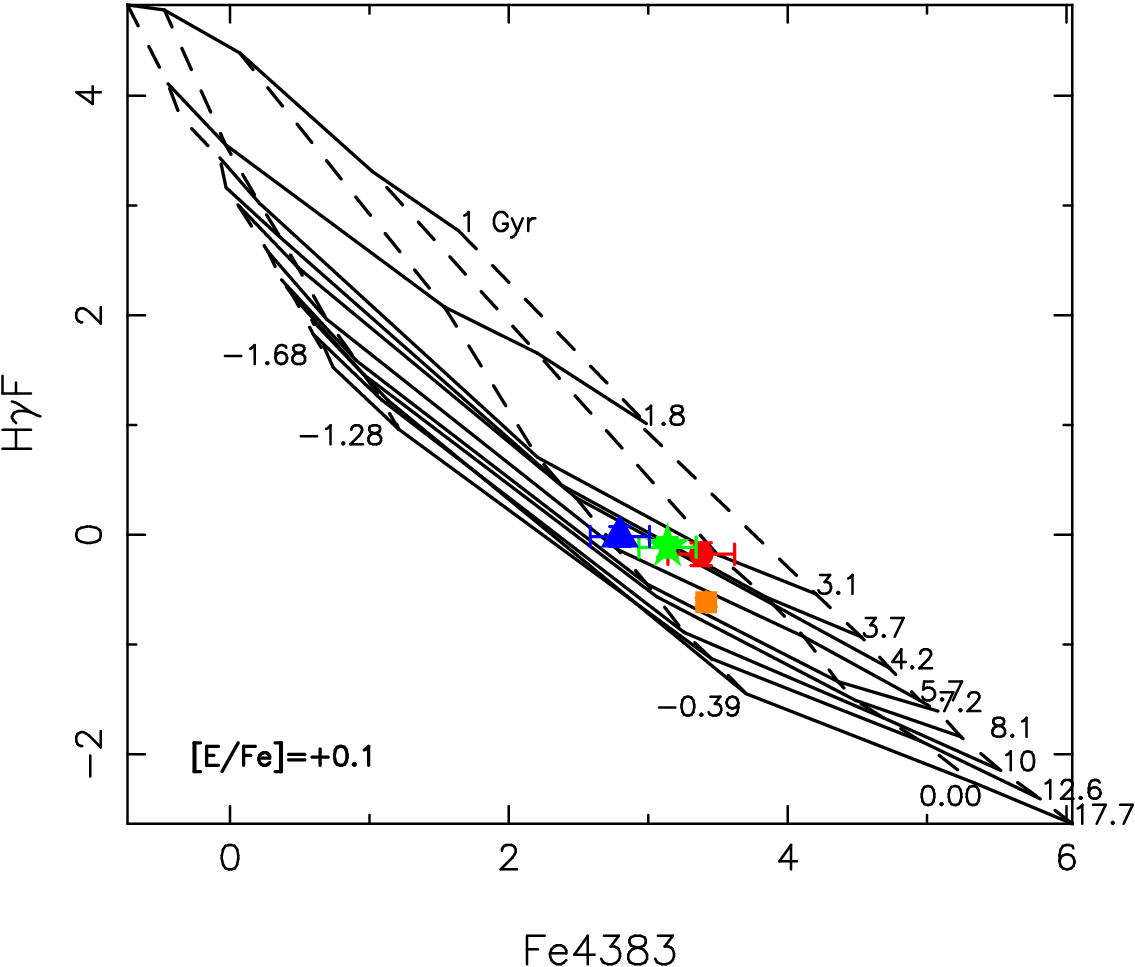}}
\caption{Index-index diagrams for the indices measured in the stacked
spectra at different redshifts. Left panels (big symbols) show
measurements on the stacked spectra of galaxies with $\sigma$ larger than  175
km~s$^{-1}$ while right panels (small symbols) show the same measurement for galaxies with
$\sigma$ smaller than 175 km~s$^{-1}$.  The different redshifts bins are
represented as different colours (and different symbol shapes);
$z=0.75$: blue triangles; $z=0.65$: green stars; $z=0.45$: red
circles; $z=0.02$: orange squares. Model grids of constant age (solid
lines) and constant metallicity (dashed lines) from V08 with
[E/Fe]=+0.35 (for the most massive galaxies ) and [E/Fe]=+0.1 (for the
lower sigma bins) are superimposed. The ages and metallicities of the
different models are indicated in the Figure.}
\label{index.index.v07}
\end{figure*}

We derive age, metallicity and the ratio of $\alpha$-element
enhancement [E/Fe] for our stacked spectra comparing our 
four selected indices  (H$\delta_A$, H$\gamma_F$, CN$_2$, and Fe4383) with the models
by V08  using a $\chi^2$ minimization routine.
 We find that the
derived ages are independent of the chosen Balmer index (H$\delta_{\rm A}$
or H$\gamma_{\rm F}$). The resulting ages, metallicities, and [E/Fe]
are listed in Table~\ref{alphayz}. Index-index 
diagrams comparing the indices measured in the stacked spectra with 
the predictions for single stellar population models
by V08 are shown Fig.~\ref{index.index.v07}.  In each panel, [E/Fe] is chosen to be as close as
possible to the values given in Table~\ref{alphayz}.

Although the selection of our sample has been done using colours, many
previous studies of the red-sequence were restricted to
morphologically classified early-type galaxies. To compare with those,
the analysis in this section has been repeated for 
the subset of red-sequence galaxies that are morphologically
classified as E or S0 galaxies.  For the clusters imaged with the ACS
(see Table \ref{tb:clusterlist}), we use their visual classification
\citep{Des07}. For the others, we apply the selection criterion of
Simard et al. (2008), based on the bulge fraction derived from our VLT
I-images (see Sec.\ref{sec.morph} for the criteria to select
early-type galaxies).

The last three rows of Table~\ref{alphayz} show the SSP-parameters derived 
for the red-early-type galaxies.  The first thing that can be seen is that the SSP-parameters 
for the whole red-sequence and for the morphologically classified early-type galaxies are
compatible within the errors. Therefore, we will not discuss further the possible differences
and will restrict our analysis to the complete red-sequence sample.

\begin{table*}
\centering
\begin{tabular}{lrrr|rrr}
\hline

              & \multicolumn{3}{c}{$\sigma> 175$ km~s$^{-1}$}   &  \multicolumn{3}{c}{$\sigma < 175$ km~s$^{-1}$}  \\
              & Age~~~~~~    & [Z/H]~~~~~~  &    [E/Fe]~~~~~    & Age~~~~~~   &   [Z/H]~~~~~~ &  [E/Fe]~~~~~\\
              & Gyr~~~~~~    &               &                  & Gyr~~~~~~   &              &              \\
\hline
\hline
 Coma          & $7.90\pm 1.81$ & $-0.03\pm 0.06$&$0.40\pm 0.03$ & $5.80\pm 0.36$ &$-0.11\pm 0.03$&$0.31\pm 0.04$\\
$z=0.45$       & $5.60\pm 1.05$ & $-0.01\pm 0.07$&$0.35\pm 0.09$ & $3.20\pm 0.64$ &$-0.05\pm 0.13$&$0.16\pm 0.12$\\
$z=0.55$       & $3.40\pm 0.52$ & $ 0.11\pm 0.07$&$0.35\pm 0.06$ & $3.40\pm 0.81$ &$-0.13\pm 0.09$&$0.11\pm 0.13$\\ 
$z=0.75$       & $2.70\pm 0.38$ & $ 0.18\pm 0.11$&$0.46\pm 0.06$ & $3.10\pm 0.83$ &$-0.08\pm 0.14$&$0.30\pm 0.21$\\
\hline
$z=0.45$       & $5.70\pm 2.42$&  $ 0.02\pm 0.12$&$0.45\pm 0.10$ &$2.80\pm 0.46 $ &$ 0.01\pm 0.11$& $0.23\pm 0.14$\\
$z=0.55$       & $3.10\pm 0.62$&  $ 0.11\pm 0.12$&$0.30\pm 0.10$ &$2.90\pm 0.42$ & $-0.04\pm 0.10$& $0.25\pm 0.11$ \\
$z=0.75$       & $4.60\pm 2.32$&  $-0.07\pm 0.14$&$0.55\pm 0.10$ &$2.50\pm 0.34$ &$ 0.11\pm  0.13$& $0.24\pm 0.15$\\ 
\hline
\end{tabular}
\caption{First 4 rows: Ages (measured in Gyr), metallicities [Z/H] and [E/Fe] 
abundances for the stacked spectra, in different redshift bins, 
of the red-sequence galaxies, measured using V08 models and the indices H$\gamma_F$, Fe4383 and
CN$_2$. The spectra have been stacked separating galaxies with 
$\sigma$ lower than 175 km~s$^{-1}$ (right side) and larger than 175
km~s$^{-1}$ (left side). The parameters have been obtained iteratively as
described in the text. Last three rows: Ages, metallicities
 and [E/Fe] derived for the stacked spectra, at different redshifts, of
morphologically classified E and S0 galaxies.\label{alphayz}}
\end{table*}

\subsection{Massive galaxies}

It can be seen in Table~\ref{alphayz} and Fig.~\ref{index.index.v07}
that the age difference between the redshift bins of the most massive
galaxies ($\sigma > 175$ kms$^{-1}$) corresponds to the expected age
difference of the Universe at those redshifts.  The absolute age
obtained from the models for these galaxies corresponds to a redshift
of 1.4.  In other words, they are compatible with being formed at
$z>1.4$ and evolving passively since then. The metallicity measured
from CN$_2$ does not evolve either, as expected in a passive evolution
scheme.

When we look at the panels using Fe4383 instead of CN$_2$ we can see
that the most massive galaxies at $z=0.75$ seem to have a lower Fe4383
than predicted by passive evolution. \citet{J05} similarly found a
weaker Fe4383 index for more than half of the galaxies in a sample of
red-sequence galaxies (including those with $\sigma$ larger than 175 km~s$^{-1}$)
in the cluster RX J101522.7-1357 at redshift $z=0.83$.  This result,
if confirmed, could be indicating that at least some massive galaxies
have experienced some chemical enrichment since $z=0.75$.  It will be
necessary to measure other Fe-sensitive indices to confirm this
trend. Unfortunately, the wavelength range covered by our spectra does
not allow us to do this  and, therefore, we will not discuss this further.

\subsection{Less-massive galaxies}

Contrary to their massive counterparts, galaxies with $\sigma <$
175~km~s$^{-1}$ do not show any evolution, within the errors, in age
or in metallicity between $z=0.75$ and $z=0.45$.  They also evolve
less than expected from a pure passive scenario between $z=0.45$ and
$z=0.0$.  A similar result was reported by \citet{Sch06} with a {\it
field} galaxy sample.  Comparing galaxies from the DEEP2 survey around $z=0.8$ with an
SDSS local galaxy sample, they showed that the H$\delta_F$ variation
was less than predicted by passively evolving models. However, they do
not group galaxies by mass.  Along the lines discussed by the authors,
our results suggest that either individual low-mass galaxies
experience continuous low levels of star formation, or that the
red-sequence is progressively built-up with new and younger small
galaxies.  This latter hypothesis is supported by a number of recent
works \citep[e.g.][]{Bell04, Pog06, dL07, Har06, Fab07}.  This
differential evolution for massive and less massive galaxies implies
that the {\it mean} difference in luminosity-weighted mean-age between
massive and intermediate-mass red-sequence galaxies increases with
time.  This should be taken into account when studying the evolution
of the colour-magnitude relation with redshift.

Low-$\sigma$ galaxies show a lower value of [E/Fe] than high-$\sigma$
galaxies, in agreement with the results at low redshift. Most of the
$"$enhanced$"$ (E) elements we are considering here arise from massive
stars and the bulk of their mass is released in Type II supernova
explosions, while at least 2/3 of the Fe is released to the
interstellar medium by Type Ia supernovae, with a delay of $\sim$ 1
Gyr. Therefore, the ratio [E/Fe] has classically been used as a cosmic
clock to measure the duration of the star formation.  The lower [E/Fe]
in low-$\sigma$ galaxies can be interpreted as a more extended star
formation history, just as found locally.

Table~\ref{alphayz} also shows that, within the errors, [Z/H] and
[E/Fe] do not vary with redshift for low-$\sigma$ galaxies.
Whether galaxies have experienced low-level star formation or new
galaxies have entered the red-sequence from the blue cloud, it is
still unclear how galaxies that have been forming stars until recently
have the same chemical composition as those where star formation was
quenched 2 Gyr previously (the lookback time between z$\sim$0.45 and
z$\sim$0.75).  Answering this question is not trivial, as there are
multiple paths by which a galaxy may enter and leave the red-sequence.
Furthermore, one has to be careful when interpreting single-stellar
population equivalent parameters. While galaxy mean ages are biased
towards their last episode of star formation, the chemical composition
depends more strongly on the oldest population (see Serra \& Trager
2007 for a quantitative study).  Therefore, galaxies with different
star formation histories can have similar [E/Fe] and [Z/H] if they
have formed the {\it bulk} of their stars on similar time-scales and with a
similar efficiency.

\section{Evolution of the index-$\sigma$ relations with redshift}
\label{sec.indices.mv}

The outcome of the previous section seems to contradict earlier
studies, in which the evolution of the index-$\sigma$ relation was
found compatible with a high-redshift formation and subsequent passive
evolution for all red-sequence galaxies, irrespective of their mass
range (Kelson et al.  2001).  We will now demonstrate that our
index-$\sigma$ relation is also consistent with this scenario, but
that this analysis cannot provide robust constraints on the
star-formation histories of red galaxies. In fact, the evolution of
the index-$\sigma$ relation is also compatible with a more extended
star formation history for the intermediate-mass galaxies if the
progenitor-bias (i.e. the fact that the galaxies with more extended
star formation history (or quenched at later times) drop out of the red-sequence at high redshift
\citet{2000ApJ...541...95V}) is taken into account.
 
\begin{figure*}
\resizebox{0.5\textwidth}{!}{\includegraphics[angle=-90]{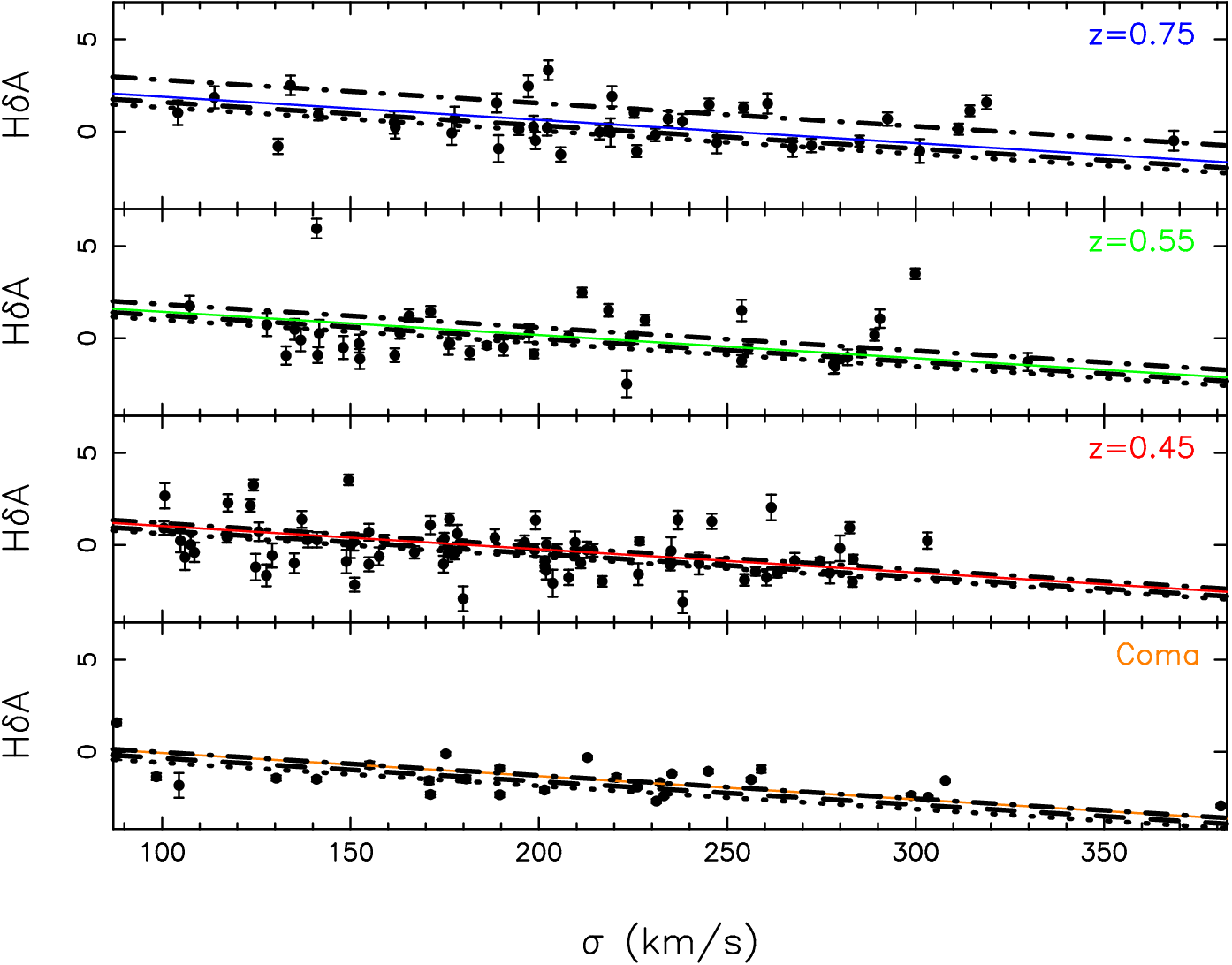}}
\resizebox{0.5\textwidth}{!}{\includegraphics[angle=-90]{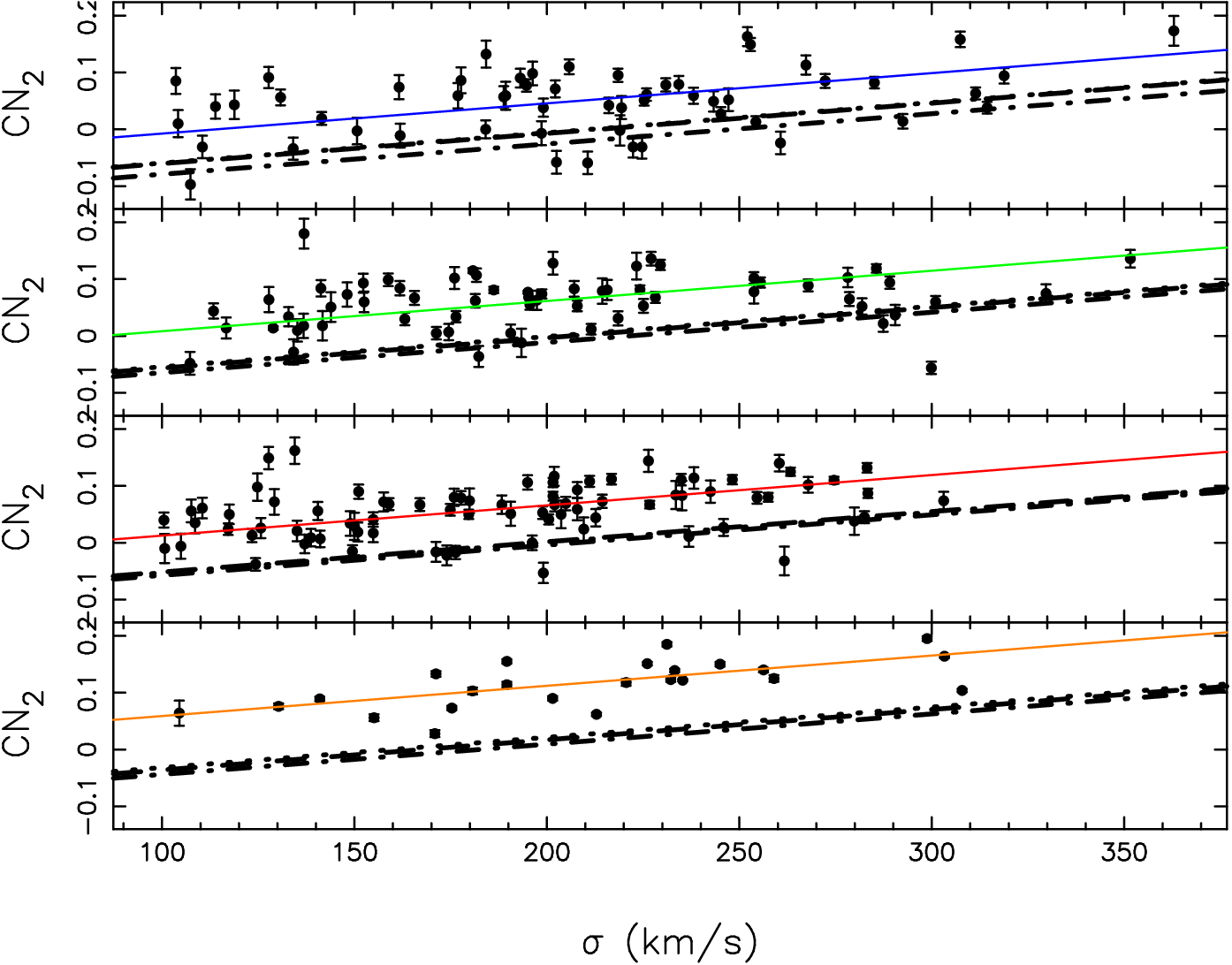}}
\resizebox{0.5\textwidth}{!}{\includegraphics[angle=-90]{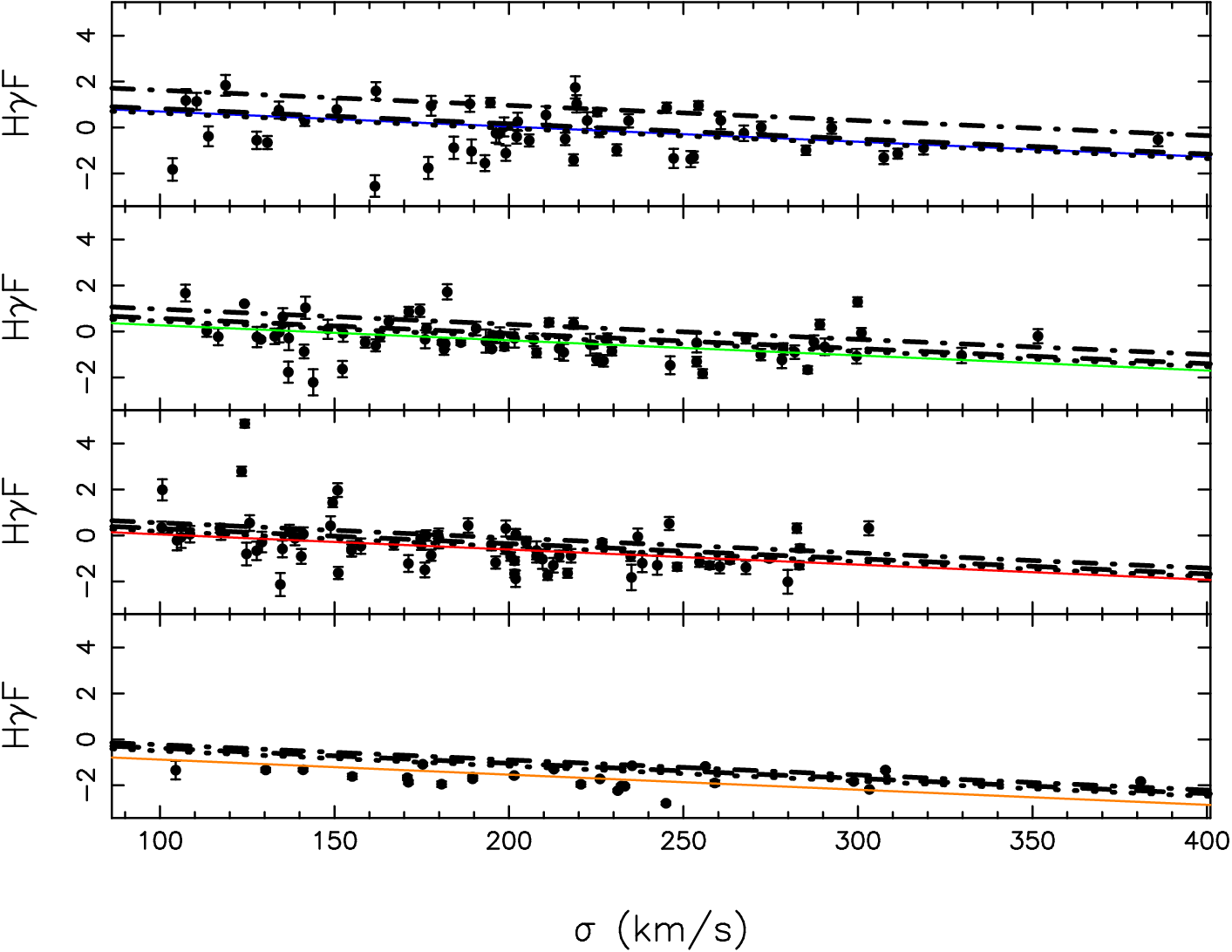}}
\resizebox{0.5\textwidth}{!}{\includegraphics[angle=-90]{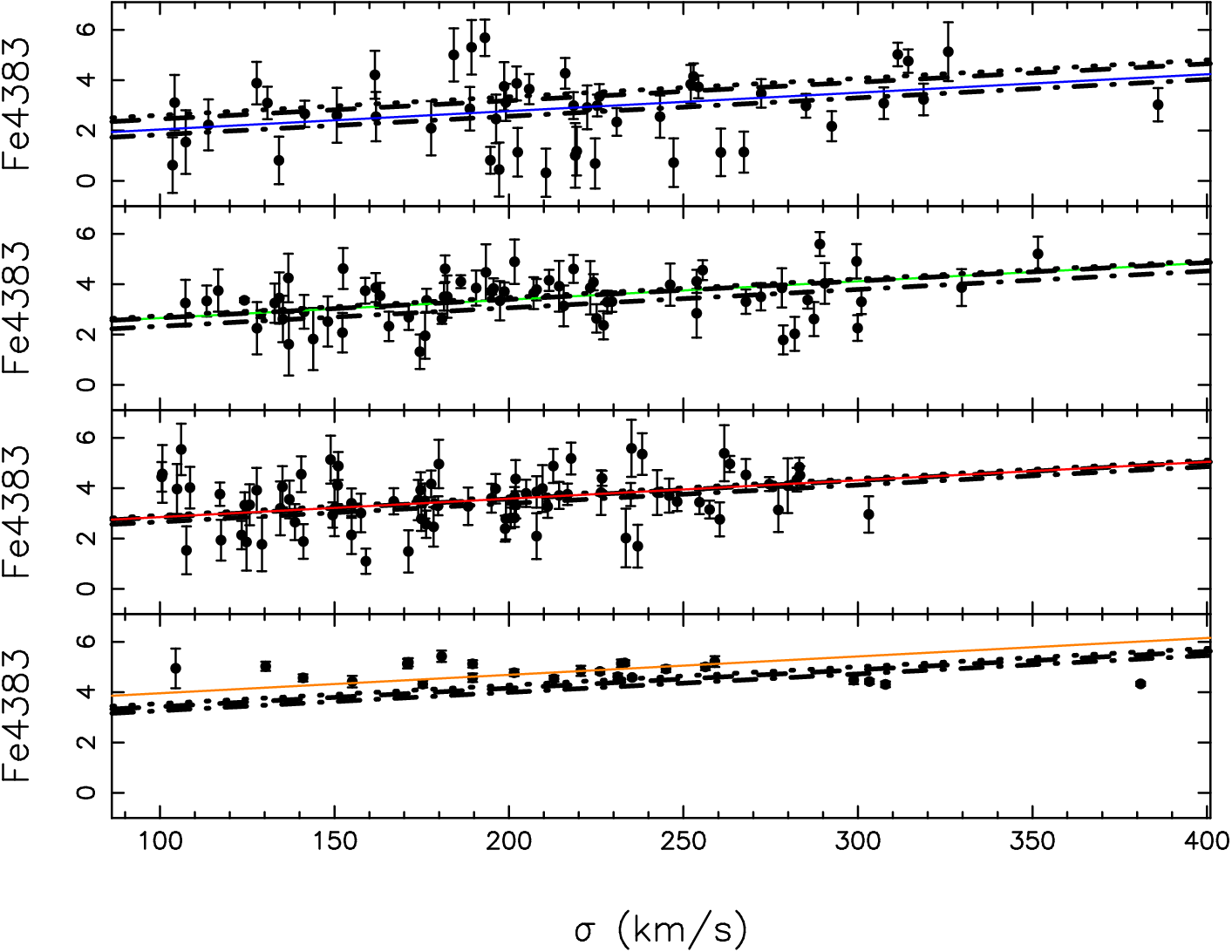}}
\caption{Relation between line-strength indices and the velocity
dispersion for the galaxies in the three different bins. Solid
coloured lines represent the result of a linear fit to the data
forcing the slope to be the same as the one obtained for the galaxies
in the redshift bin at $z=0.45$.  The black lines represent the
expected relation assuming that galaxies formed at $z_{\rm f}$=1.4
(dotted-dashed line), $z_{\rm f}=2$ (thick, dashed line), and $z_{\rm
f}$=3 (dotted line), and have evolved passively since then and
assigning a solar metallicity for galaxies with velocity dispersion of
300~km~s$^{-1}$.  As can be seen, the evolution of the index-$\sigma$
relation is compatible with a scenario in which the stars formed at
$z> 1.4$ and evolved passively since then, except for the CN$_2$ index, 
which cannot be reproduced by any of the proposed models (see text for details)}. However, the evolution of
these relations are also compatible with more complex scenarios (see
Sec.~\ref{ops}).
\label{indices.sigma}
\end{figure*}

Figure~\ref{indices.sigma} displays the relation between H$\gamma_A$,
H$\gamma_F$, CN$_2$, and Fe4383 and the galaxy velocity dispersion,
for the Coma cluster and our EDisCS sample.

Following \citet{Kel01} and \citet{J05}, among others, we first
calculate the best linear fit to the data using the galaxy sample at
$z=0.45$.  Then, keeping the slope of the relation fixed, we evolve
its zero point with look-back-time. 
This approach assumes that, if galaxies were coeval and evolving passively, the
slope of the relation would not change.  In reality, {\it if} all the
galaxies in the red-sequence were coeval and evolving passively, we
would expect a small variation in the slope of the index-$\sigma$
relation, as the variation of the indices is not completely linear
with age.  However, this variation would be {\it very} small over the
range of ages considered. 
 To calculate the lines of passive
evolution, we used the V08 models, assigning solar metallicity to the
galaxies with $\sigma=300$~km~s$^{-1}$, to match the observed
value of the metal-sensitive index Fe4383 at $z=0.45$.  We show the best linear fits in
Fig.~\ref{indices.sigma}, obtained by minimizing the residuals in the
$y$-direction.  We also show the expected evolution of the
index-$\sigma$ relation assuming three different formation redshifts:
$z_{\rm f}=$1.4, 2, and $3$.  The value $z_{\rm f}=1.4$ is chosen as
it corresponds to the mean age measured in the stacked spectra of the
massive galaxies in the 0.75 galaxy group (see
Sec.~\ref{sec.stacked}).  We only plot models with solar-scaled
chemical abundances.

Figure~\ref{indices.sigma} shows that a scenario where all stars form
at $z_{\rm f}>1.4$ and evolve passively afterwards is compatible with the
observations.  The exception is CN$_2$ which we discuss below.  In
order to analyze this more quantitatively, we performed a $t-$test,
comparing the linear fit values at $\sigma=200$~km~s$^{-1}$ (which is
approximately the mean of the distribution of $\sigma$) with the ones
predicted by passive evolution.  The results are shown in
Table~\ref{table.t}.  $t$-values higher than 1.9 would indicate that
the probability that we have rejected the null-hypothesis (prediction
and measurement are equal) by chance is less than $\sim$5\%, i.e., a
$t$-value higher than 1.9 implies that the passive scenario does not
reproduce our relations within the errors.  We do not obtain such
$t$-values, implying that indeed a passive evolution model can
reproduce the evolution of the zero-point of the index-$\sigma$
relations from $z=0$ to, at least, $z$$\sim$0.75.  Although we obtain
a slightly better agreement between model and data if we assume a
formation redshift above $2$, we do not find any statistically
significant difference from a formation redshift of 1.4.  Noticeably,
formation redshifts lower that 1.4 produce an evolution of the
index-$\sigma$ relation significantly larger than the one observed.

The relation of the other indices with $\sigma$ is presented in
Appendix~\ref{appendix.indices}. The analysis of these indices gives
consistent results without adding any new information to our study,
but we show it for comparison purposes with eventual following
studies.  We did not include CN$_2$ in the $t$-test procedure as the
stellar population models with solar-scaled chemical abundances cannot
reproduce the absolute values of this index.  This is a well known
effect in nearby early type galaxies \citep{W98, SB03, Kel06, Sch06,
Grav07} and it is attributed to differences in the chemical
composition of these galaxies compared to the solar values. (Note that
in Sect.~\ref{sec.stacked} we used models where C/Fe and N/Fe are
enhanced with respect to the solar values in order to reproduce this
index).

Summarizing, we show here that, in agreement with previous works, the
evolution of the index-$\sigma$ relation is compatible with a scenario
where the red galaxies formed their stars at z$>$1.4 and have evolved
passively since.  However, this conclusion has been reached assuming
that all red galaxies formed at the same time independently of their
mass.  The results of Sec.~\ref{sec.stacked} argue against this.
Furthermore, numerous works \citep{dL04,dL07,Kod04} indicate that the
red sequence was not yet fully in place at z$\sim$1 and that it has
been growing since then. If this effect is taken into account, a more
complex star formation history is allowed while keeping the slope and
the tightness of the index-$\sigma$ relations. This was proposed by \citet{vDF01}
 to explain the evolution of the magnitude
and the constancy and scatter of the colours on the red sequence.

\citet{Har06} showed that quenched models, where a constant star
formation is truncated at evenly spaced time intervals can explain the
evolution of the {\it mean} H$\delta$ of all the field red galaxies
between $z=1$ to $z=0$.  However, they did not explore the dependence
of this evolution on galaxy mass.  In order to study this aspect, we
have used the \citet{BC03} population synthesis models with a Salpeter
IMF and solar metallicity. We built a series of star-formation
histories starting at z$_{\rm f}$=2 where a constant rate of star
formation of 1 M$_{\odot}$/yr is quenched at different look-back
times.  Whenever a galaxy satisfies our criteria to belong to the
red-sequence, we measure its indices.

\begin{figure*}
\resizebox{0.5\textwidth}{!}{\includegraphics[angle=-90]{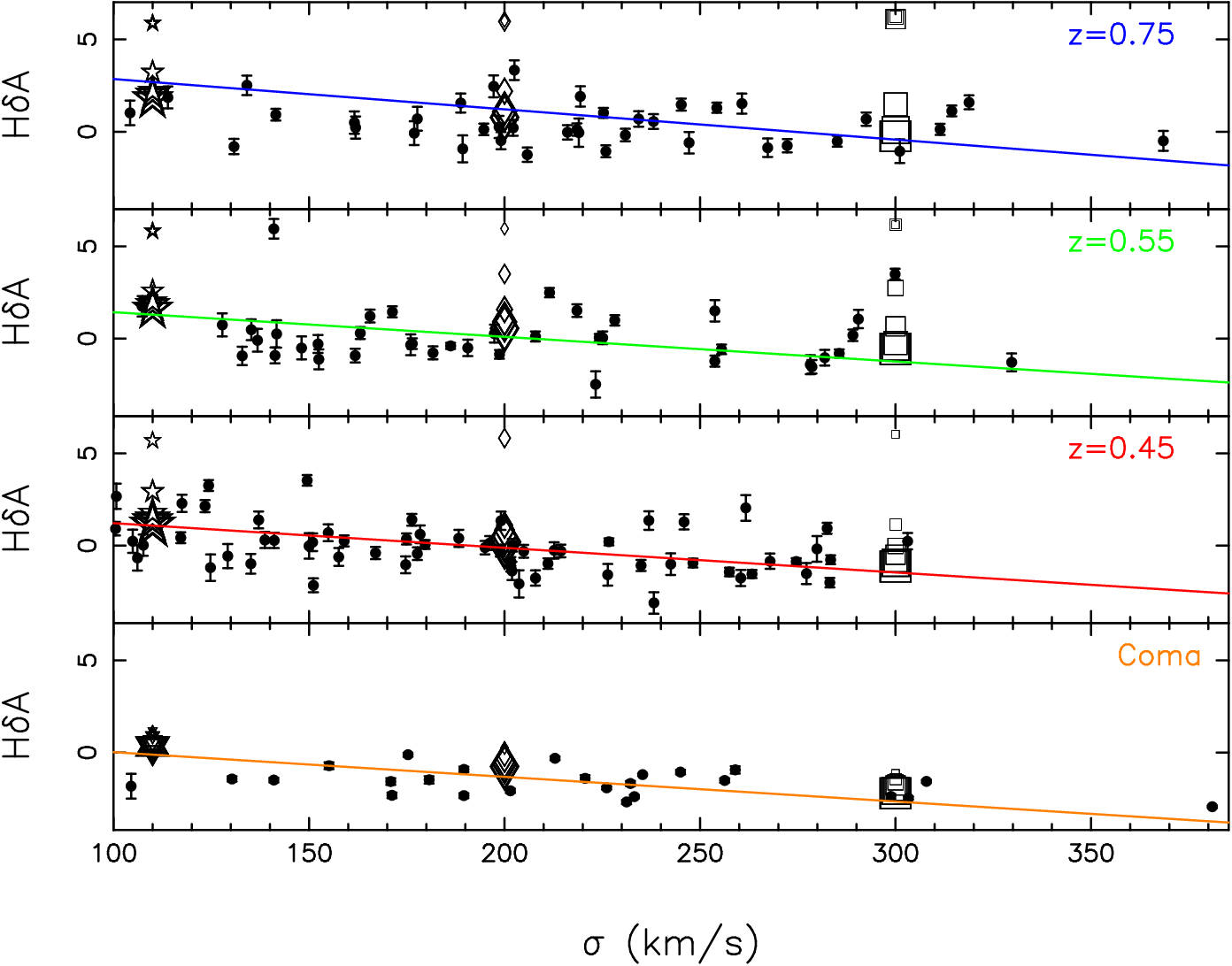}}
\resizebox{0.5\textwidth}{!}{\includegraphics[angle=-90]{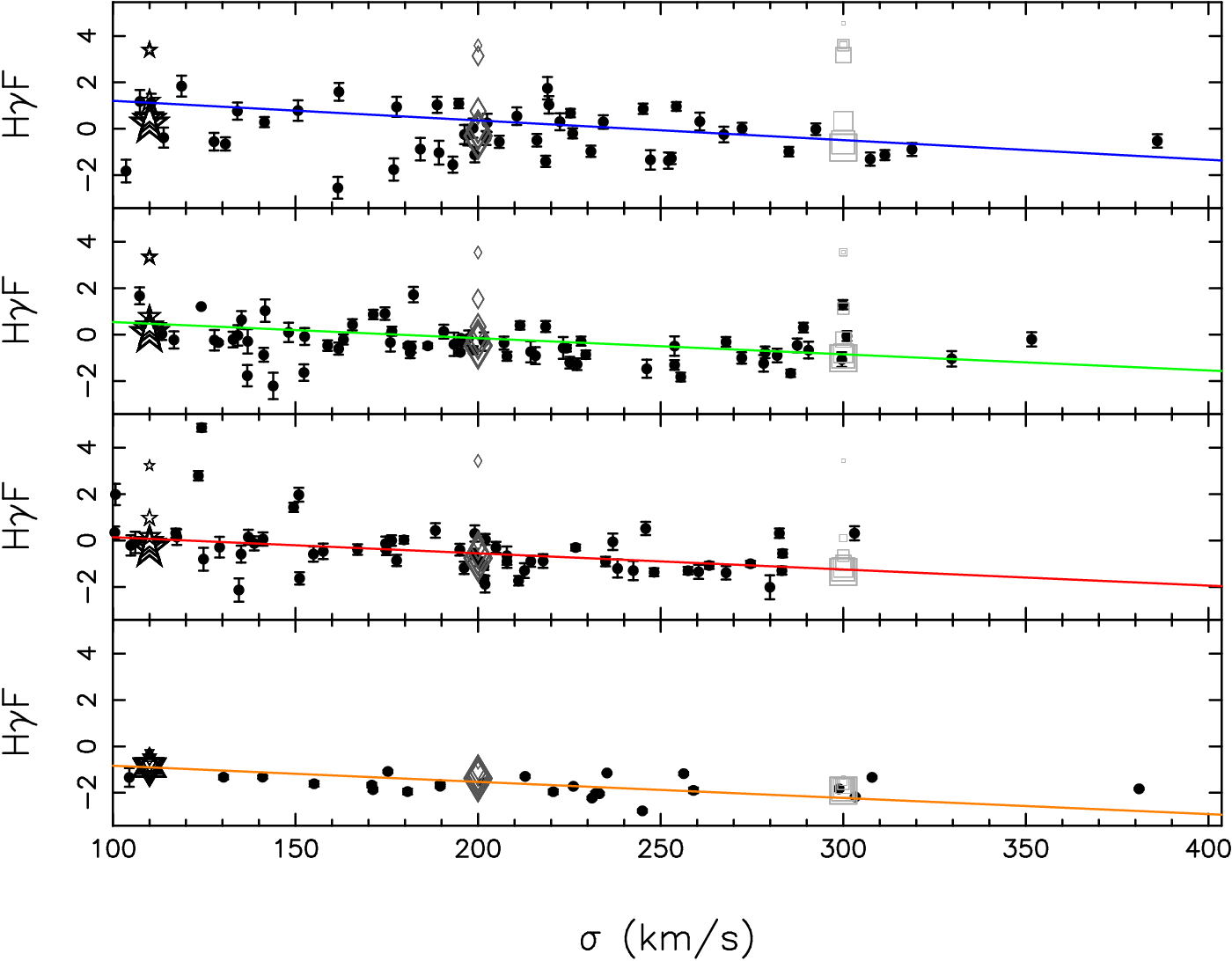}}
\resizebox{0.5\textwidth}{!}{\includegraphics[angle=-90]{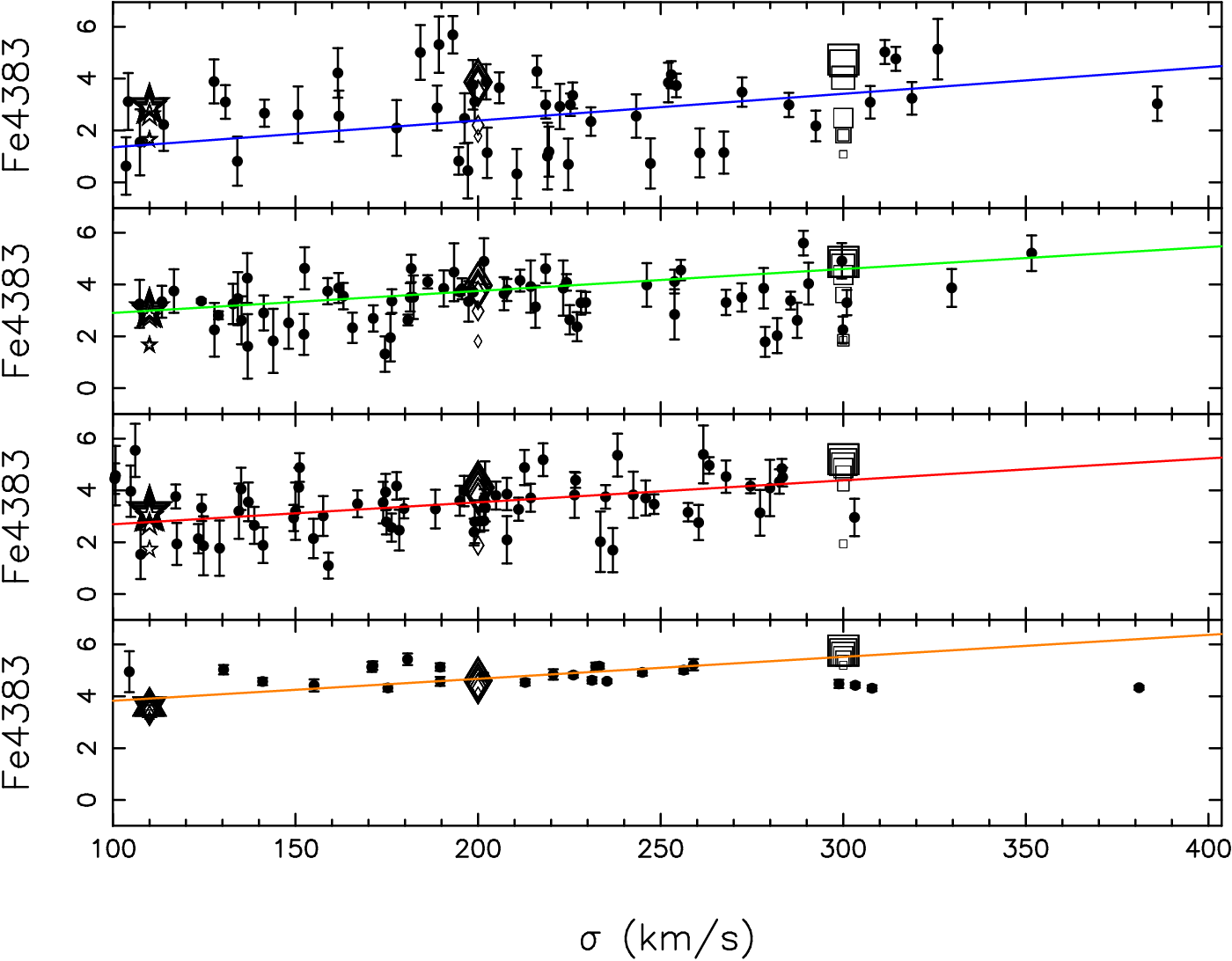}}
\caption{Evolution of the relation of line-strength indices with
$\sigma$. We show with coloured, solid lines, the best linear fits to
the observations, fixing the slope to that obtained for the redshift
bin $z=0.45$ (see Sec.~\ref{sec.indices.mv}).  Squares, diamonds,
and black stars represent the predicted indices for a constant star
formation rate starting at $z_{\rm f}=2$, for stellar populations with
metallicities Z=0.001,0.004 and 0.02, respectively. At each of these
metallicities, the sta r formation is truncated after 1 (largest
symbols), 2, 3, 4 and 5 (smallest symbols) Gyr.}
\label{SFH}
\end{figure*}

\begin{figure*}
\resizebox{0.5\textwidth}{!}{\includegraphics[angle=-90]{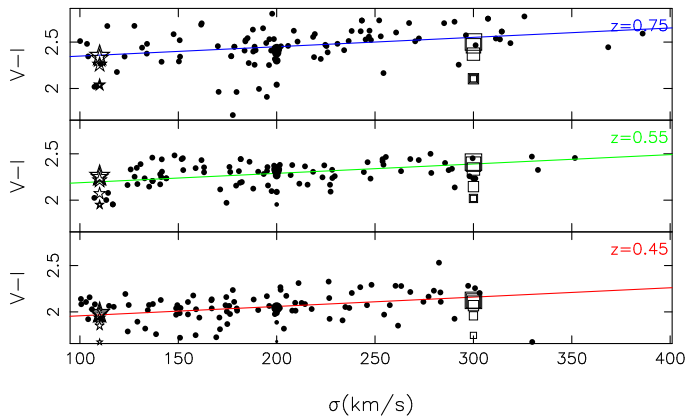}}
\caption{Measured $V-I$ vs $\sigma$ for the galaxies in the different
redshift bins.  Symbols are the same as in Fig.~\ref{SFH}.}
\label{colors.sfh}
\end{figure*}

Figure~\ref{SFH} shows the resulting line-strength indices assuming 3
different metallicities, $Z=$0.001, 0.004 and 0.02.  We display, again,
the relation of the indices with $\sigma$ for each redshift bin
discussed in Sec.~\ref{sec.indices.mv}.  While the index-$\sigma$
relations of the most massive galaxies are only well reproduced by a
scenario where star formation is truncated after 1 Gyr (i.e.  the rest
of the star formation histories produce relations systematically
shifted compared to the observed one), less massive ones can be
described by a variety of star formation histories, in agreement with
our results of Sec.~\ref{sec.stacked}.  Furthermore, this variety of
star formation histories is supported by the larger scatter in
metallicity and age values at a given mass of low-$\sigma$ galaxies
compared to that of the most massive ones \citep[e.g.][]{CRC00}.  Some
studies have claimed that star formation histories of this nature
would have problems reproducing the colours of red-sequence galaxies.
As a sanity check (see Fig.~\ref{colors.sfh}) we verify that these
star formation histories reproduce the colours of the red-sequence.

This, of course, does not prove that different galaxies have different
star formation histories, but it does demonstrate that more
complicated scenarios than pure passive evolution are compatible with
the index-$\sigma$ relations.

\begin{table*}
\centering
\begin{tabular}{rr  rr| rr| rr|rr}
\hline
          &    & \multicolumn{2}{c}{linear fit}&\multicolumn{2}{c}{ $z_{\rm f}=1.4$} &\multicolumn{2}{c}{ $z_{\rm f}=2$} 
&\multicolumn{2}{c}{ $z_{\rm f}=3$}\\
Index         & redshift &  @200~km/s        & $\sigma$  &  @200~km/s    & $t$&  @200~km/s &  $t$  &  @200~km/s & $t$  \\
\hline
\hline 
Fe4383          & 0.02    &$ 4.70$ & 0.41  & $  4.09$ & 1.5  &      $ 4.33$      & 0.9&  $ 4.38$  &  0.8\\ 
                & 0.45    &$ 3.58$ & 0.82  & $  3.39$ & 0.2  &      $ 3.57$      & 0.0&  $ 3.65$  &  0.1\\
                & 0.55    &$ 3.73$ & 0.38  & $  3.06$ & 1.7  &      $ 3.40$      & 0.8&  $ 3.47$  &  0.8\\  
                & 0.75    &$ 2.77$ & 1.24  & $  2.56$ & 0.2  &      $ 3.19$      & 0.3&  $ 3.34$  &  0.4\\
H$\delta$A      & 0.02    &$-1.40$ & 0.79  & $ -1.37$ & 0.0  &      $-1.85$      & 0.6&  $-1.93$  &  0.7\\
                & 0.45    &$-0.23$ & 1.12  & $ -0.08$ & 0.1  &      $-0.47$      & 0.2&  $-0.64$  &  0.4\\ 
                & 0.55    &$ 0.09$ & 1.09  & $  0.58$ & 0.4  &      $-0.02$      & 0.1&  $-0.26$  &  0.3\\
                & 0.75    &$ 0.63$ & 1.10  & $  1.54$ & 0.8  &      $ 0.33$      & 0.3&  $ 0.05$  &  0.5\\
H$\gamma$F      & 0.02    &$-1.48$ & 0.44  & $ -0.86$ & 1.4  &      $-1.09$      & 0.9&  $-1.13$  &  0.8\\
                & 0.45    &$-0.61$ & 0.66  & $  0.10$ & 1.1  &      $-0.36$      & 0.4&  $-0.48$  &  0.2\\
                & 0.55    &$-0.48$ & 0.66  & $  0.31$ & 1.1  &      $-0.09$      & 0.6&  $-0.21$  &  0.4\\
                & 0.75    &$0.04$  & 0.97  & $  0.96$ & 0.9  &      $ 0.16$      & 0.12& $-0.04$  &  0.0\\
\hline
\end{tabular}
\caption{Comparison between the indices predicted by the linear fit
to the data at $\sigma=200$~km~s$^{-1}$ and the ones predicted -- at the
same velocity dispersion-- in a passively evolving model with redshift
formations $z_{\rm f}=1.4$, $z_{\rm f}=2$ and $z_{\rm f}=3$.  Column  3 shows the predicted
value by  the linear  fit at  $\sigma=200$~km~s$^{-1}$, column  4  the
standard  deviation among    the  relation;  columns 5, 7 and 9 show  the   value
predicted by the  model with redshift formations of 1.4, 2 and 3 respectively.
Columns 6, 8 and 10  shows the $t$-parameter
of comparing   these two  values (fitted and predicted)  using a Student's t-test.  A  $t$ value higher   than 1.96 would
indicate that  the probability  that the two  values are different by chance
is less than 5\% (for samples larger than 40 and 6\% for smaller samples).  As can be seen, none of the 
values show significant
differences from the model.\label{table.t}}
\end{table*}

\section{Morphological content of the red-sequence}
\label{sec.morph}
Table~\ref{tb:clusterlist} lists the EDisCS structures wich have been
observed with the HST/ACS in the F814W band, i.e, all six in the 0.75
redshift bin, six at $z=0.55$, and three at $z=0.45$.  
In the following, we use the visual classifications of \citet{Des07}, which
has been performed down to $I_{auto}$\footnote{SExtractor Kron magnitudes}=23 mag.  
Since we concentrate on the EDisCS spectroscopic sample, the actual magnitude cuts for the
morphological classification are I (r $\le$ 1$"$) $\sim$ 22 mag at
$z=0.45$ and I(r $\le$ 1$"$) $\sim$ 23~mag at z=0.75.  This translates
into restframe magnitudes of Mv$\sim -18.7$ and $\sim -$19.3 mag at
z=0.45 and z=0.75, respectively.  Once we account for the dimming due
to passive evolution (0.5 mag from z=0.75 to 0.45 assuming an age of
3.1 Gyr at z=0.75), the intrinsic magnitude cut is  the same at
all redshifts.  We note that some secondary structures at low redshift
were discovered in high-redshift targeted fields.  Consequently, they
were observed down to lower magnitude limits than the primary targets
at similar redshifts.  As such, they are exceptions to the above
general rule.  However, they represent a small percentage of the 
total number of galaxies.

Interestingly, only 15 E/S0 galaxies out of 168 (i.e., 9\%) are not on
the red-sequence.  This is similar to the fraction of blue early-type
galaxies found at low redshift \citep{Bam08}, which is highly
dependent on local density but varies between 12 and 2\%.

\begin{figure}
\resizebox{0.5\textwidth}{!}{\includegraphics[angle=-90]{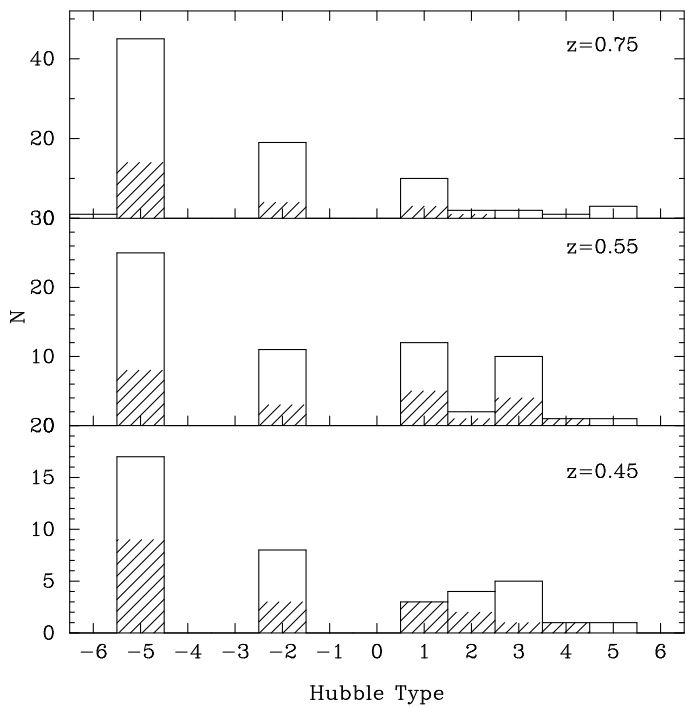}}
\caption{The distribution of Hubble types among the N+W emission line
red-sequence galaxies in redshift bins: E(T=-5), S0(T=$-2$), Sa
(T=1), Sab(T=2), Sd(T=7). The solid line histograms trace the
morphological distribution for the entire sample. The dashed areas
trace the histograms for the galaxies with velocity dispersions lower
than 175 km~s$^{-1}$.}
\label{Hubble.Sigma.Type12}
\end{figure}

We concentrate on the N+W red-sequence galaxies, the same for which we
have analyzed the stellar population. Figure~\ref{Hubble.Sigma.Type12}
shows their Hubble type distribution.  We also indicate the
distribution of galaxies with $\sigma$ $<$ 175km/s.  There is no
apparent morphological segregation as a function of galaxy mass, i.e,
whatever the galaxy mass range, the dominant morphological types on
the red-sequence are clearly E and S0, although the entire Hubble
sequence is covered.

In order to look for possible evolution with time, we now restrict the
sample to the EDisCS clusters (galaxy structures with $\sigma$ larger than 400
km~s$^{-1}$) and fix the area of analysis to the region falling inside
a radius of 0.6R$_{200}$ from the cluster centers, where R$_{200}$ is
the radius within which the average cluster mass density is equal to
200 times the critical density.  R$_{200}$ values were derived using
equation (8) in \citet{Finn05} and are listed in Table
\ref{tb:clusterlist}.  These restrictions are governed by the need to
have the same spatial coverage in all structures
given that a morphology-radius relation is observed in clusters
(e.g. Whitmore \& Gilmore 1991; Postman et al. 2005).  
 The fraction of galaxies from each morphological type 
have been corrected for magnitude and spatial
incompleteness \citep{Pog06}. Errors in these fractions are calculated using the formulae in
\citet{1986ApJ...303..336G}, since Poissonian and binomial statistics
apply to our small samples.

The fraction of early-type galaxies (E and S0) at 
redshifts $z=0.45$, 0.55 and 0.75 are:
56$^{+12}_{-13}$\%, 61$^{+9}_{-9}$\% and 75$^{+9}_{-9}$\%, 
respectively.  In other words,  our calculations suggest that, 
in the red-sequence, the fraction of early-type galaxies 
{\it decreases} from $z=0.75$ to $z=0.45$.

 In order to check the
robustness of this result, we also consider another technique of
morphological classification. In particular
 the GIM2D decompositions presented in Simard et al.  (2008, in
preparation), where early-type galaxies are identified by their bulge
fraction ($B/T$$\ge$ 0.35) and the ACS image smoothness within two
half-light radii ($S2$ $\le$ 0.075).  This method leads to
33$^{+14}_{-12}$\%, 62$^{+8}_{-9}$\%, and 80$^{+5}_{-6}$\% of
early-type galaxies at $z=0.45$,0.55 and 0.75, respectively, in
agreement with our previous finding.  As mentioned earlier, only 3
clusters have ACS data in the 0.45 bin, and only 22 galaxies within
0.6R$_{200}$, as these clusters did not constitute our primary targets
for HST.  This most likely explains the large difference found at this
redshift between the two classification techniques.  Nevertheless, the
decrease of early-types fractions in the red-sequence
 is confirmed.  It is worth noting that even when
strong emission line galaxies are kept in the sample, the fractions of
early-type galaxies are very much the same, i.e., 62$^{+10}_{-9}$\%
and 72$^{+5}_{-7}$\% at z=0.55 and z=0.75.

\begin{figure}
\resizebox{0.5\textwidth}{!}{\includegraphics[angle=-90]{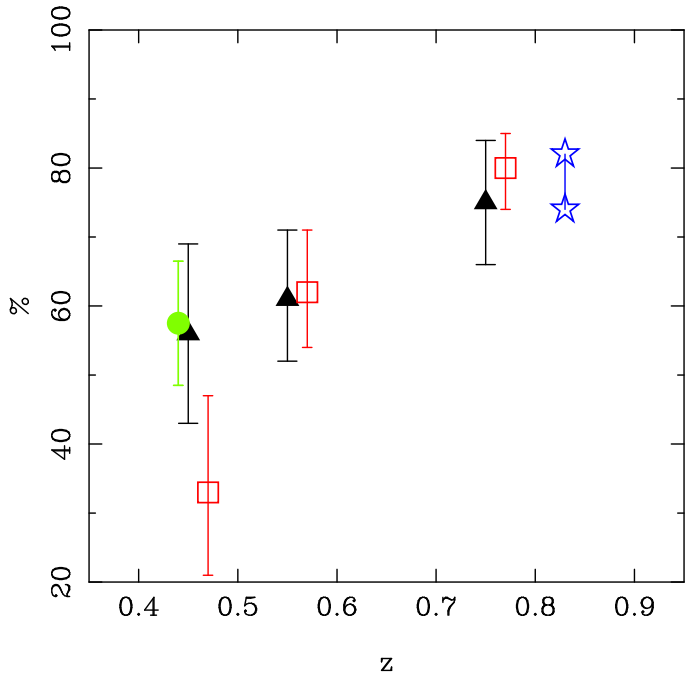}}
\caption{Variation of the fraction of early-type galaxies in the red-sequence with
redshift.  Black and red symbols show the fraction derived from the
EDisCS data.  Black triangles show the results for the morphological
classification based on visual inspection.  Red squares stand for the
GIM2D bulge and disk decomposition method, where early-type galaxies
are selected upon their bulge-to-total light fractions and image
smoothness.  The green circle represents MORPHS and the blue star
indicates the results of Holden et al. (2006), for their two X-ray
selected clusters.}
\label{Hubble.Fraction}
\end{figure}

The mean redshift of the MORPHS sample is 0.45 \citep{Dress99}, hence
allowing direct comparison with our lowest redshift group and
increasing the statistical significance (10 clusters).  After
selection of the red galaxies (g$-$r$>$1.4) without emission lines,
i.e, a selection comparable to ours, the fraction of MORPHS E+S0s is
57$\pm$9\% (B.  Poggianti, private communication), i.e, very close to
the initial value coming from our visual classification.  Additional
evidence of a decrease in early-type fraction with time on the
red-sequence comes from \citet{2006ApJ...642L.123H} who measured a
total fraction of E/S0 galaxies in two clusters at z=0.83 (MS
1054.4--0321 and Cl 0152.7--1357) of 74\% and 82\%, for their
selection in mass ( $ > 10^{10.9}$ M$_{\odot}$ ), best corresponding
to our red-sequence criteria.  Note that our selection extends to
lower masses or accordingly to fainter limits, i.e., $3-6\times
10^{10}$ M$_{\odot}$ (derived from stellar M/L with the calibration by
\citet{2001ApJ...550..212B} and the re-normalization proposed by
\citet{dJB07} for elliptical galaxies).  However, in this fainter
regime, Holden et~al.\ take into account all cluster galaxies, rather
than only those on the red-sequence, making impossible a direct
comparison.  Anyhow, their figure~2 shows that their early type
fraction can not vary much along the red-sequence.
Figure~\ref{Hubble.Fraction} summarizes our results.

 The decrease, on the red-sequence, of the early-type fraction  represents a
remarkable synthesis between the results of \citet{Des07} who detect
no evolution in the cluster {\it total} fraction of E/S0 galaxies
between redshift 0.4 and 1.4, and those of \citet{dL07} who measure a
decrease of the {\it red-sequence} faint-to-luminous galaxy ratio of
$\sim$40 \% between redshift 0.45 and 0.75.  Indeed, if the
red-sequence keeps building up, as suggested by \citet{dL07}, it must
find its supply in the cluster blue galaxies.  These EDisCS blue
galaxies are composed of only 20$\pm$7\% early-types and of 80\%
spiral galaxies.  This distribution in types is constant with
redshift.  Hence, provided that galaxy colours redden before any
morphological transformation occurs, an increase in the number of
spirals on the red-sequence is naturally expected when new galaxies
reach it.  By restricting our analysis to the EDisCS spectroscopic
sample, we do not reach as faint magnitudes as \citet{dL07}.  As a
consequence, our 20\% variation in the fraction of E/S0 galaxies on
the red-sequence between redshifts 0.45 and 0.75 most certainly
constitutes a lower limit.

It is interesting to estimate the subsequent evolution of the
red-sequence population, i.e., from $z=0.45$ to $z=0$.  \citet{Dres97}
and \citet{Fas00} found that, while the elliptical fraction does not
evolve in the redshift range 0.5 to 0.0, the fraction of lenticular
galaxies doubles. This increase in the number of S0s is accompanied by
a drop in the fraction of (Sp+Irr) type galaxies.  Combining these
results together with those of \citet{Post05} and EDisCS,
\citet{Des07} conclude that $z\sim 0.4$ must constitute a special
epoch after which the {\it total} fraction of S0-type galaxies in
clusters begins to increase. Meanwhile, \citet{dL07} measure only
$\sim 10$ \% variation of the red-sequence faint-to-luminous galaxy
ratio between redshift 0.45 and 0.  As a consequence, at the time the
red sequence reduces its growth ($ z \le 0.45$), one can expect a rise
in its E/S0 fraction as well.

\section{Discussion}
\label{ops}

We have shown that the rate at which red-sequence galaxies evolve
depends on their mass range.  In particular, less massive galaxies show
evidence for a more extended star formation history.  Two possibilities
are considered: i) the sequence is in place at redshift $z\sim 0.8$,
but a low level of star formation continues in galaxies on the faint
end \citep[e.g.][]{CC02}; or ii) the red-sequence is continuously
built-up by new incomers, preferentially selected among faint systems.
We examine both hypotheses.

\citet{Geb03} explore, in the context of the first model, the maximum
amount of low-level star formation that the galaxies can experience
without leaving the red-sequence.  They found that, to keep the
colours of the red-sequence galaxies almost constant since $z=1$, in
agreement with observations, only 7\% \footnote{the percentage varies
between 4 and 10\% depending on the chosen metallicities} of the mass
could have been formed after the initial burst (assumed to be at
$z=$1.5) using exponential star formation histories with e-folded time
of 5 Gyr.  We did a similar experiment, and calculated the maximum
amount of low-level star formation that the galaxies could experience
from $z=0.75$ to $z=0.45$ in order to reproduce the evolution of the
index-$\sigma$ relations with redshift measured in
Sect.~\ref{sec.indices.mv}.  The exact percentage depends on the
chosen metallicity but we also find that this fraction has to be
always less than $\sim$10\% (note that our redshift baseline is
smaller).  However, recent results of the evolution of the UV colours
in early-type galaxies suggest that this percentage is higher:
\citet{Kav08} found that, at intermediate luminosity (M$_{\rm v} <
-20.5$), early-type galaxies in the red-sequence have formed 30-60\%
of their mass since $z=1$.  However, this study does not sample
exclusively galaxies in dense clusters, but in a wide range of
different environments.

\citet{Fab07, Pog06} proposed a mixed scenario where red-sequence
galaxies form through by two different channels. \citet{Pog06}
discussed a scenario where the red-sequence is composed of primordial
galaxies formed at $z_{\rm f}>$2.5 on the one hand and of galaxies
that quenched their star formation due to the dense environment of
clusters on the other hand.  They calculate that, of the 80\% of
passive galaxies at z=0, 20\% are primordial and 60\% have been
quenched.  With regards to this later aspect, \citet{Har06} could
reproduce the evolution of $U-B$ and the H$\delta$, as well as the
galaxy number density, by assuming models of constant star formation
histories truncated at evenly spaced intervals of 250 Myr up to now.

Along this line, we calculate the number of new quenched galaxies that
have to enter the red sequence to reproduce the results obtained in
Sec.~\ref{sec.stacked}.  We start with a population of galaxies with a
formation redshift of $z_f=1.4$ and solar metallicity, both derived
for the low-mass galaxy bin at $z=0.75$.  Then we consider galaxies
that started forming stars at $z=2$ at a constant rate of
1M$_{\odot}$/yr and start quenching them at regular intervals of 250
Myr.  Once a galaxy is red enough to pass our red-sequence selection
criteria at $z=0.45$ and 0.55, we add its normalized spectrum to the
that of the initial population.  If we assume that all the new
galaxies have velocity dispersions lower than 175 km~s$^{-1}$, we need
$\sim$40\% of new arrivals in the interval z=0.75 to z=0.45 in order
to reproduce the constant luminosity-weighted age of the low-$\sigma$
galaxy bins with redshift.  This is in surprisingly exact
correspondence with the result of \citet{dL07}, who measure a decrease
of the red-sequence faint-to-luminous galaxy ratio of $\sim$40 \%
between redshifts 0.45 and 0.75.  Note that \citet{dL07} deal with
photometric data and therefore their dataset reaches fainter
magnitudes than our spectroscopic sample.  The exercise presented here
is of extremely simple nature and is not intended to represent the
true star formation histories of the galaxies.  Apart form quenching,
disk galaxies can merge and have bursts of star formation before
becoming red. By merging with other galaxies, our objects can also
move from the low-$\sigma$ bin to the high-$\sigma$ bin, which is
something not considered here. However, simple models similar to this
one have proved to be very successful in reproducing the evolution of
the colours and H$\delta$ index and the evolution of the luminosity
function (e.g., Harker et al. 2007).

 Linking these constraints to our morphological results, it appears
that the mechanism that quenches star formation does not necessarily
produce a morphological transformation {\it at the same} time, but
certainly, it favors it. Most likely, the mechanism that quenches star
formation also produces the morphological transformation, but on a
different time-scale.  Indeed, the fraction of spiral galaxies evolves
by 20\% and not by the 40\% calculated in the spectroscopic analysis.
Splitting between stellar population and morphologies was also raised
by MORPHS \citep{Dress99,Pog99}.


\section{Summary}
\label{sec.summary}

We  addressed the questions of  the epoch of  formation of the reddest
galaxies in clusters, the extent of their period of star formation and
the link between morphological and  stellar population evolution  time
scales.

Our analyses are primarily based on 215 red-sequence galaxies,
selected from the EDisCS spectroscopic database, that we divided into
three redshift bins, $z=0.75$, $0.55$, and $0.45$. We considered their
mass range, via the proxy of their velocity dispersions; their stellar
population properties derived from absorption line features; and their
morphologies, thanks to HST/ACS imaging.  We have been able to trace
the evolution of the red-sequence from $z=0.75$ to $z=0.45$ in a
homogeneous dataset, the largest to date, extracted from a single
survey, hence avoiding a mix of different systematic errors.

$\bullet$ Before discarding red-sequence galaxies with strong emission
lines from our absorption line analysis, we investigated the nature of
their ionizing sources. Most of the EDisCS red-sequence galaxies with
emission lines seem to be forming stars.  The proportion of dusty star
forming galaxies among our total sample is larger than the fraction
reported by \citet{Yan06} at $z=0$ in the SDSS field red
population. Whether this difference is due to evolution or to
environment is worth clarifying in the future.

$\bullet$ We measured 12 Lick/IDS indices, carefully visually
inspecting the spectra in order to eliminate those possibly affected
by sky subtraction residuals or any other systematic effect. To
compare with our local sample of Coma galaxies, we derived new
aperture corrections.  We used state-of-the art stellar population
models to derive ages, metallicities, and chemical abundance ratios.
These models make predictions over the whole spectral energy
distribution, allowing us to analyze our data at a resolution of 325
km/s, avoiding any spurious correction for the galaxy velocity
dispersion.

$\bullet$ After selecting on secure redshifts and signal-to-noise
ratios, we stacked the galaxy spectra in redshift bins.  In each bin,
we also distinguished galaxies with respect to their velocity
dispersions, dividing at $\sigma = 175$~km~s$^{-1}$.  We derived the
age, metallicity (Z), and $\alpha$-element abundance ([E/Fe]) of each
redshift and velocity dispersion group.

\begin{itemize}

\item Massive galaxies ( $\sigma > 175$~km~s$^{-1}$ ) show a variation
in age corresponding to the expected cosmological variation between
redshifts, a mean solar metallicity and an overabundance [E/Fe] with
respect to the solar values.  Therefore, they are well represented by
a scenario of formation at high redshift, followed by passive
evolution.  Conversely, the properties of less massive galaxies
($\sigma< 175$ kms$^{-1}$) require longer star formation, at low
level.  Indeed, their luminosity weighted ages is found constant with
time in our redshift range.  An immediate consequence is that the age
difference between low- and high-$\sigma$ galaxies increases with
time.  This need to be taken into account in studies of the evolution
of the colour-magnitude diagrams.

\item Values of [E/Fe] and [Z/H]  are  constant  with time, independent
of  $\sigma$.    This implies that  the  bulk  of  the  galaxy stellar
population is  formed on a time-scale  and  with efficiency fixed by  the
galaxy mass. Possible subsequent episodes   of star formation,   which
change  the  age of the less-massive  galaxies,  can not account for a
large fraction of the galaxy mass.

\end{itemize}

$\bullet$ We confirm that the evolution of the zero-points of the
index-$\sigma$ relationships with redshift is compatible with a
scenario in which galaxies formed all their stars at high redshift and
evolved passively since then.  However, we demonstrate that it is also
compatible with more complex star formation histories.  In particular,
galaxies can progressively enter the red-sequence as their star
formation is quenched. In this scenario, galaxies with lower $\sigma$
enter the red-sequence at lower redshift than more massive galaxies.\\

The two more important results of this work are:

$\bullet$ The morphological analysis, based on visual or automated
classification, indicates that red sequences are composed of types
covering the entire Hubble sequence, irrespective of the redshift,
with, however, a very clear plurality of Es and S0s. The fraction of
red-sequence early type galaxies decreases between $z=0.75$ and
$z=0.45$ by about 20\%. This means that spiral galaxies get first
redder, stopping their star formation, before they turn into earlier
morphological types.  This evolution is however lower than the rate at
which new galaxies enter the red-sequence, meaning that quenching and
morphological transformation operate on different time-scales.

$\bullet$ We found that the evolution of the line-strength indices with 
redshift can be reproduced if 40\% of the galaxies with $\sigma$ lower than 175 kms$^{-1}$
entered the red-sequence between $z=0.75$ and $z=0.45$, in agreement
with the fraction derived in studies of the luminosity function.

In summary, a number of works dedicated to the evolution of the luminosity
function of red galaxies in clusters (Rudnick et al.  2008, in
preparation; De Lucia et al.\ 2004; De Lucia et al.\ 2007) and in the
field (Faber et al.\ 2007; Brown et al.\ 2007) report an increase of
the galaxy number density since $z\sim$1.  Conversely,
colour-magnitude diagrams and index-$\sigma$ relations of cluster
red-sequence galaxies at intermediate- and high-redshift usually
describe them as in place very early, passively evolving,
independently of their mass \citep[e.g.,][]{Ell97, Dres97, vD98,
SED98, K00c, Zie01, Blak03, Wuy04, Tran07}.  The present work
reconciles the luminosity functions and stellar content outcomes.

\begin{acknowledgements} We thank the anoymous referee for useful suggestions
that have improved the final presentation of the paper.
PSB is supported by a Marie Curie
Intra-European Fellowship within the 6th European Community Framework
Programme. The Dark Cosmology Centre is funded by the Danish National
Research Foundation. \end{acknowledgements} \bibliographystyle{aa}
\bibliography{aa3}{}

\begin{appendix}

\section{Transformation to the Lick system}
\label{lickoffsets}
Although the new generation of stellar population models is starting
to make predictions, not only of individual features, but of the whole
spectral energy distribution \citep[e.g.][]{V99, v03, BC03}, some of
the most popular ones \citep[e.g.][]{w94, TMB03} still base their
predictions on the original fitting functions released by the Lick
group \citep{Gor93, Worea94}. In order to compare with these models
and with the bulk of the data in the literature, which is usually
transformed onto this system, one has to follow two steps. The first
step is to degrade the resolution of the spectra into the wavelength
variable resolution of the Lick system \citep[see]{WO97} and then
apply a correction due to the velocity dispersion of the galaxies.
The second step is to correct small systematic differences in the
indices due to differences in the shape of the spectra of the Lick/IDS
stars and a fully-flux calibrated spectrum. To do that, one usually
observes stars in common with the Lick library with the same
instrumental configuration as the objects one wants to analyze and
compare the indices after degrading their spectra to the Lick
resolution.  This cannot be done, though, in the growing number of
work analyzing Lick indices in galaxies at high redshift. For these
cases when stars in common with Lick are not present, it is still
possible to correct for the instrumental response of the Lick/IDS
stars by applying an offset derived by comparing flux calibrated
spectra with Lick/IDS spectra previously degraded to the Lick/IDS
resolution.  This experiment has been attempted before by \citet{WO97}
and \citet{BC03} using the Jones (1999) and STELIB \citep{Leborg03}
libraries respectively.  Here we present new offsets using the MILES
library \citep{SB06_miles}.  The advantage of this calibration is that
the MILES library contains a large number of stars (242) in common
with the Lick/IDs library.  Furthermore, the flux calibration is very
accurate (better than 1\%)

To obtain the offsets we broadened the MILES spectra to the
wavelength-dependent resolution of the Lick/IDS spectra using figure~7
in \citet{WO97} and measured the Lick/IDS indices with the definition
given in \citet{T98}.  In some cases, differences between the indices
measured in both datasets can be described with a linear fit, better
than with a single offset. Therefore, we have also derived a linear
transformation between both libraries obtained with a simple linear
regression eliminating galaxies deviating more than 3 times the rms
from the relation.

\begin{table}
\centering
\begin{tabular}{lrr}
\hline
Index  &  offset     &        Linear relation~~~~~~~~~~         \\
       &(MILES-Lick) &  (MILES=a+ b $\times$ Lick)            \\
\hline
\hline

CN$_1$     &$  0.007$      &$  0.0082 (0.0017) + 0.9112 (0.0124)$ \\
CN$_2$     &$  0.006$      &$  0.0101 (0.0021) + 0.9040 (0.0136)$ \\
Ca4227     &$ -0.081$      &$  0.0627 (0.0256) + 0.9046 (0.0119)$ \\
G4300      &$ -0.032$      &$  0.5795 (0.0663) + 0.8480 (0.0126)$ \\
Fe4383     &$ -0.587$      &$ -0.1660 (0.0798) + 0.9321 (0.0154)$ \\
Ca4455     &$ -0.391$      &$ -0.0752 (0.0340) + 0.7849 (0.0204)$ \\
Fe4531     &$ -0.109$      &$  0.3201 (0.0544) + 0.8541 (0.0164)$ \\
Fe4668     &$  0.050$      &$  0.5664 (0.0649) + 0.8873 (0.0110)$ \\
H$\beta$   &$  0.068$      &$  0.1535 (0.0283) + 0.9755 (0.0108)$ \\
Fe5015     &$ -0.629$      &$  0.1940 (0.0543) + 0.8573 (0.0088)$ \\
Mg$_1$     &$ -0.005$      &$ -0.0024 (0.0001) + 0.9263 (0.0082)$ \\
Mg$_2$     &$ -0.020$      &$ -0.0043 (0.0012) + 0.9186 (0.0050)$ \\
Mgb        &$ -0.208$      &$  0.2201 (0.0351) + 0.8934 (0.0089)$ \\
Fe5270     &$ -0.151$      &$  0.1311 (0.0393) + 0.9008 (0.0147)$ \\
Fe5335     &$ -0.060$      &$  0.1059 (0.0361) + 0.9268 (0.0150)$ \\
Fe5406     &$  0.045$      &$  0.2190 (0.0242) + 0.8543 (0.0151)$ \\
Fe5709     &$ -0.029$      &$  0.0135 (0.0227) + 0.9162 (0.0242)$ \\
Fe5782     &$ -0.029$      &$  0.0157 (0.0208) + 0.9361 (0.0267)$ \\
Na5895     &$ -0.148$      &$ -0.1842 (0.0352) + 1.0038 (0.0107)$ \\
TiO$_1$    &$ -0.014$      &$ -0.0066 (0.0006) + 0.9128 (0.0065)$ \\
TiO$_2$    &$ -0.010$      &$  0.0015 (0.0007) + 0.9119 (0.0045)$ \\
H$\delta_A$&$  0.083$      &$  0.0286 (0.0512) + 0.9354 (0.0115)$ \\
H$\gamma_A$&$ -0.073$      &$ -0.2029 (0.0565) + 0.9607 (0.0082)$ \\
H$\delta_F$&$ -0.013$      &$  0.0069 (0.0329) + 0.9585 (0.0147)$ \\
H$\gamma_F$&$ -0.051$      &$ -0.0750 (0.0258) + 0.9763 (0.0091)$ \\
\hline
\end{tabular}
\caption{Mean index offsets between he stars in common between MILES and 
the Lick/IDS. Some times the differences can be better described with 
a linear relation than a simple offset. Last column display the best linear relation describing 
the relation of the indices measured in MILES and Lick/IDS spectra. \label{offset.licks}}
\end{table}

In principle, the offsets between the indices measured in the Lick and
other libraries are due to differences in the continuum
shape. However, as was noted before by \citet{WO97}, some of the
indices with larger differences have very narrow definition bands and
show a very small dependence with the flux shape. Therefore, the
reason for the differences is not clear.

The mean offsets and linear relations obtained in the comparison
between the MILES and Lick/IDS libraries are listed in
Table~\ref{offset.licks}.

\begin{figure*}
\resizebox{0.15\textwidth}{!}{\includegraphics[angle=-90]{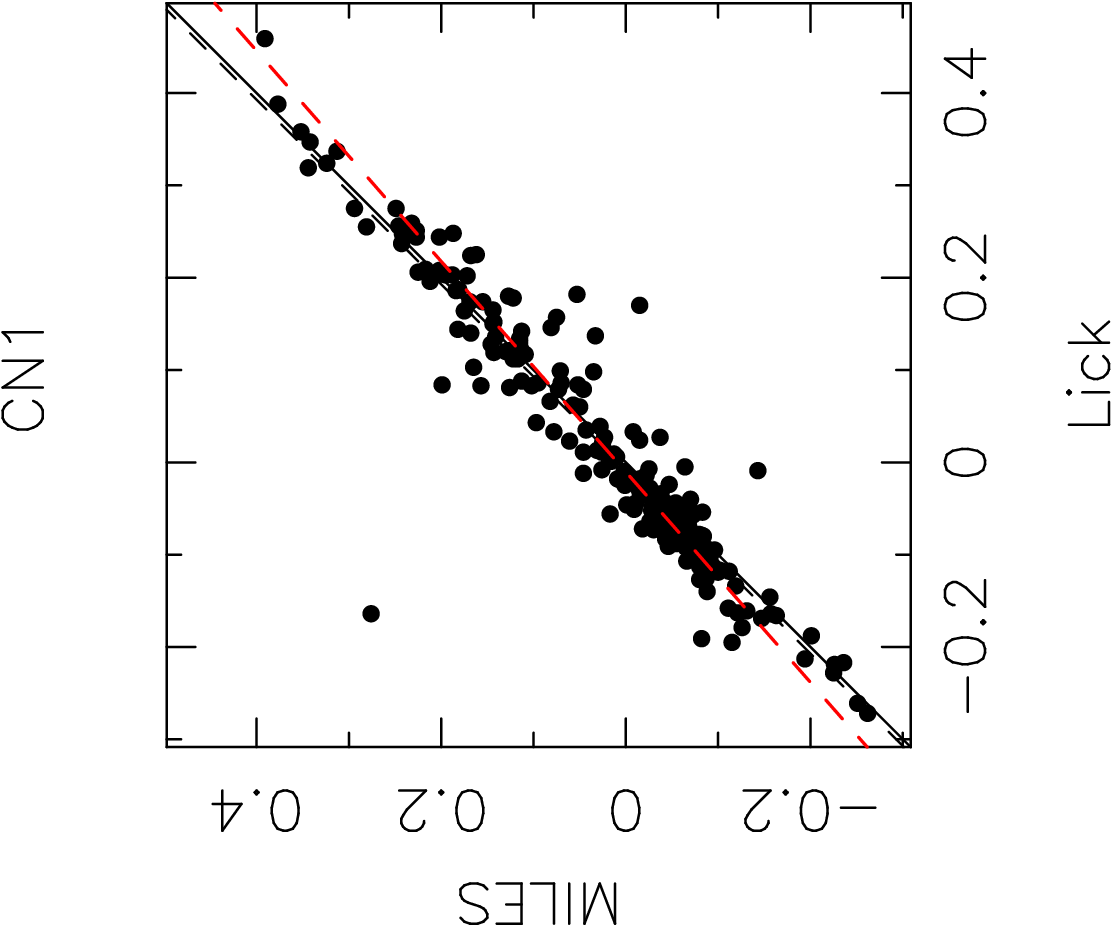}}
\resizebox{0.15\textwidth}{!}{\includegraphics[angle=-90]{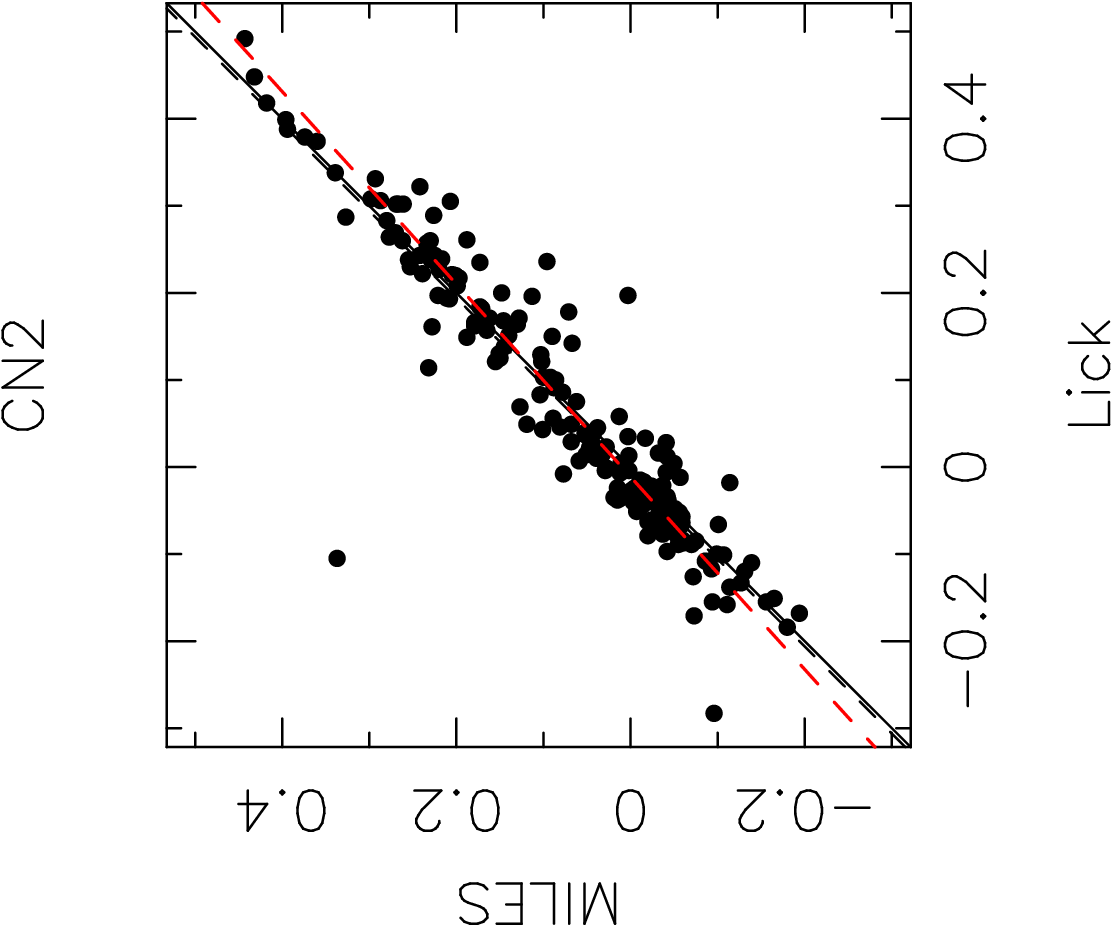}}
\resizebox{0.15\textwidth}{!}{\includegraphics[angle=-90]{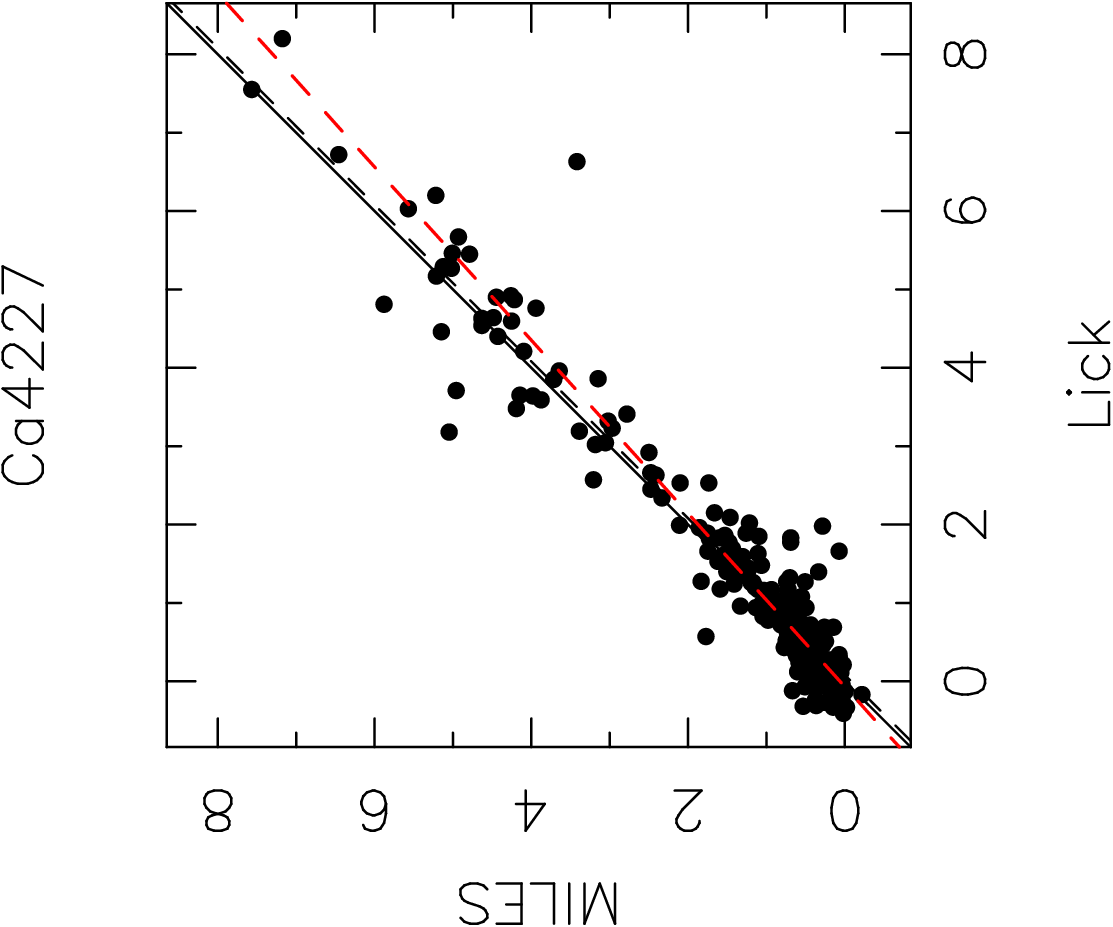}}
\resizebox{0.15\textwidth}{!}{\includegraphics[angle=-90]{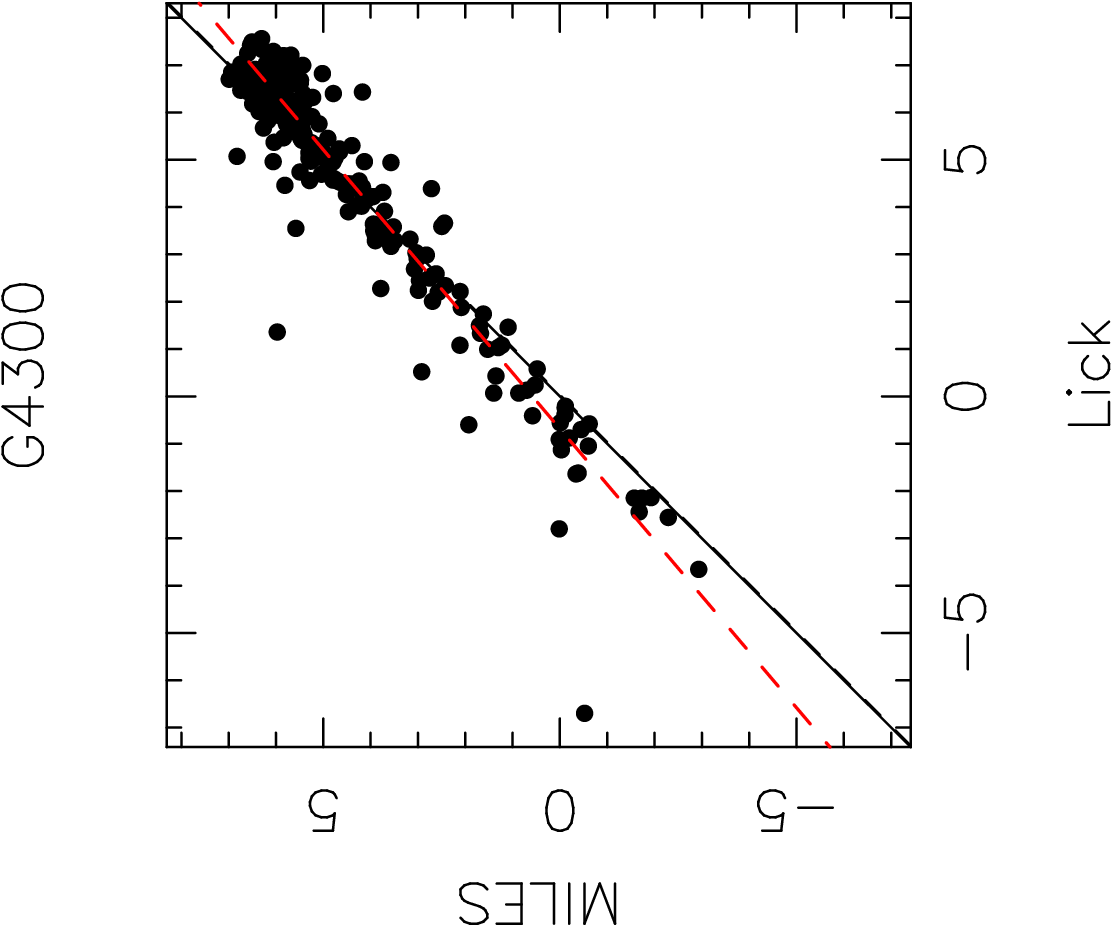}}
\resizebox{0.15\textwidth}{!}{\includegraphics[angle=-90]{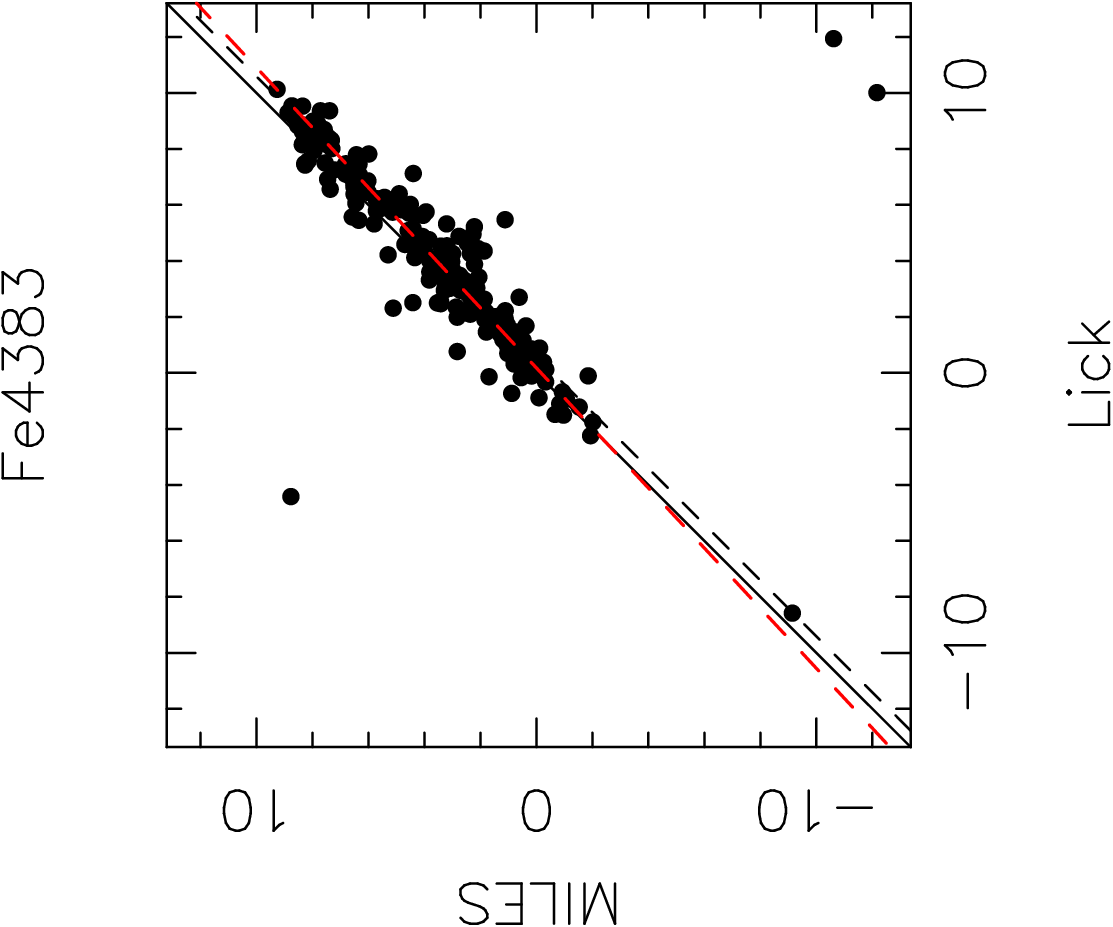}}
\resizebox{0.15\textwidth}{!}{\includegraphics[angle=-90]{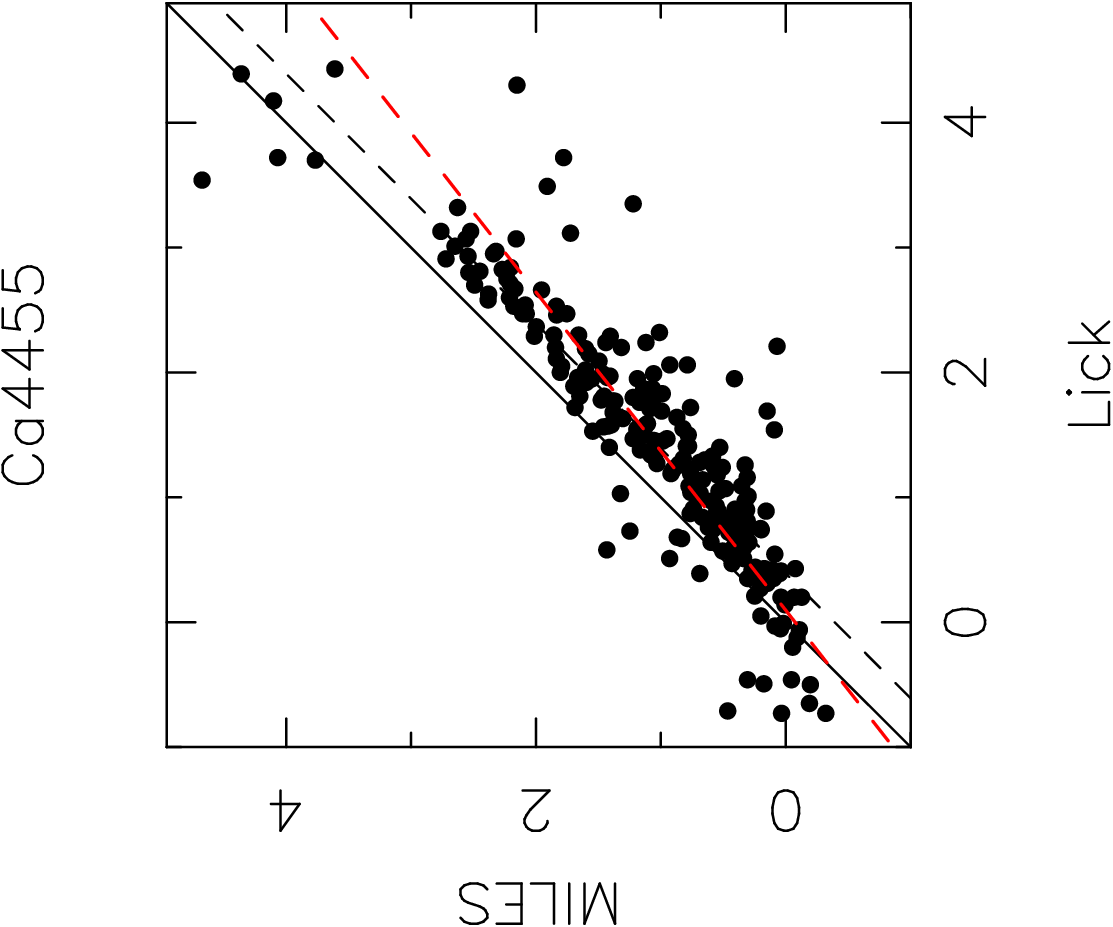}}
\resizebox{0.15\textwidth}{!}{\includegraphics[angle=-90]{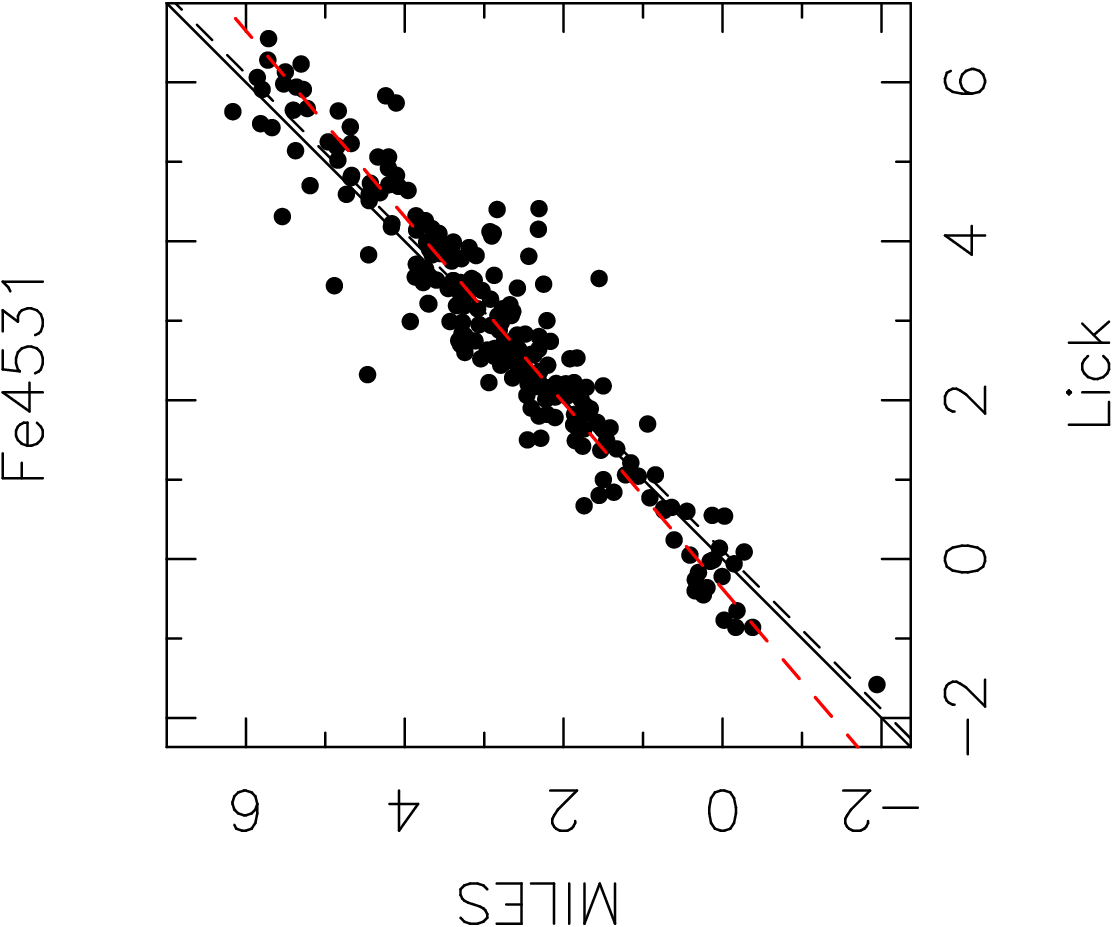}}
\resizebox{0.15\textwidth}{!}{\includegraphics[angle=-90]{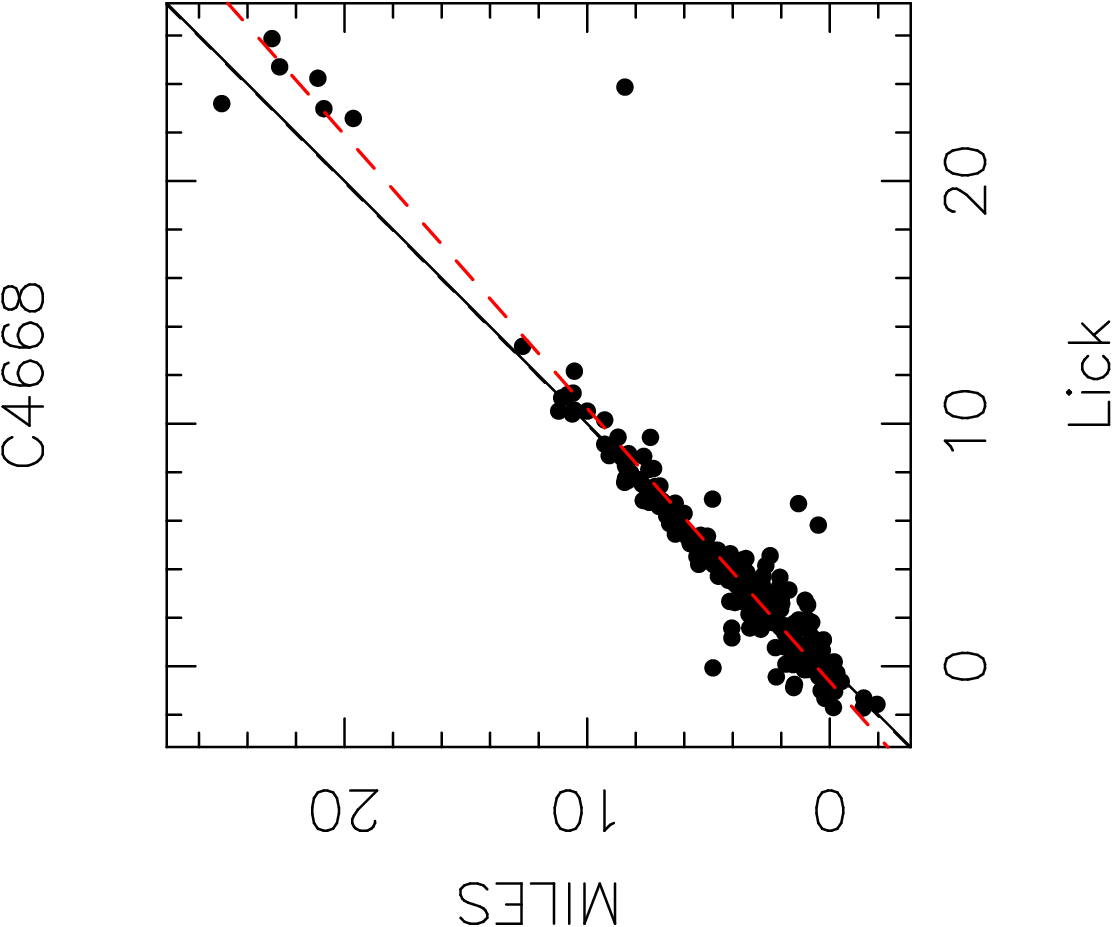}}
\resizebox{0.15\textwidth}{!}{\includegraphics[angle=-90]{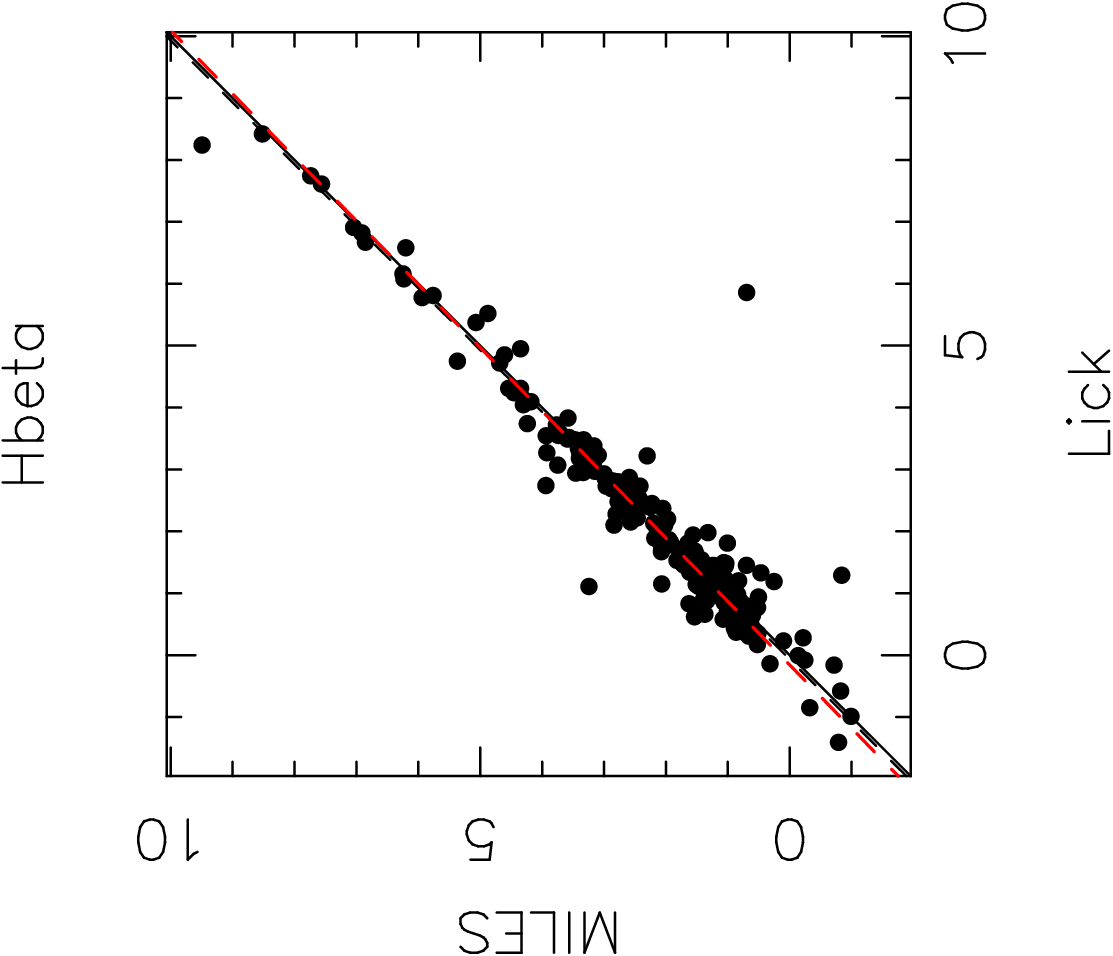}}
\resizebox{0.15\textwidth}{!}{\includegraphics[angle=-90]{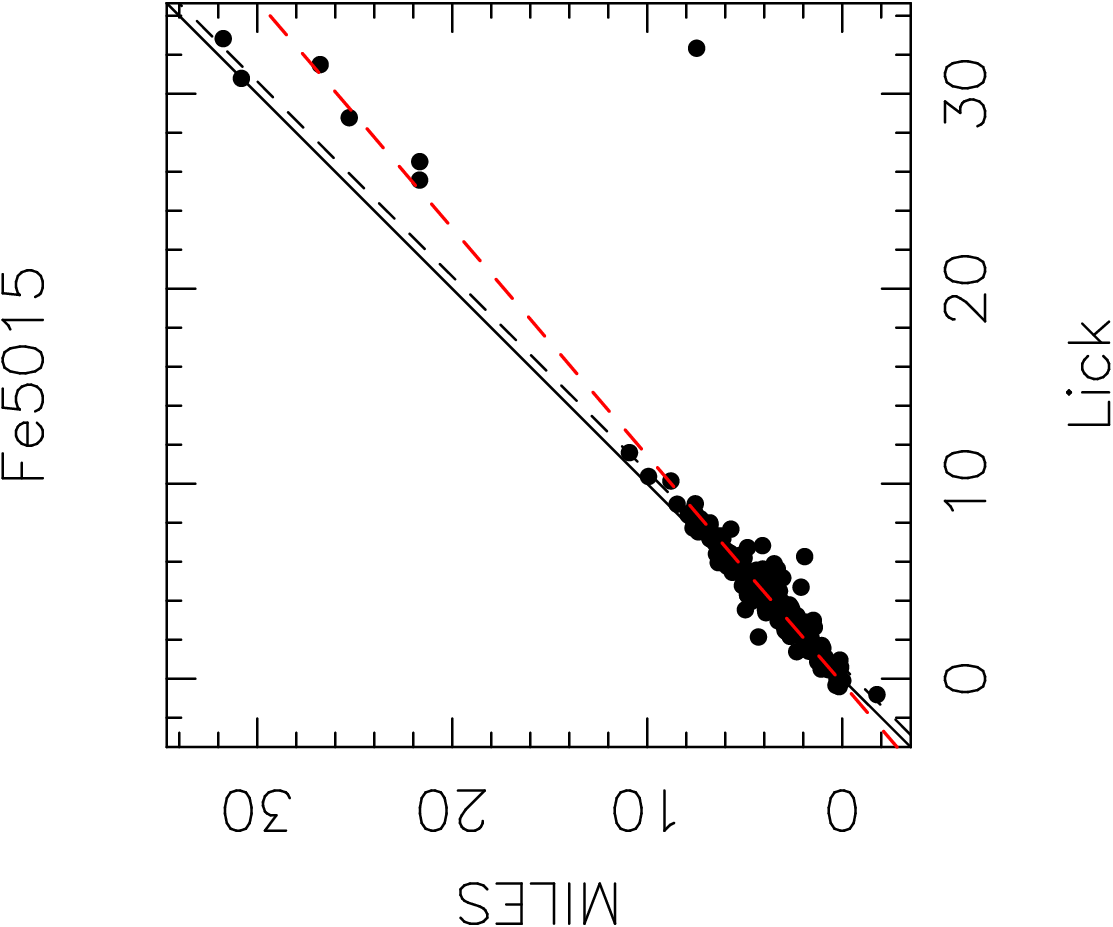}}
\resizebox{0.15\textwidth}{!}{\includegraphics[angle=-90]{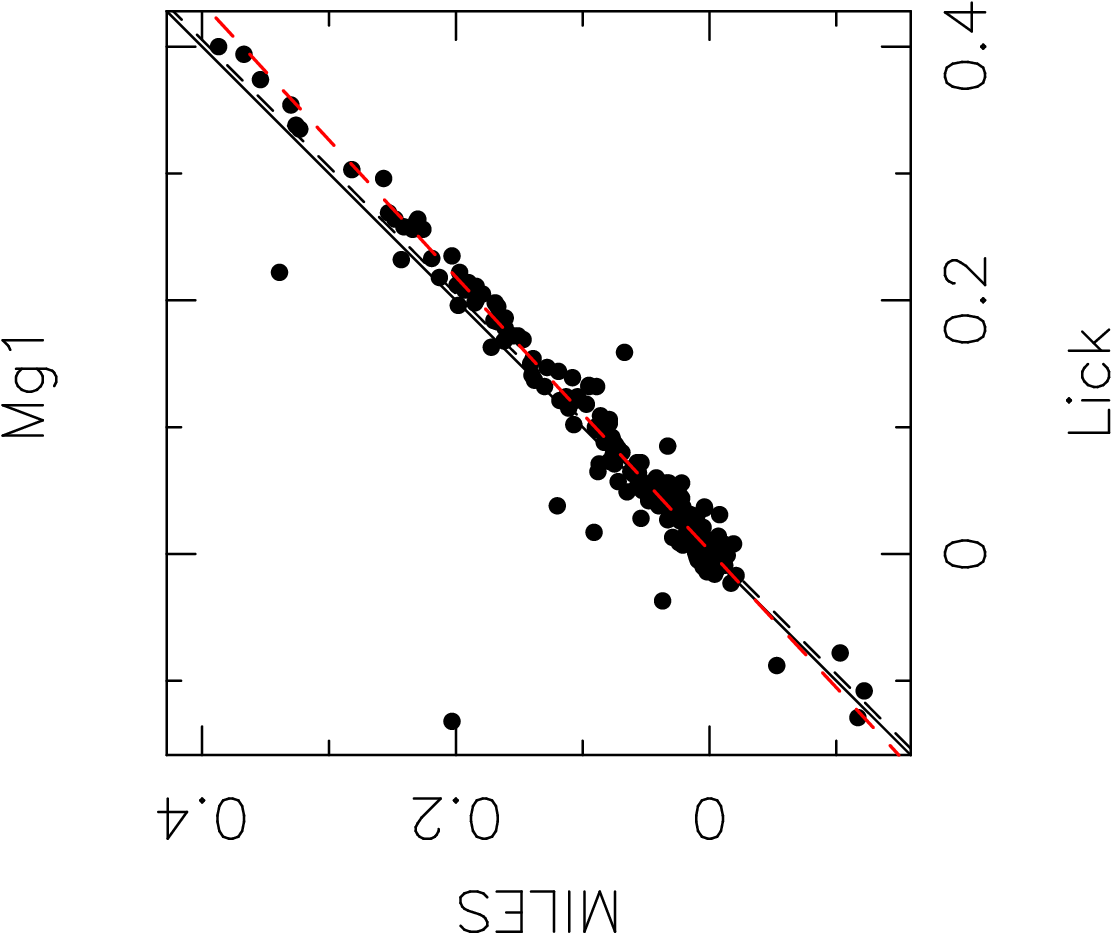}}
\resizebox{0.15\textwidth}{!}{\includegraphics[angle=-90]{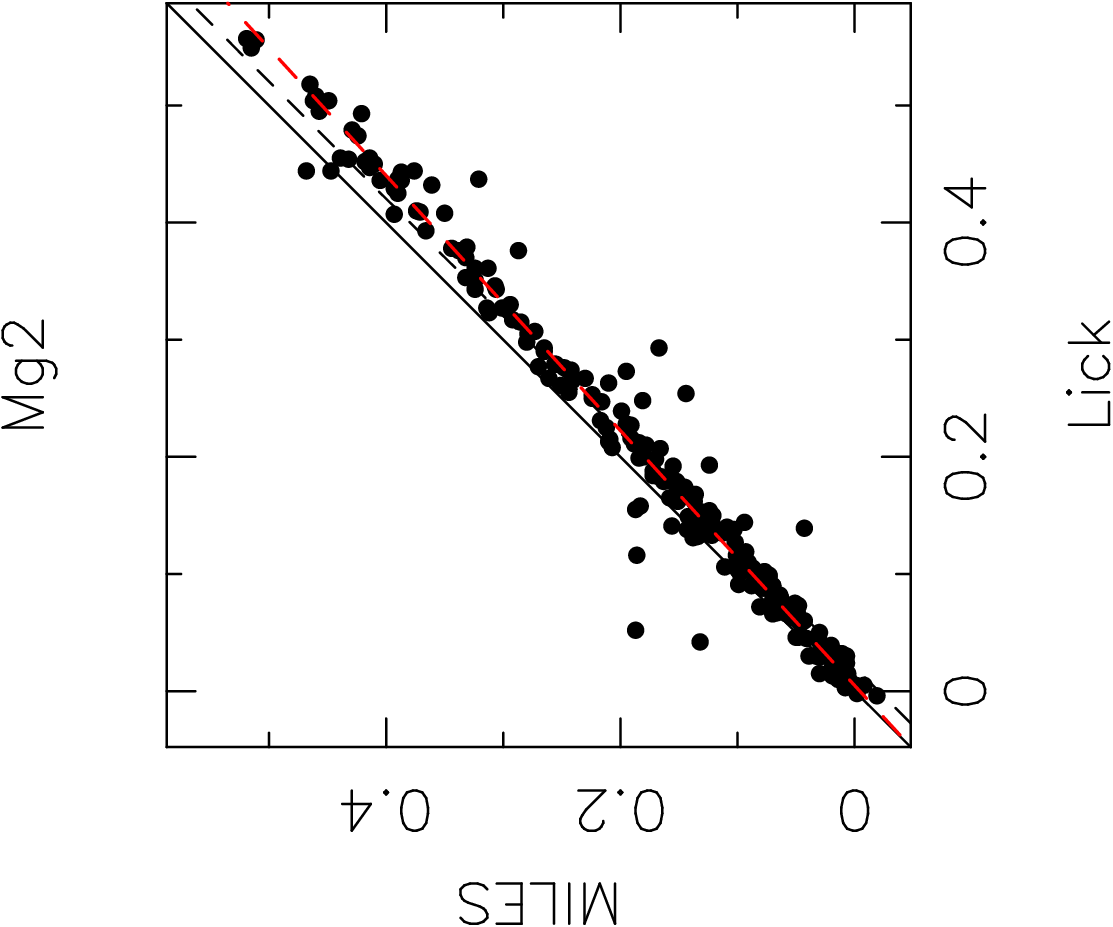}}
\resizebox{0.15\textwidth}{!}{\includegraphics[angle=-90]{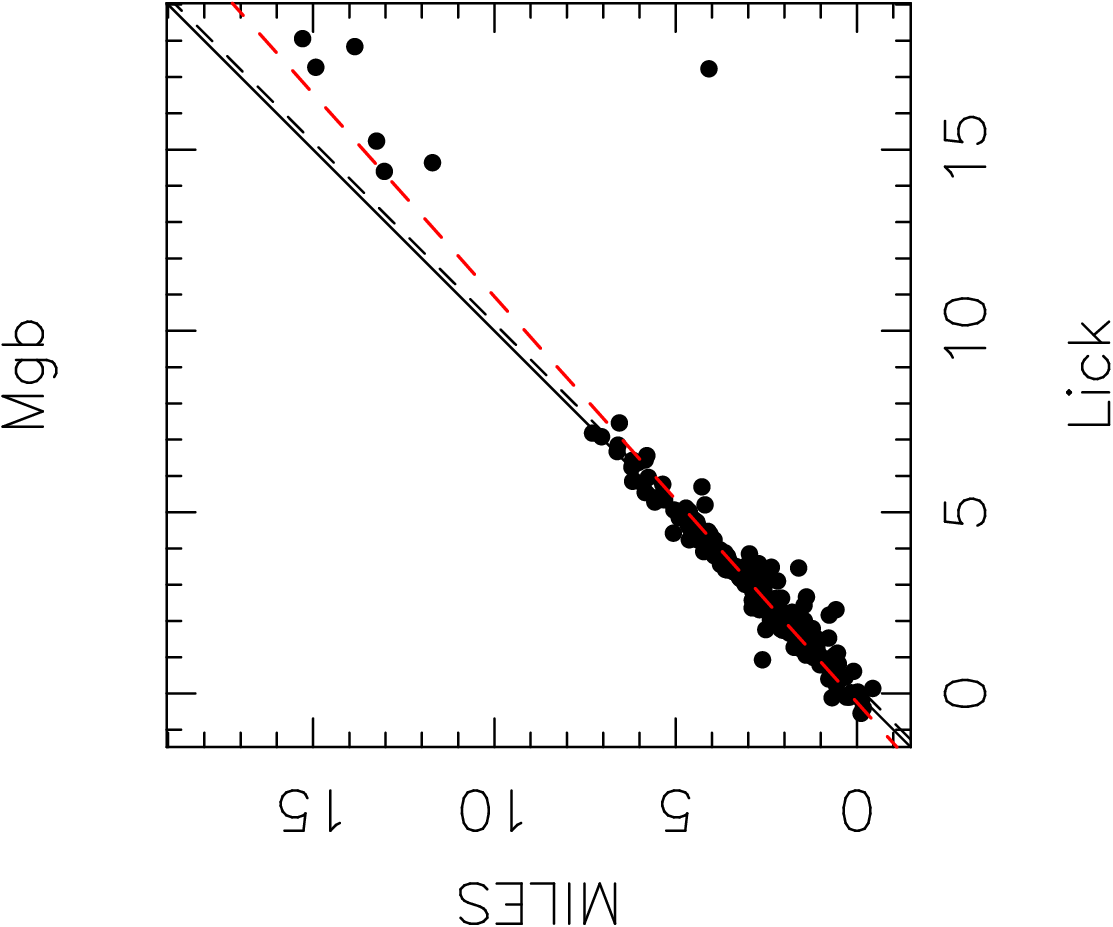}}
\resizebox{0.15\textwidth}{!}{\includegraphics[angle=-90]{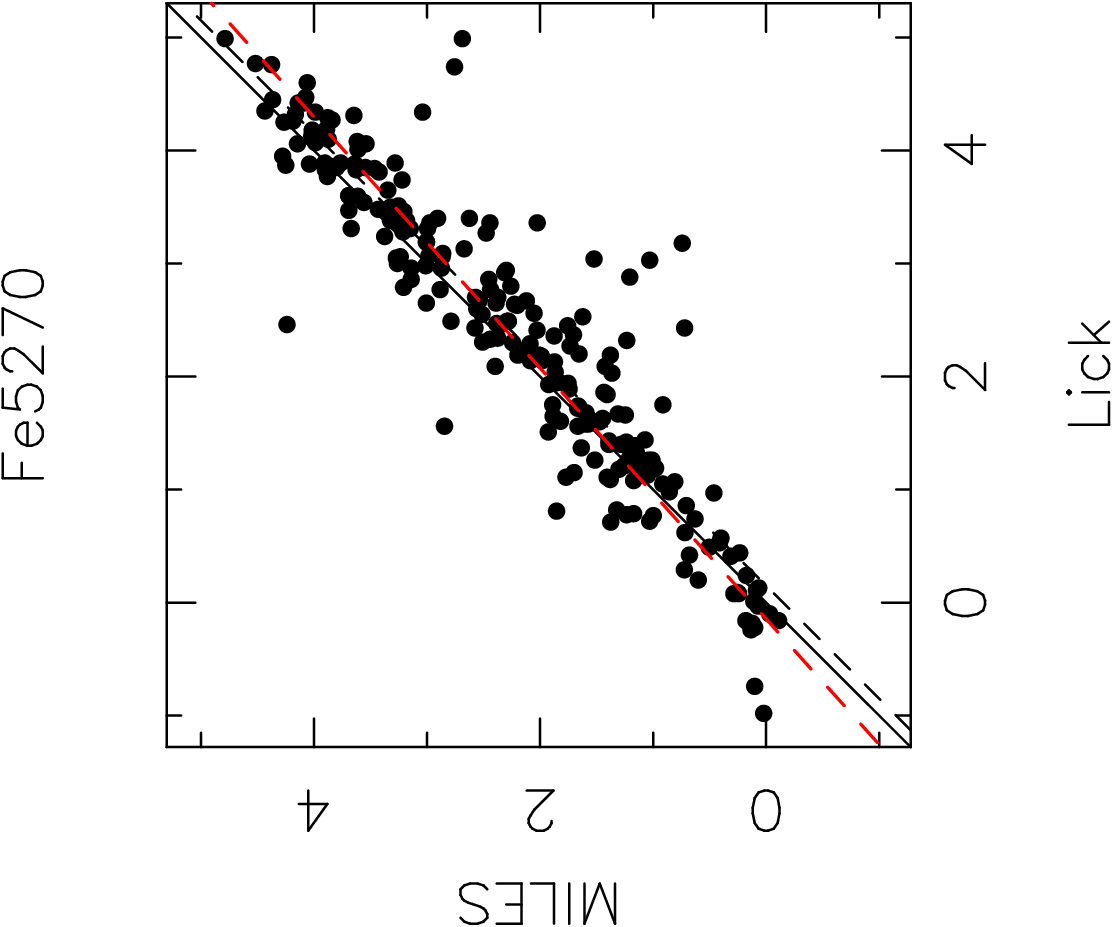}}
\resizebox{0.15\textwidth}{!}{\includegraphics[angle=-90]{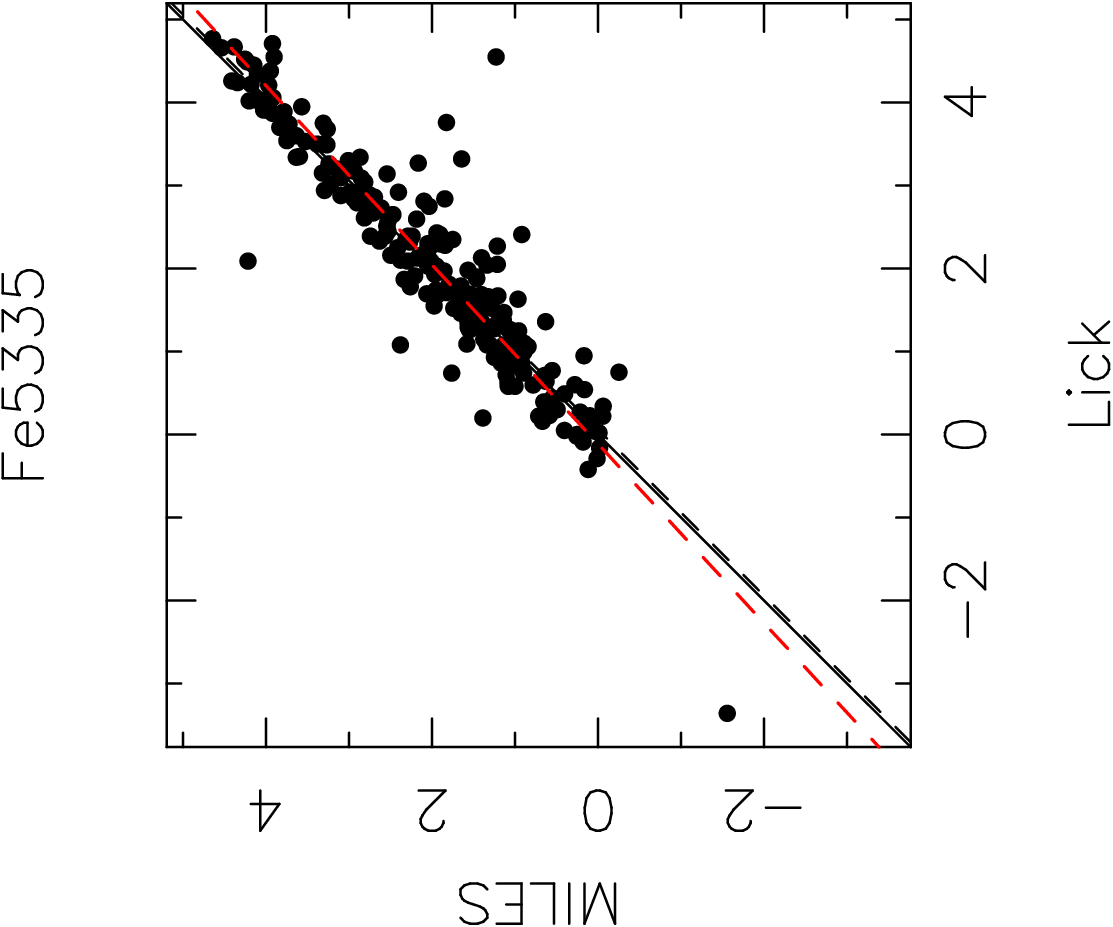}}
\resizebox{0.15\textwidth}{!}{\includegraphics[angle=-90]{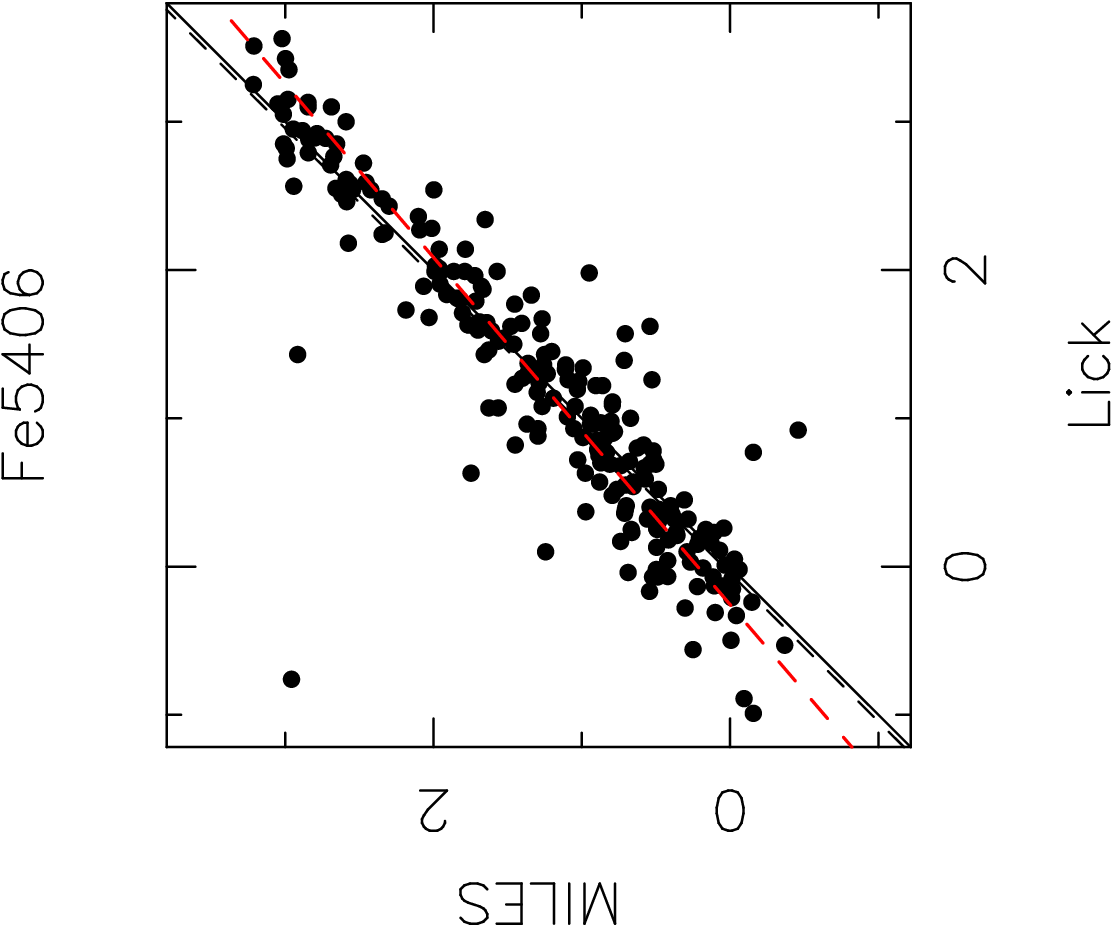}} 
\resizebox{0.15\textwidth}{!}{\includegraphics[angle=-90]{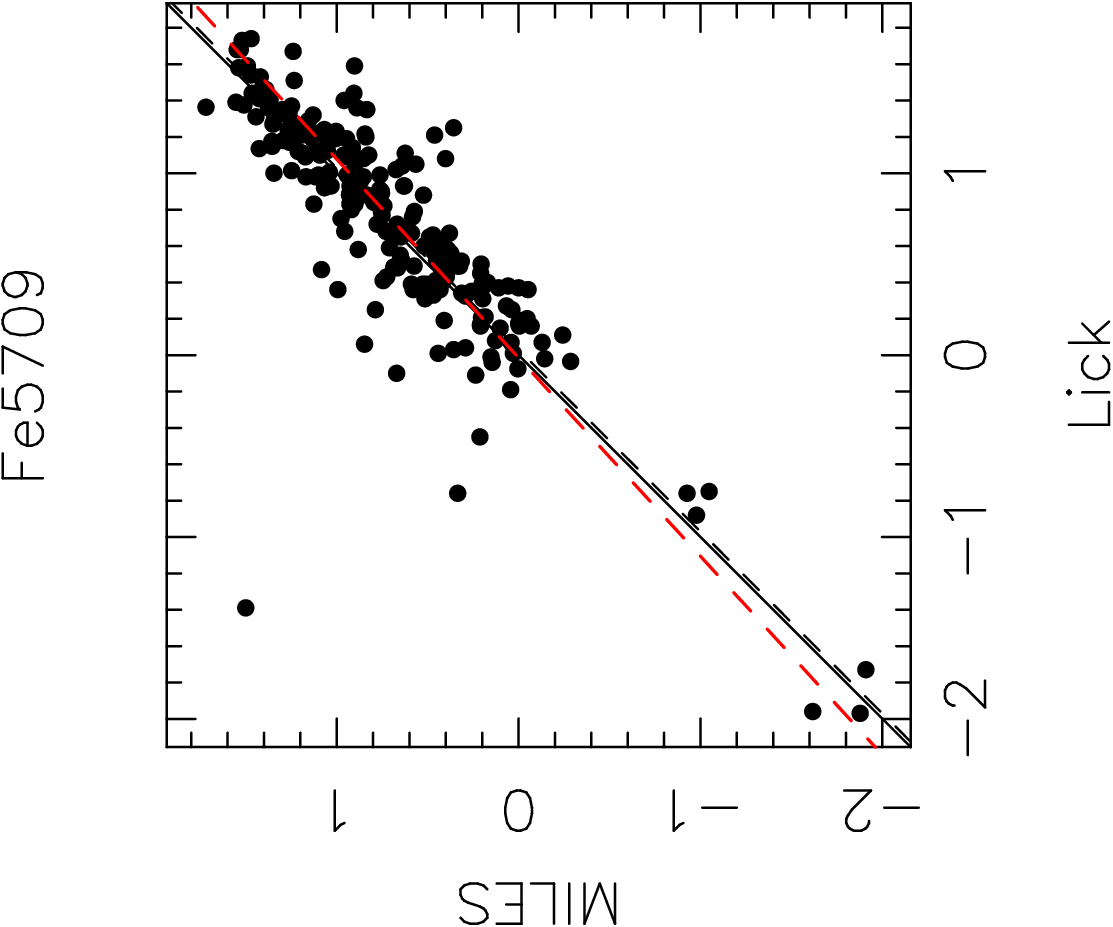}}
\resizebox{0.15\textwidth}{!}{\includegraphics[angle=-90]{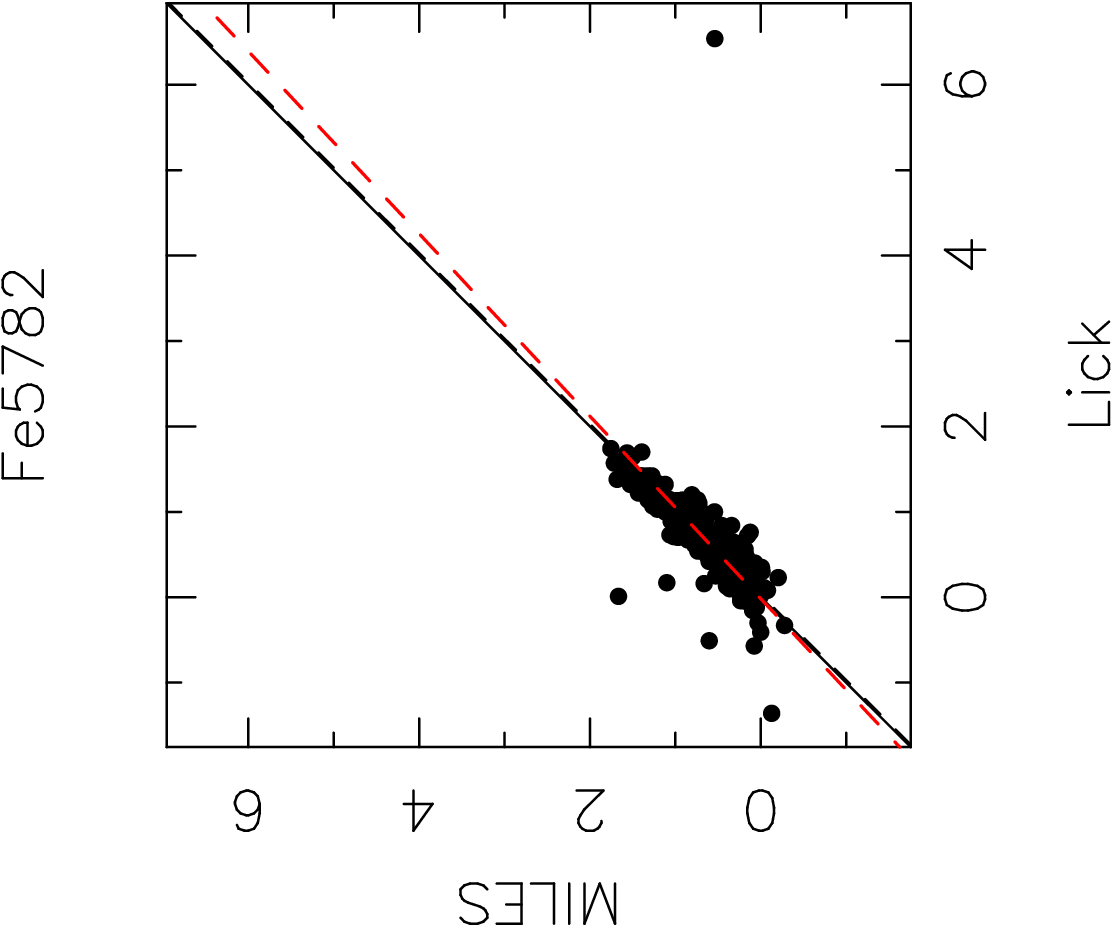}}
\resizebox{0.15\textwidth}{!}{\includegraphics[angle=-90]{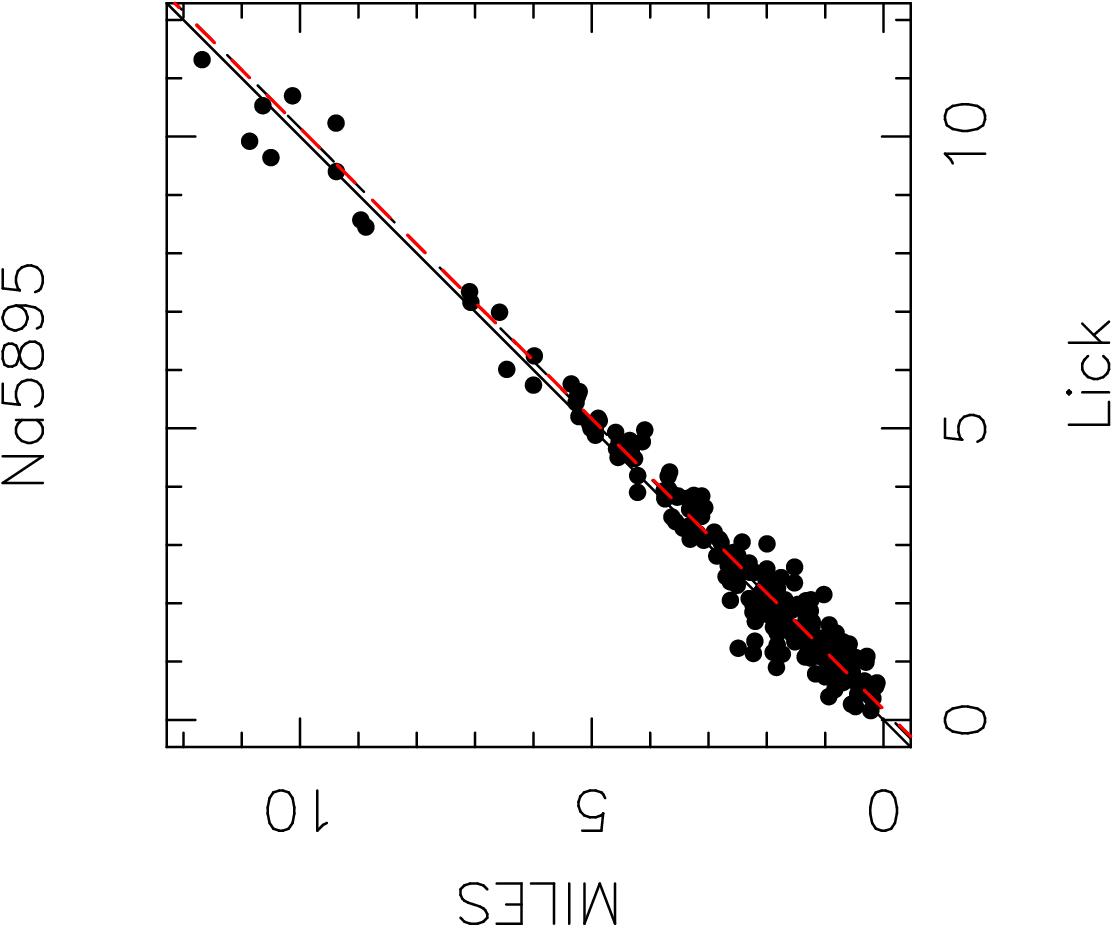}}
\resizebox{0.15\textwidth}{!}{\includegraphics[angle=-90]{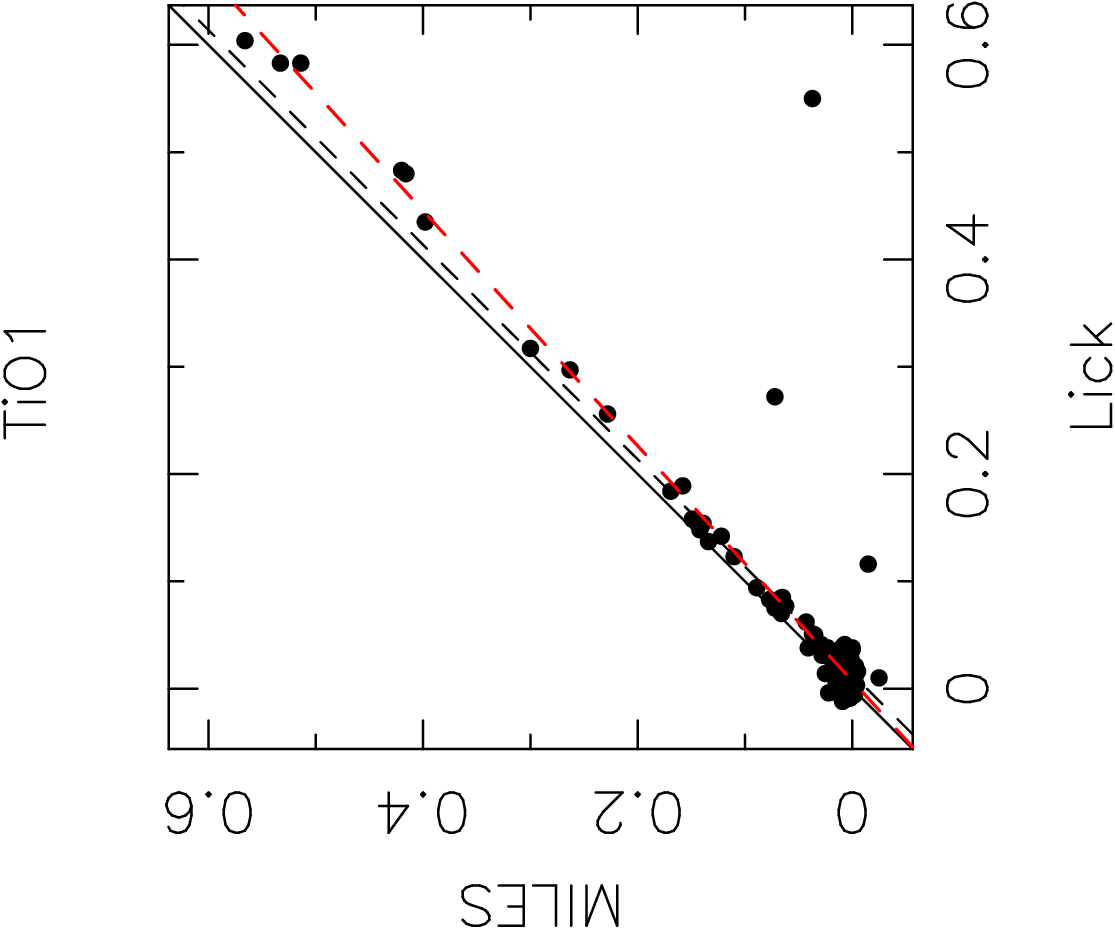}}
\resizebox{0.15\textwidth}{!}{\includegraphics[angle=-90]{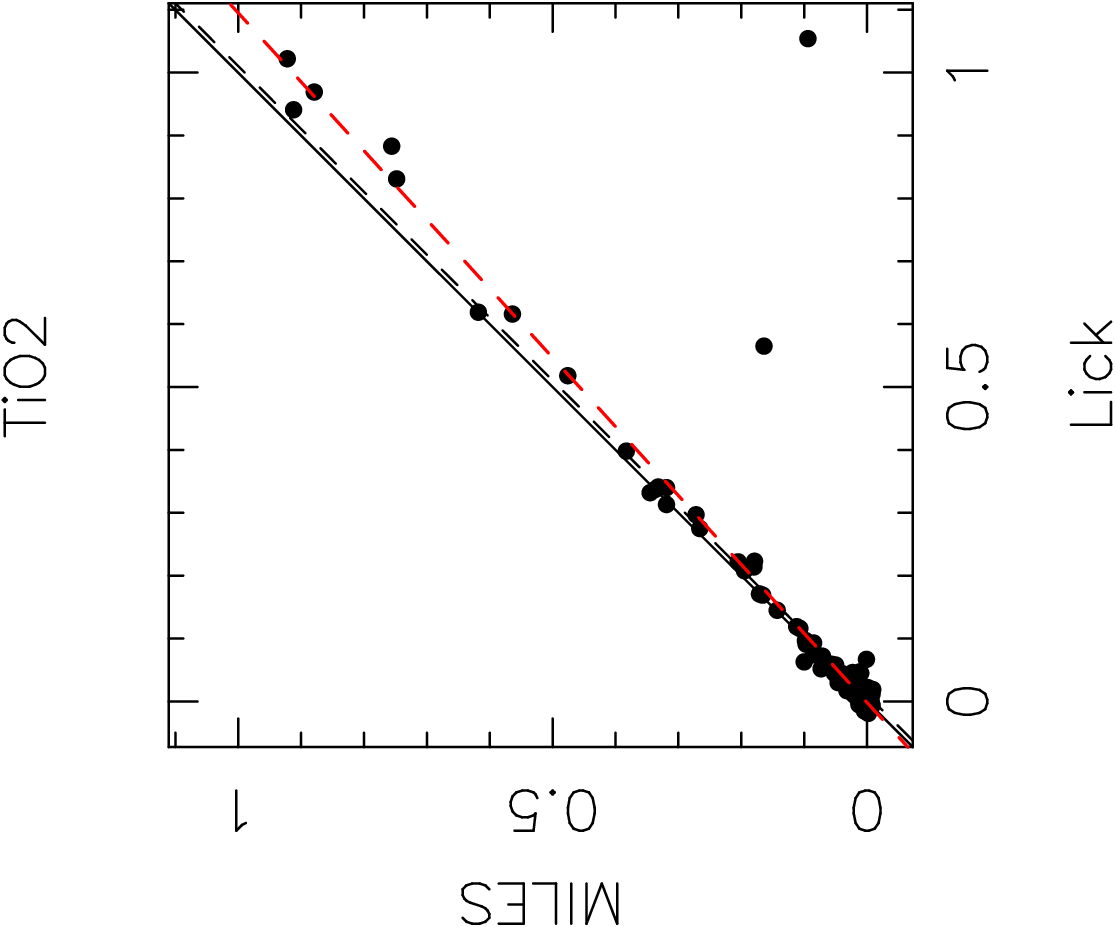}}
\resizebox{0.15\textwidth}{!}{\includegraphics[angle=-90]{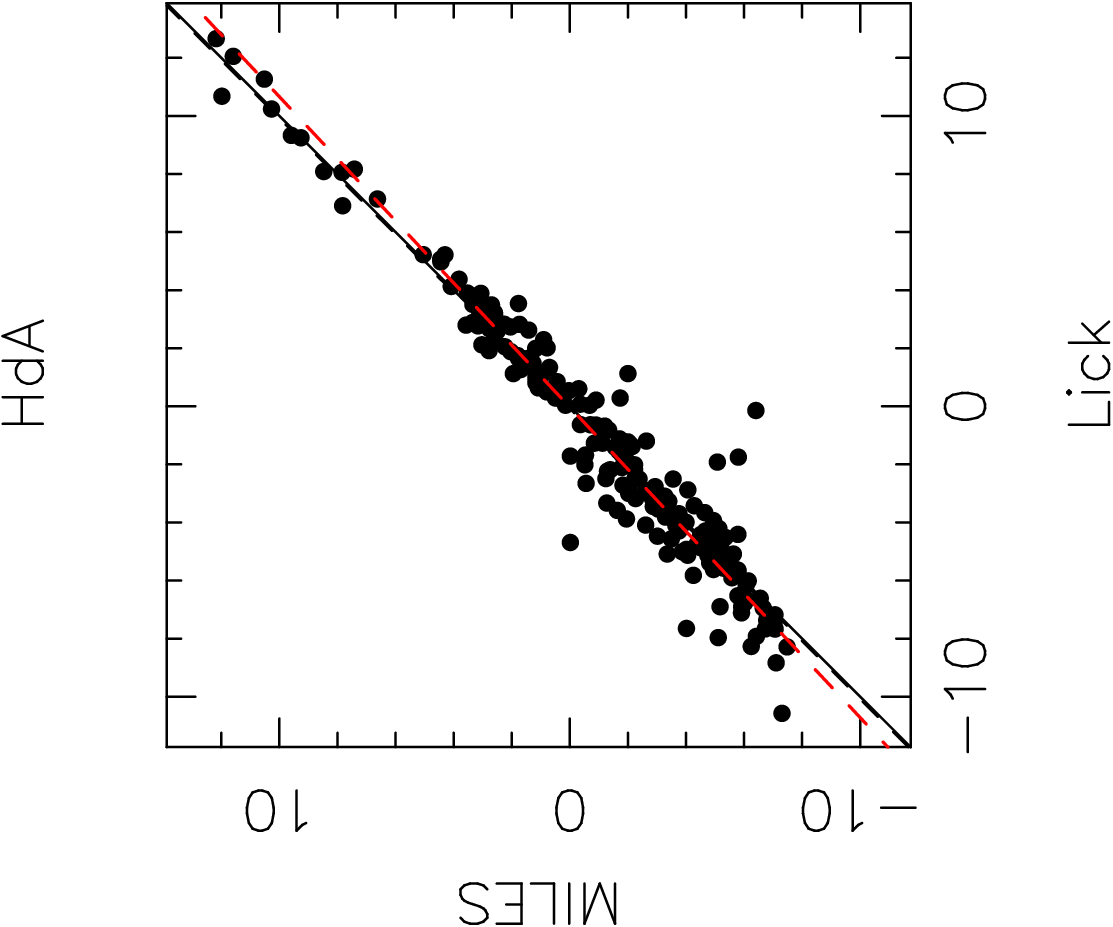}}
\resizebox{0.15\textwidth}{!}{\includegraphics[angle=-90]{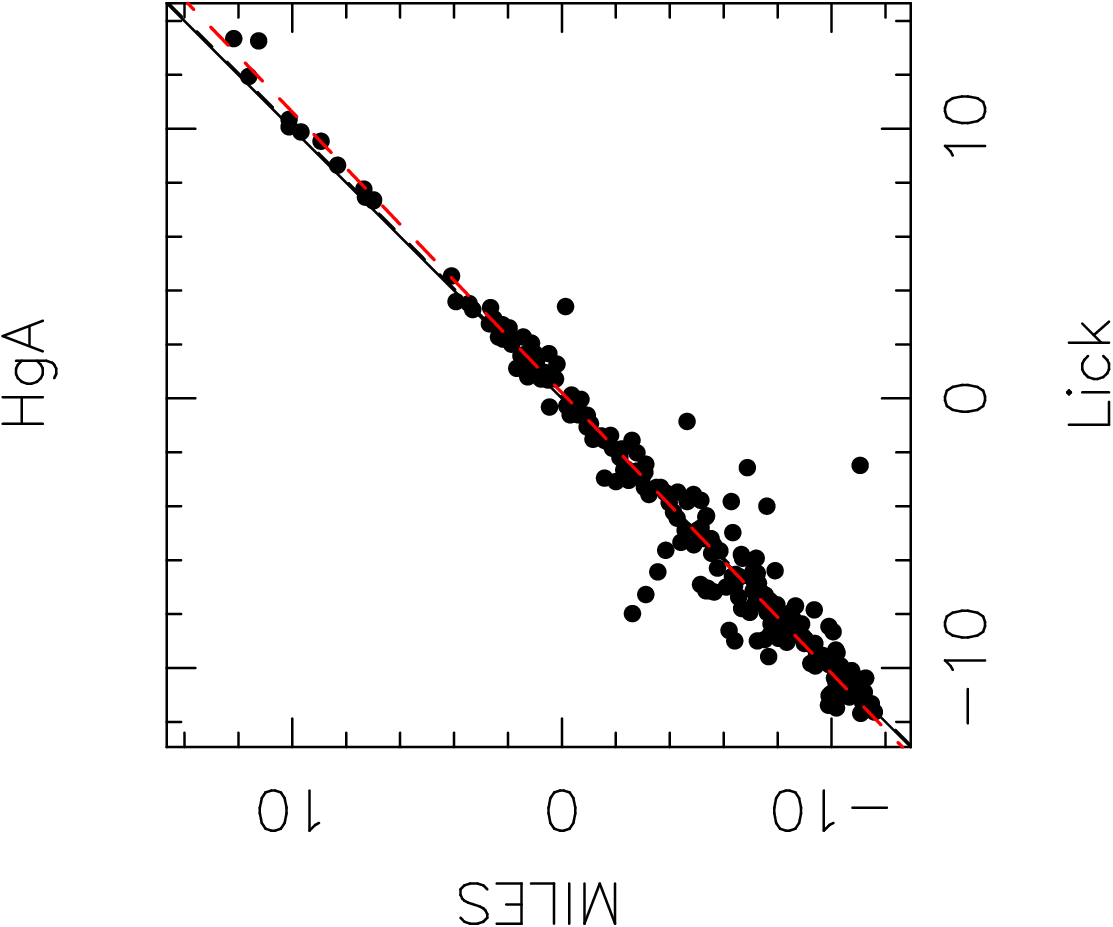}}
\resizebox{0.15\textwidth}{!}{\includegraphics[angle=-90]{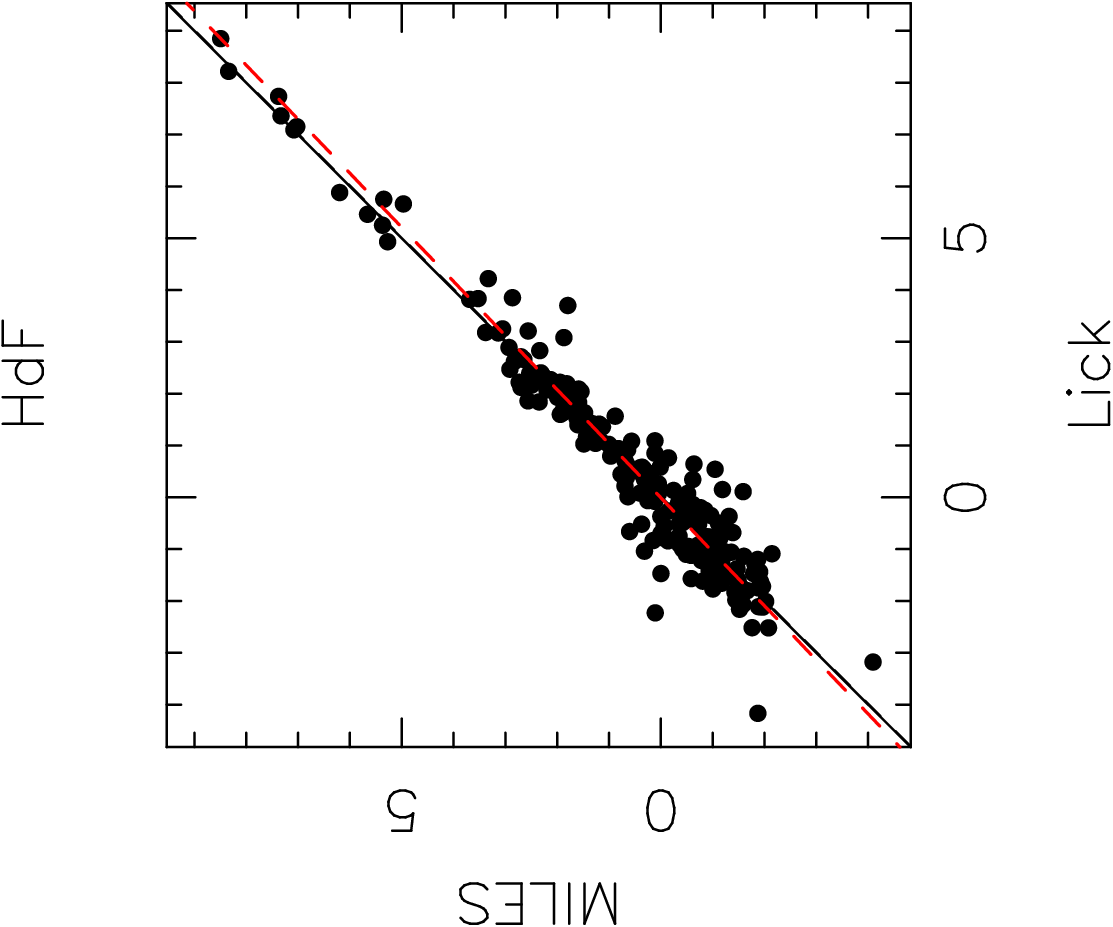}}
\resizebox{0.15\textwidth}{!}{\includegraphics[angle=-90]{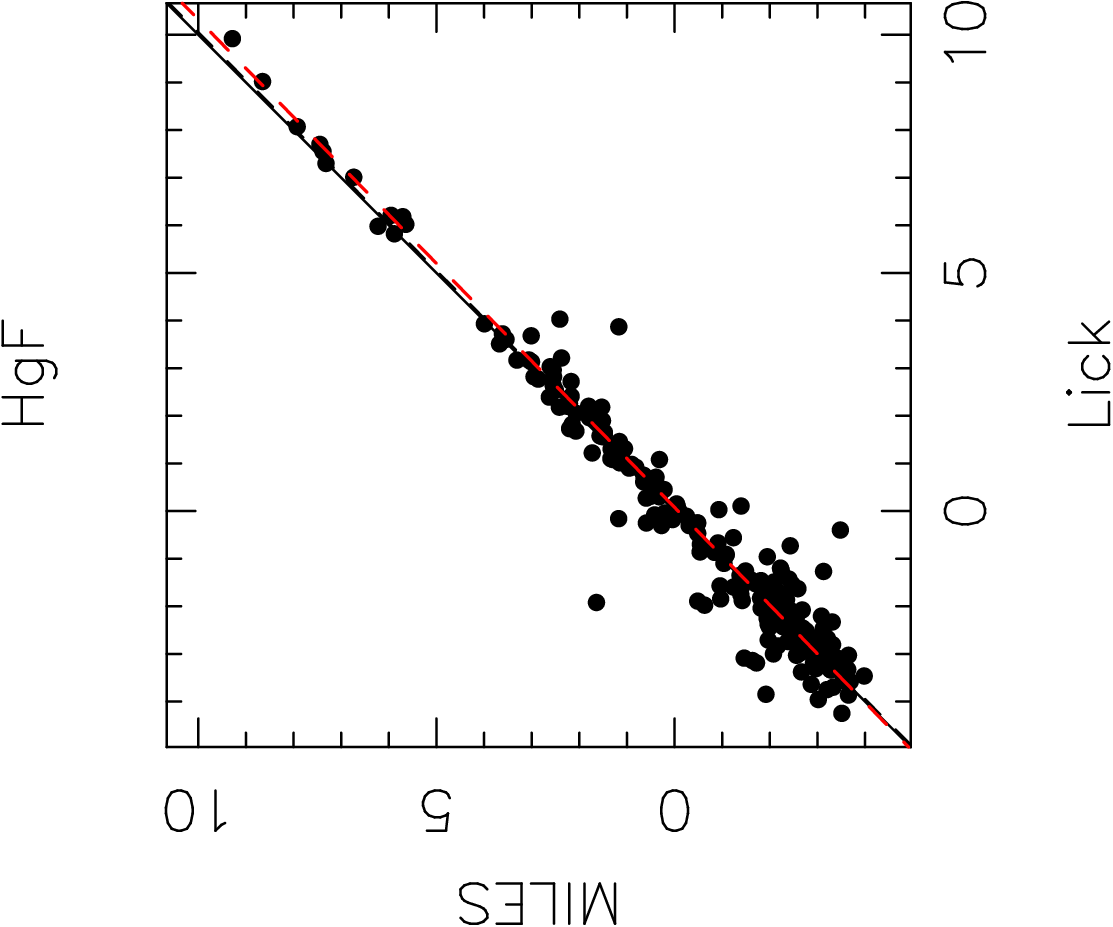}}
\caption{Comparison between the Lick/IDS indices measured in the stars
in common between the MILES and Lick/IDS libraries. The solid line
represents the 1:1 relation.  The dashed line shows the calculated
offset, while the red dashed line indicates the best linear fit to the
data.}
\label{fig.lick.offset}
\end{figure*}   

\section{Mean gradients and aperture correction}
\label{appendix.aperture}

Galaxies have radial gradients in the line-strength indices and,
therefore, when comparing galaxies at different redshifts, the indices
need to be aperture corrected.  \citet{Jor97} and \citet{J05} present
a series of corrections based on the radial gradients from
\citet{Vaz97}, \citet{Car98}, and \citet{DC94}.  However, those works
do not include many of the Lick indices, particularly the higher order
Balmer lines.

We have derived new aperture corrections for all the Lick/IDS indices
+ the higher order Balmer line indices \citep{WO97}, taking advantage
of the large sample of nearby galaxies presented in \citet{SB06c} and
\citet{SB07}.
To calculate
the $\alpha$ parameter,
we assume that the light
in our galaxies follows a de Vaucouleur profile. Using the 
gradients for each individual galaxy, we obtain
the final index inside an aperture as the luminosity-weighted
index. We then calculate how the final indices depends on
aperture size. 
The variation of the luminosity-weighted
indices with different aperture sizes can be described as power law.
Therefore,
aperture corrections (those measured in \AA) can be written as 
\citep[see][]{Jor95,Jor97}:

\begin{equation}
\log(index)_{corr} = \log(index)_{ap}+\alpha \log \frac{r_{ap}}{r_{nor}},
\end{equation}
and for molecular indices and higher-order Balmer lines\footnote{The correction 
for the higher-order Balmer lines is additive as these indices can have both positive
and negative values.}:
\begin{equation}
index_{corr}= index_{ap}+\alpha\log \frac{r_{ap}}{r_{nor}},
\end{equation}
where $r_{ap}$ is the equivalent circular aperture, obtained as
$2r_{ap}=1.025\times 2(xy/\pi)^{1/2}$; and $x$ and $y$ are the width
and the length of the rectangular aperture.  

The $\alpha$ coefficients
are listed in Table~\ref{mean.grad} and can be used by any other
study.  In our case, we were more interested in selecting a fraction
of the total light of the galaxy instead of a fixed aperture and
therefore $y$ is not fixed but is chosen to be a FWHM of the spatial
profile.  However, we have fixed the $y$ for all the spectra to be the
physical aperture equivalent to 1 arcsec (the width of the slit) at
$z=0.75$. The correction for the galaxies at $z > $ 0.4 was very small and
we did not apply any.  The aperture correction was applied only to the
galaxies in the Coma cluster.

Elliptical galaxies show a great variety of gradients that do not seem
to be correlated with any other properties of the galaxies, like mass
or colour.  For this reason, we used the mean gradient to compute the
aperture corrections. We calculated an error due to the dispersion
among the mean of all the measured gradients. To do this, we performed
a series of Monte Carlo simulations where each gradient was perturbed
following a Gaussian distribution with $\sigma$ equal to the typical
RMS dispersion in the gradients.  For each new gradient, the
coefficient $\alpha$ and the final correction for the galaxies in the
Coma cluster were recomputed.  The errors in the aperture correction
calculated this way are indicated in Table~\ref{aperture.corrections}.

\begin{table}
\centering
\begin{tabular}{lrr}
Index       &$\alpha$~~~~~~~~~~~~~ &  Type\\  
\hline
\hline
$\sigma$    &$ 0.058\pm 0.047$ &   1   \\
D4000       &$ 0.027\pm 0.018$ &   1   \\
H$\delta_A$ &$-0.721\pm 0.508$ &   2    \\
H$\delta_F$ &$-0.206\pm 0.249$ &   2    \\
CN$_1$      &$ 0.058\pm 0.023$ &   2     \\
CN$_2$      &$ 0.059\pm 0.030$ &   2     \\
Ca4227      &$ 0.047\pm 0.075$ &   1     \\
G4300       &$ 0.027\pm 0.022$ &   1      \\ 
H$\gamma_A$ &$-0.850\pm 0.422$ &   2     \\   
H$\gamma_F$ &$-0.436\pm 0.321$ &   2     \\   
Fe4383      &$ 0.061\pm 0.033$ &   1     \\  
Ca4455      &$ 0.097\pm 0.075$ &   1     \\  
Fe4531      &$ 0.040\pm 0.021$ &   1    \\   
Fe4668      &$ 0.119\pm 0.049$ &   1    \\   
Fe5015      &$ 0.05 \pm 0.02$  &   1    \\   
H$\beta$    &$-0.05\pm  0.19$  &   1    \\   
Mgb         &$ 0.048\pm 0.051$ &   1    \\   
\hline
\end{tabular}
\caption{$\alpha$ coefficient and corresponding error
obtained using the samples  of SB06 and S\'anchez-Bl\'azquez et al. 2007.
\label{mean.grad}}
\end{table}

\section{Relation of the Lick/IDS indices with velocity dispersion}
Figure~\ref{indices.sigma} shows the relation between all the measured
line-strength indices and the velocity dispersion for the 4 considered
redshift bins.

\label{appendix.indices}

\begin{table*}
\centering
\begin{tabular}{lccc|cc|cc|cc}
\hline
              &          & \multicolumn{2}{c}{linear fit}                &\multicolumn{2}{c}{ $z_{\rm f}=1.4$} 
&\multicolumn{2}{c}{ $z_{\rm f}=2$} &\multicolumn{2}{c}{ $z_{\rm f}=3$}\\
Index         & redshift &  @200~km~s$^{-1}$      & $\sigma$ &  @200~km~s$^{-1}$ & $t$ &  @200~km~s${-1}$  & $t$  &  
@200~km~s$^{-1}$  & $t$  \\
\hline
\hline 
D4000       & 0.45    &  2.03  & 0.11  &  1.94  & 0.8 &  1.98 & 0.4   &  2.01 & 0.2  \\
            & 0.55    &  2.08  & 0.07  &  1.85  & 3.2 &  1.93 & 2.1   &  1.96 & 1.7 \\
            & 0.75    &  1.98  & 0.12  &  1.75  & 1.9 &  1.89 & 0.7   &  1.92 & 0.5 \\
            & 0.02    &  2.12  & 0.12  &  2.11  & 0.1 &  2.16 & 0.3   &  2.20 & 0.7 \\
H$\delta_F$ & 0.45    &  0.92  & 0.52  &  1.05  & 0.2 &  0.86 & 0.1   &  0.79 & 0.2 \\   
            & 0.55    &  0.99  & 0.89  &  1.36  & 0.4 &  1.07 & 0.1   &  0.97 & 0.0 \\
            & 0.75    &  1.40  & 0.71  &  1.80  & 0.6 &  1.24 & 0.2   &  1.10 & 0.4 \\ 
            & 0.02    &  0.70  & 0.33  &  0.50  & 0.6 &  0.38 & 0.9   &  0.28 & 1.3 \\
Ca4227      & 0.45    &  0.82  & 0.31  &  1.06  & 0.8 &  1.12 & 0.9   &  1.15 & 1.1\\  
            & 0.55    &  0.89  & 0.27  &  0.97  & 0.3 &  1.05 & 0.6   &  1.08 & 0.7\\
            & 0.75    &  0.83  & 0.32  &  0.86  & 0.1 &  0.99 & 0.5   &  1.05 & 0.7 \\ 
            & 0.02    &  1.17  & 0.16  &  1.27  & 0.6 &  1.33 & 1.0   &  1.39 & 1.3 \\
G4300       & 0.45    &  4.58  & 0.81  &  3.87  & 0.9 &  4.03 & 0.7   &  4.12 & 0.6 \\
            & 0.55    &  4.41  & 0.53  &  3.55  & 1.6 &  3.85 & 1.0   &  3.95 & 0.8 \\
            & 0.75    &  4.06  & 1.31  &  3.05  & 0.8 &  3.66 & 0.3   &  3.81 & 0.2\\
            & 0.02    &  5.25  & 0.51  &  4.35  & 1.7 &  4.41 & 1.6   &  4.45 & 1.5 \\
Ca4455      & 0.45    &  0.86  & 0.28  &  0.88  & 0.1 &  0.91 & 0.2   &  0.91 & 0.2\\
            & 0.55    &  0.96  & 0.30  &  0.83  & 0.4 &  0.88 & 0.3   &  0.90 & 0.2\\
            & 0.75    &  0.79  & 0.61  &  0.75  & 0.1 &  0.85 & 0.1   &  0.87 & 0.1 \\
            & 0.02    &  1.15  & 0.12  &  0.97  & 1.5 &  1.00 & 1.2   &  1.02 & 1.1 \\
Fe4531      & 0.45    &  2.77  & 0.43  &  2.76  & 0.0 &  2.82 & 0.1   &  2.83 & 0.1 \\
            & 0.55    &  2.93  & 1.74  &  2.66  & 0.1 &  2.75 & 0.1   &  2.78 & 0.1  \\ 
            & 0.75    &  2.51  & 0.80  &  2.53  & 0.0 &  2.69 & 0.2   &  2.74 & 0.3  \\ 
            & 0.02    &  3.04  & 0.23  &  2.92  & 0.5 &  2.98 & 0.3   &  3.03 & 0.0 \\
H$\gamma_A$ & 0.45    &$-3.85$ & 1.35  &$-2.92$ & 0.7 &$-3.38$& 0.3   &$-3.61$& 0.2 \\ 
            & 0.55    &$-3.83$ & 1.77  &$-2.12$ & 0.9 &$-2.89$& 0.7   &$-3.12$& 0.4 \\ 
            & 0.75    &$-2.56$ & 1.87  &$-0.86$ & 0.9 &$-2.41$& 0.1   &$-2.78$& 0.1 \\   
            & 0.02    &$-5.56$ & 1.04  &$-4.38$ & 1.1 &$-4.68$& 0.8   &$-4.90$& 0.6 \\
\hline
\end{tabular}

\caption{Comparison between the indices predicted by the linear fit to
the data at $\sigma=200$~km~s$^{-1}$ and the ones predicted -- at the
same velocity dispersion -- in a passively evolving model with
formation redshifts of $z_{\rm f}=1.4$, $z_{\rm f}=2$ and $z_{\rm
f}=3$.  Column 3 shows the value predicted by the linear fit at
$\sigma=200$~km~s$^{-1}$; Column 4 shows the standard deviation of the
relation; Column5 shows the value predicted by the SSP model; and the
last column shows the $t$ parameter of the comparison of these two
values. A $t$ value higher than 1.9 would indicate that the there is a
5\% probability that the two values are different by chance.  As can
be seen, none of the values show significant differences from the
model.}
\label{table.t.appendix}
\end{table*}

\begin{figure*}
\resizebox{0.3\textwidth}{!}{\includegraphics[angle=-90]{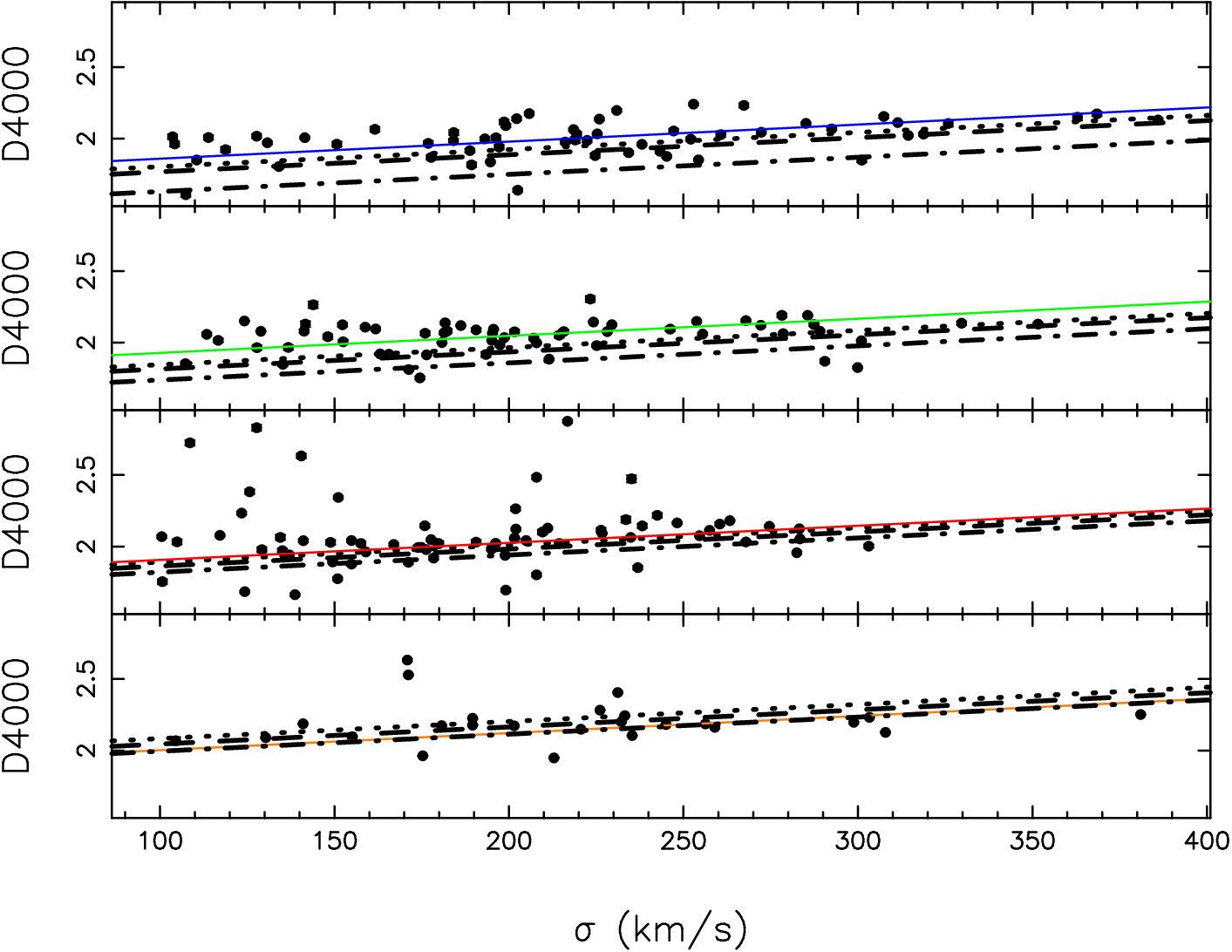}}
\resizebox{0.3\textwidth}{!}{\includegraphics[angle=-90]{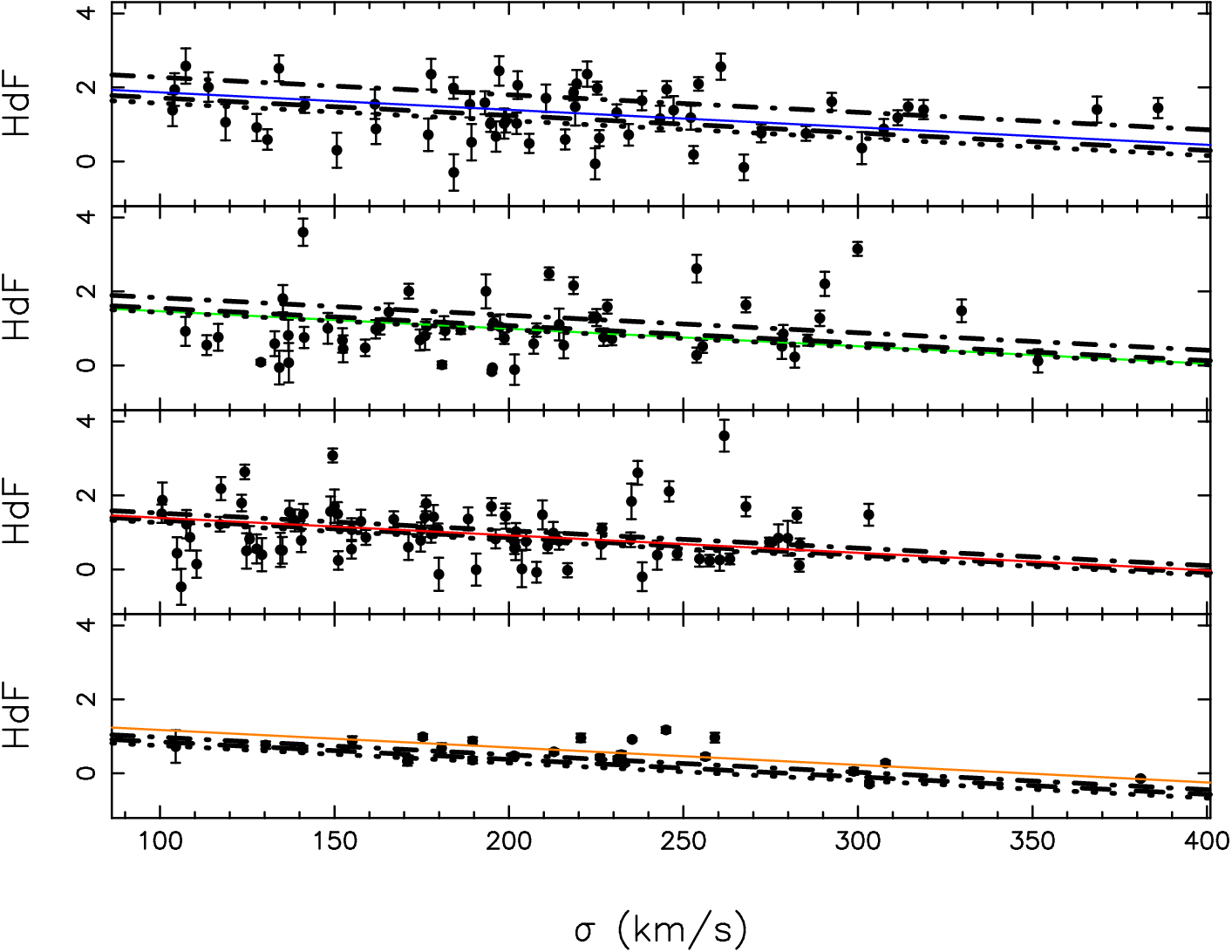}}
\resizebox{0.3\textwidth}{!}{\includegraphics[angle=-90]{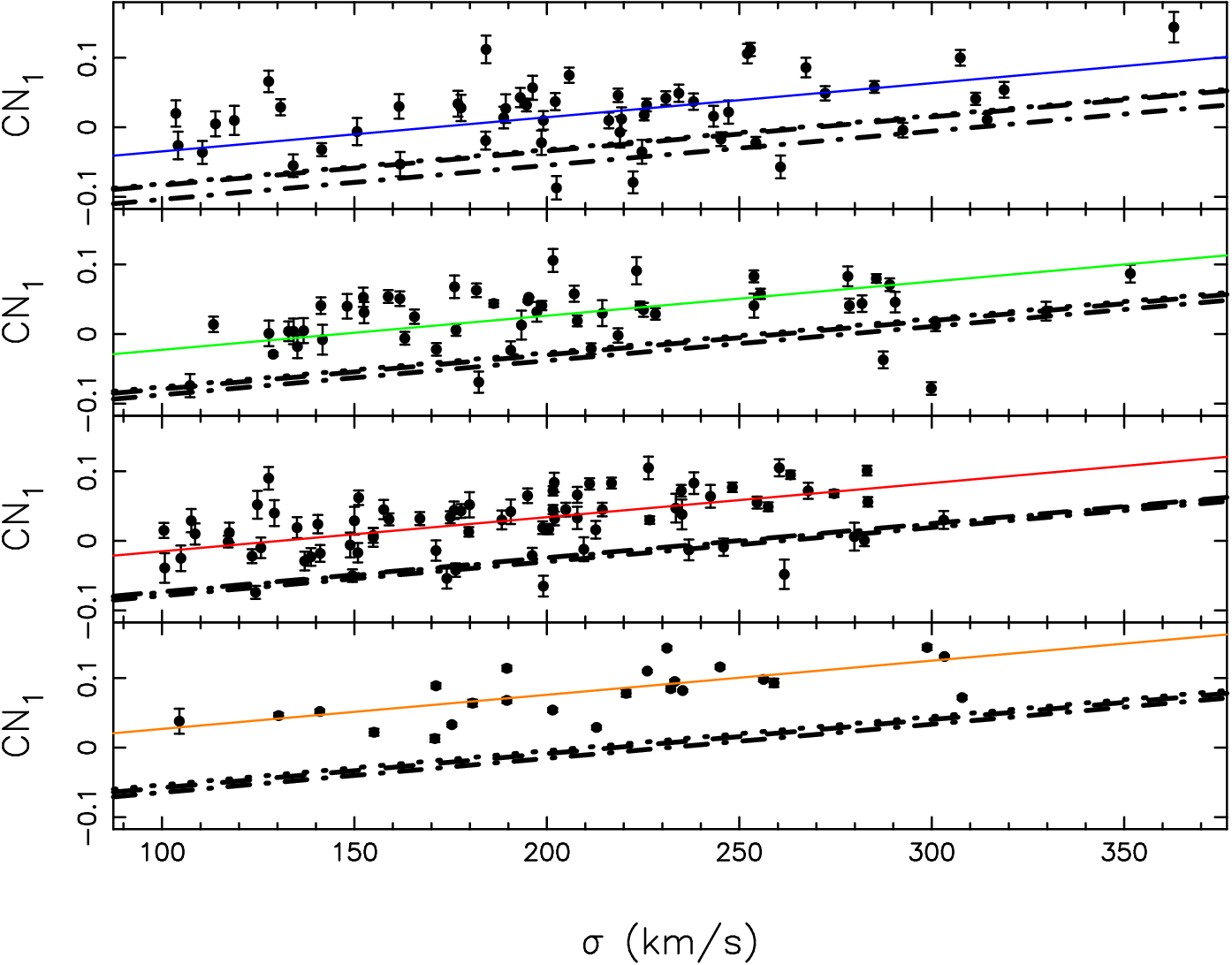}}
\resizebox{0.3\textwidth}{!}{\includegraphics[angle=-90]{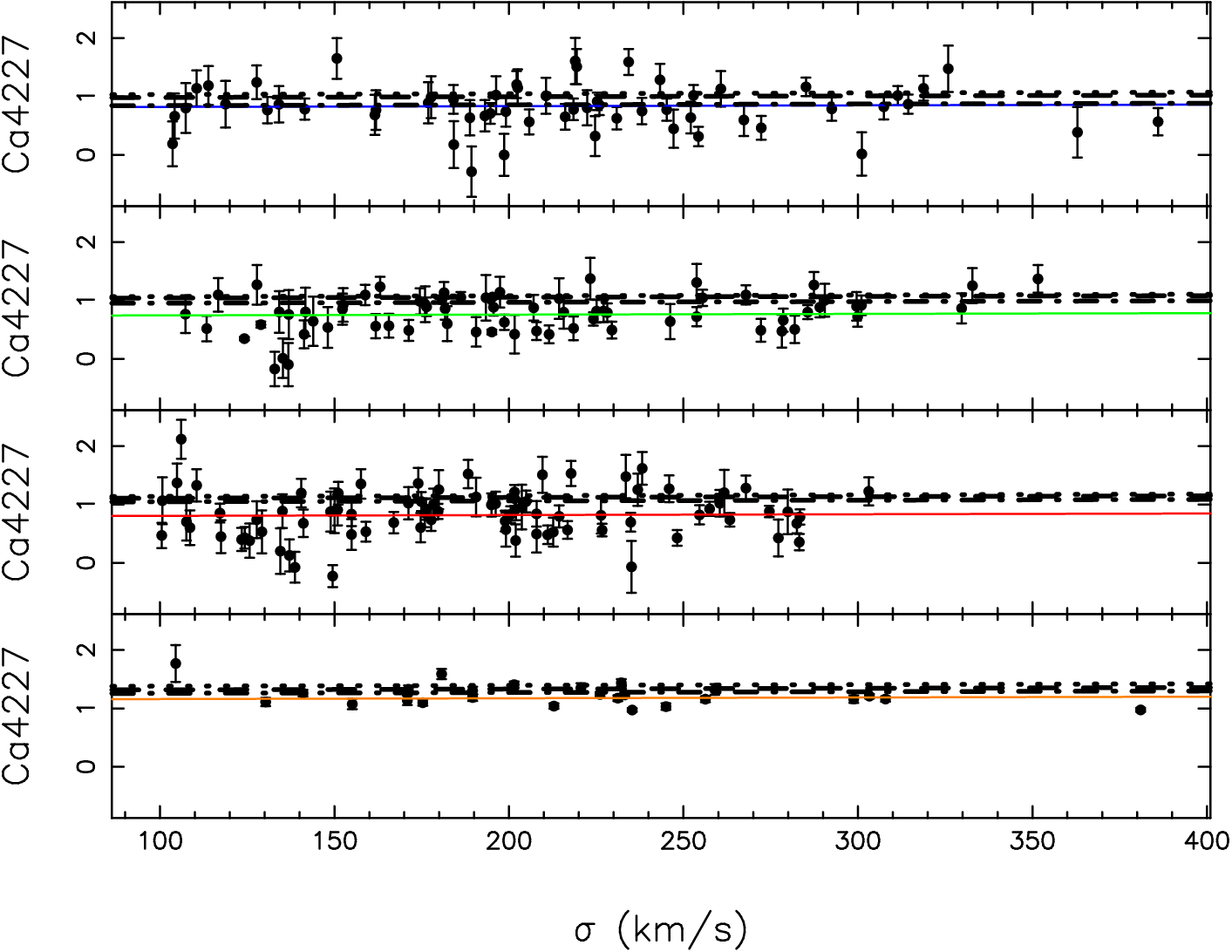}}
\resizebox{0.3\textwidth}{!}{\includegraphics[angle=-90]{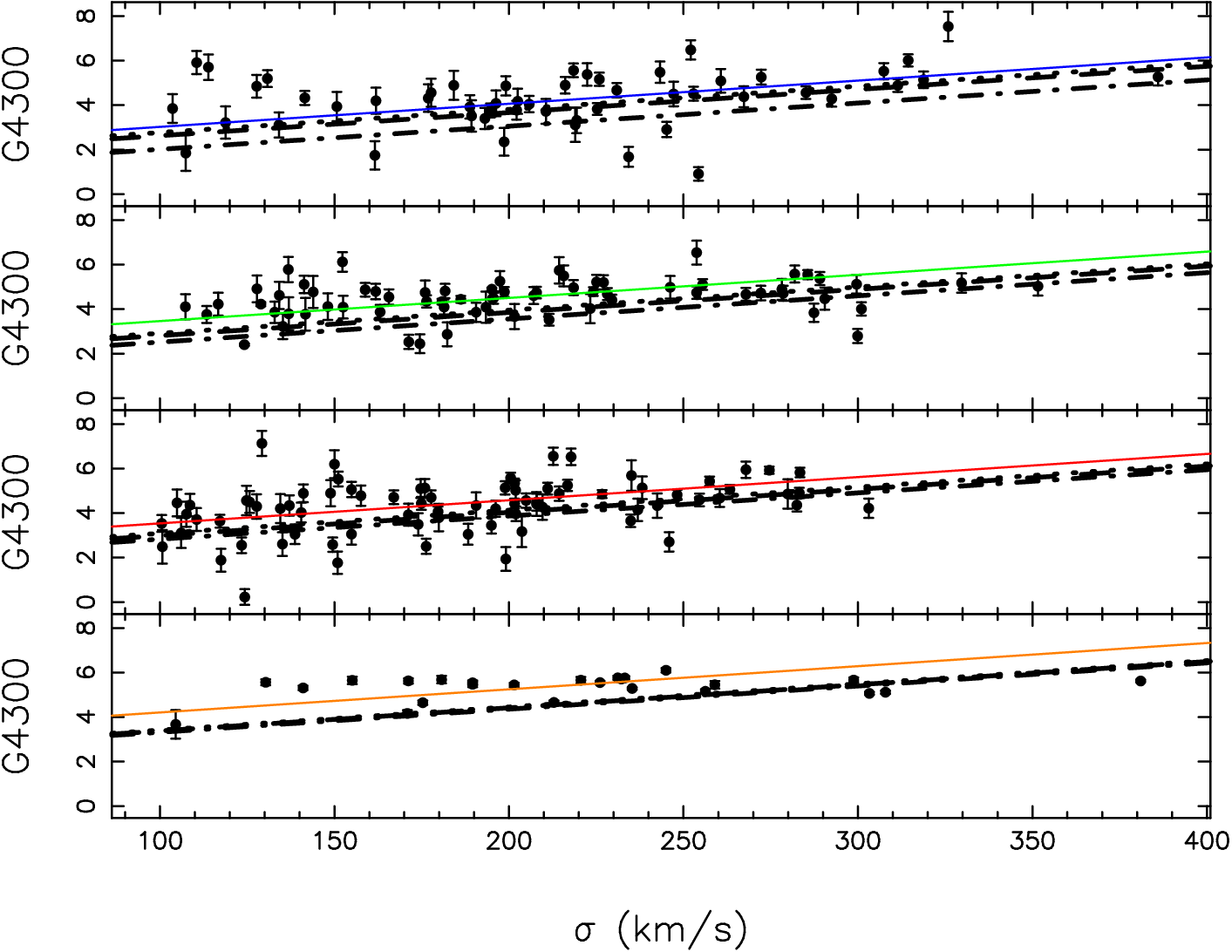}}
\resizebox{0.3\textwidth}{!}{\includegraphics[angle=-90]{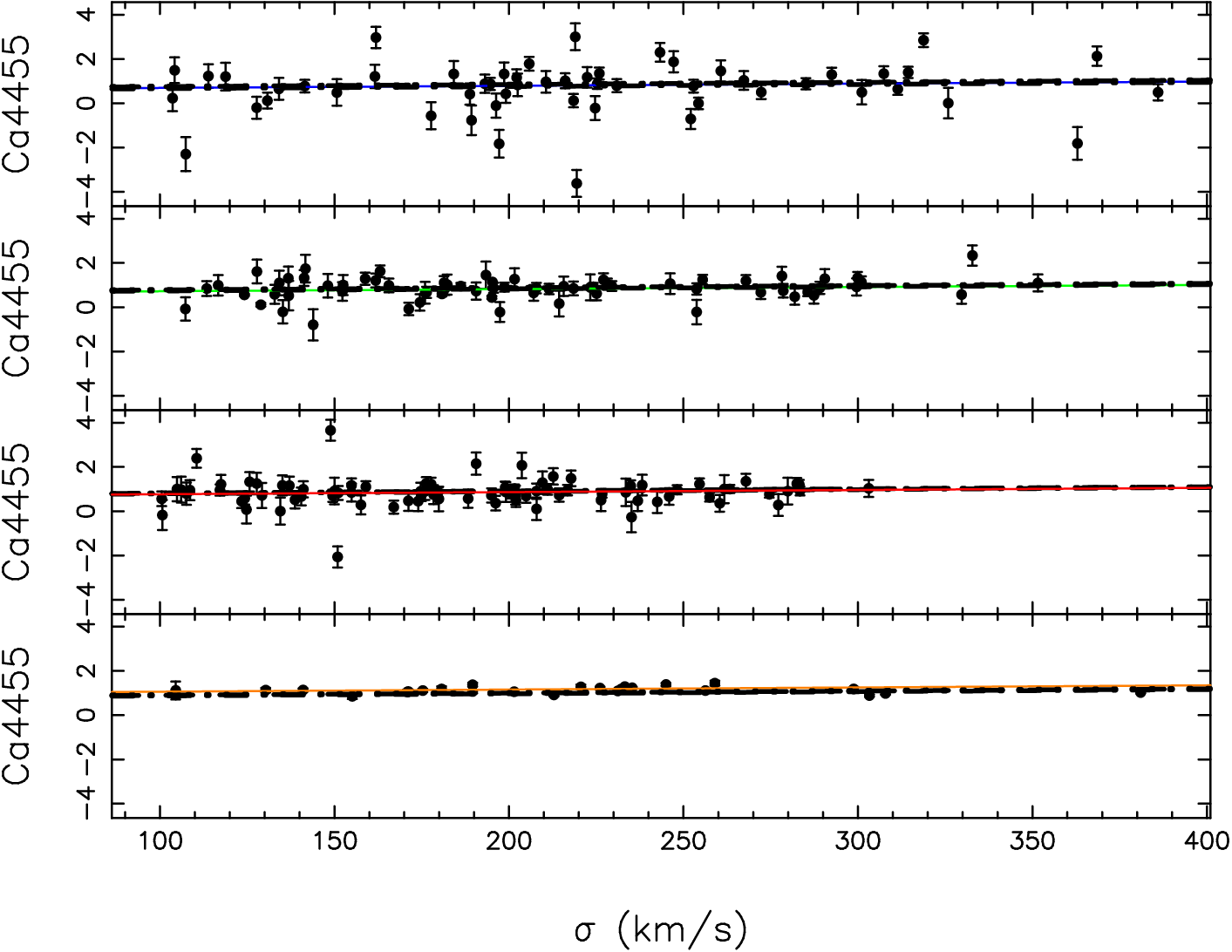}}
\resizebox{0.3\textwidth}{!}{\includegraphics[angle=-90]{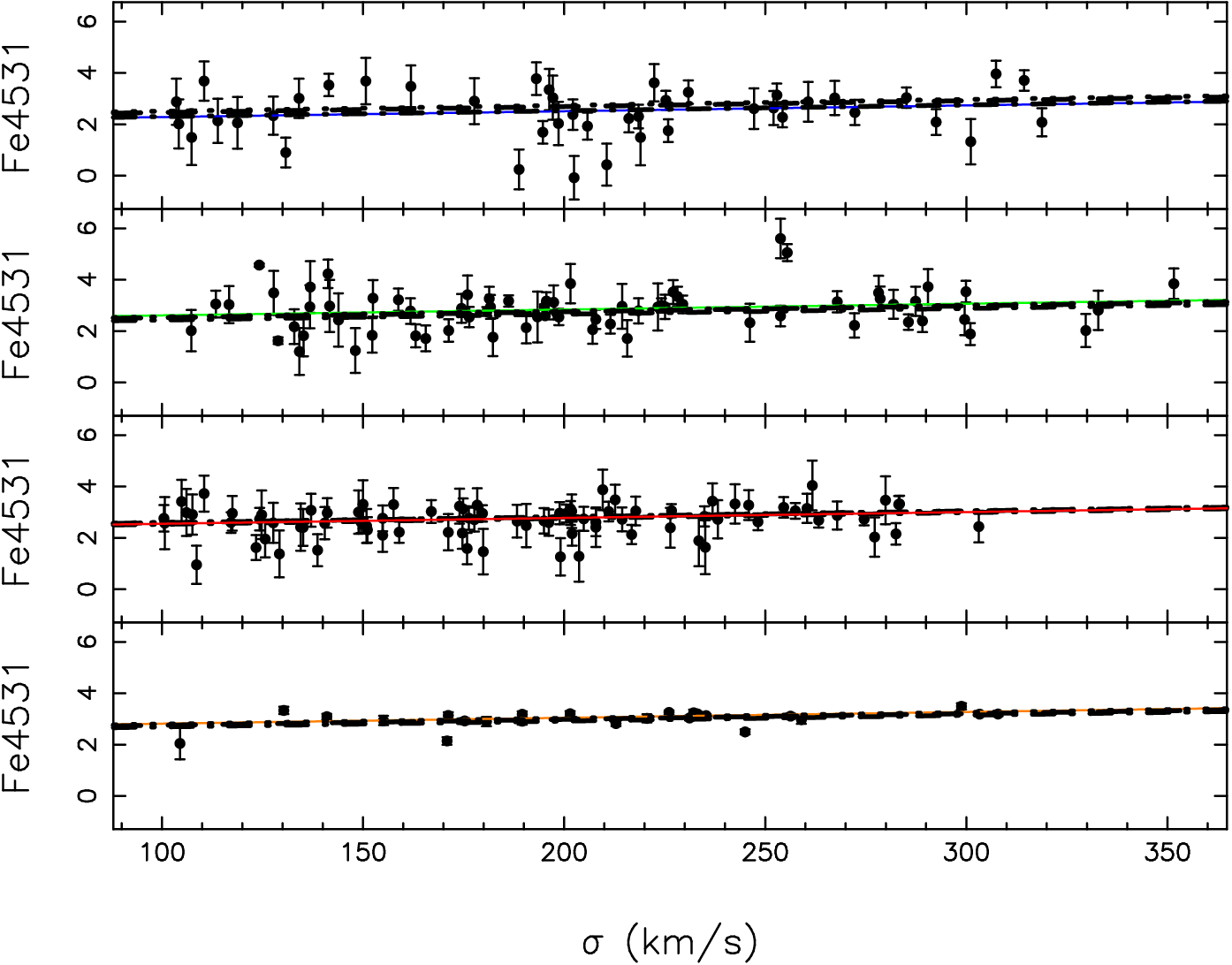}}
\hspace{0.6cm}
\resizebox{0.3\textwidth}{!}{\includegraphics[angle=-90]{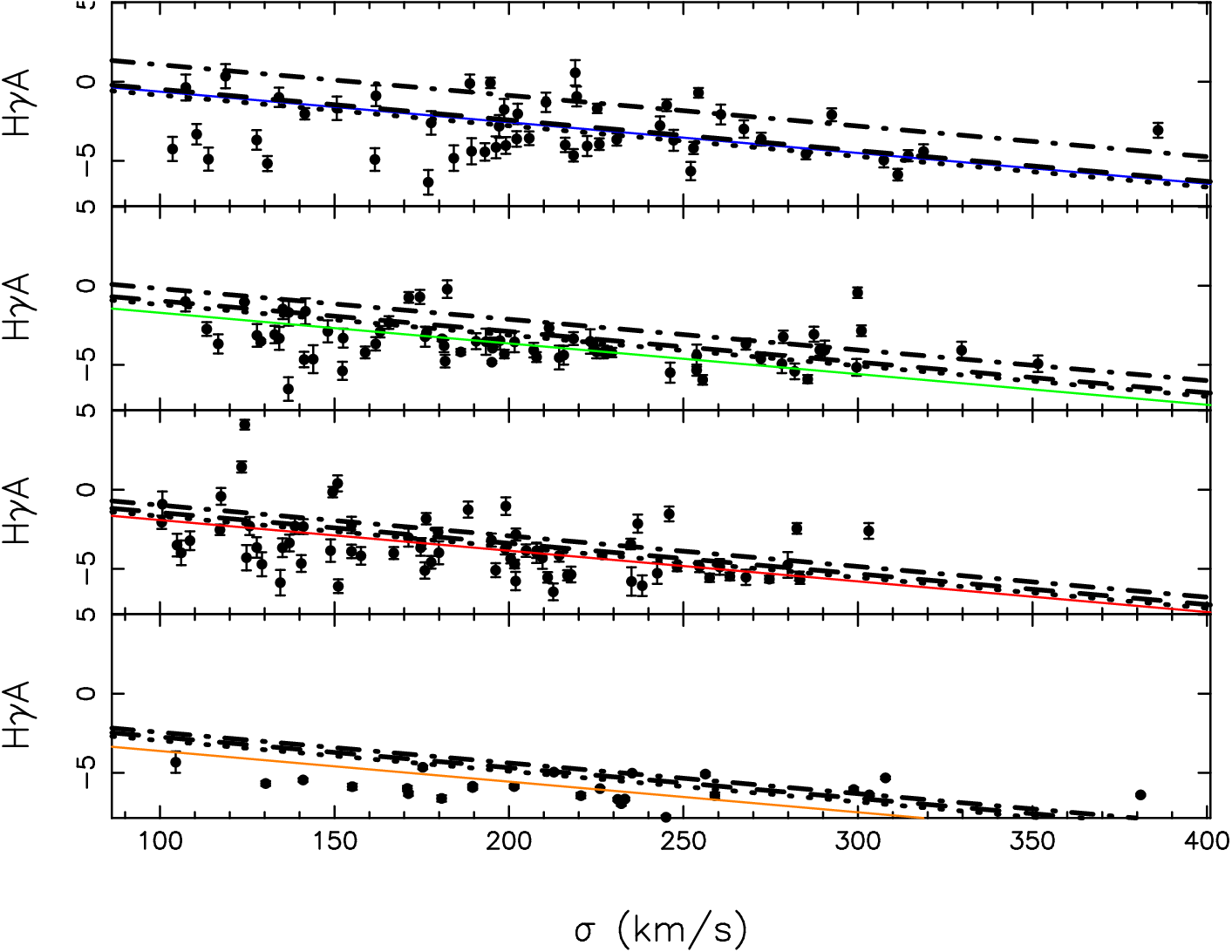}}
\caption{Relation of the indices with the velocity dispersion for the
galaxies of the sample grouped in three different redshift bins. The
meaning of the lines is the same as in Fig.~\ref{indices.sigma}. From
top to bottom: $z=0.75$, $z=0.55$, $z=0.45$, and Coma galaxies are
represented.}
\end{figure*}

\end{appendix}

\end{document}